\documentclass[a4paper,fleqn]{cas-dc}

\usepackage[numbers]{natbib}
\usepackage{booktabs}
\usepackage{tabularx}
\usepackage{array}
\usepackage{lineno}
\usepackage{threeparttable}

\def\tsc#1{\csdef{#1}{\textsc{\lowercase{#1}}\xspace}}
\tsc{WGM}
\tsc{QE}
\tsc{EP}
\tsc{PMS}
\tsc{BEC}
\tsc{DE}

\begin{document}
\let\WriteBookmarks\relax
\def\floatpagepagefraction{1}
\def\textpagefraction{.001}
\shorttitle{Quantifying thermal-signature equivalence in infrared breast thermography}

\shortauthors{R.~Muslim, R.~Fardela, T.~A.~I.~Kusuma}

\title [mode = title]{Quantifying thermal-signature equivalence in infrared breast thermography using a modified Pennes bioheat model}                      

\author[1,2]{Roni Muslim}[orcid=0000-0001-6925-5923]
\cormark[1]
\ead{roni.muslim@apctp.org}

\author[3]{Ramacos Fardela} [orcid=0000-0002-6991-4156]
\ead{ramacosfardela@sci.unand.ac.id}

\author[4]{Tista Artu Indra Kusuma}
\ead{tista@rs-jih.co.id}

\cortext[cor1]{Corresponding author}

\affiliation[1]{
    organization={Asia Pacific Center for Theoretical Physics (APCTP)},
    city={Pohang},
    postcode={37673},
    country={Republic of Korea}
}

\affiliation[2]{
    organization={Research Center for Quantum Physics, National Research and Innovation Agency (BRIN)},
    city={South Tangerang},
    postcode={15314},
    country={Indonesia}
}

\affiliation[3]{
    organization={Department of Physics, Faculty of Mathematics and Natural Science, Universitas Andalas},
    state={Sumatera Barat},
    postcode={25163},
    country={Indonesia}
}

\affiliation[4]{
    organization={JIH Hospital Yogyakarta},
    state={Yogyakarta},
    postcode={55283},
    country={Indonesia}
}

\begin{abstract}
Infrared breast thermography provides a noninvasive measurement of skin-surface temperature, but the relation between surface thermal patterns and intratumoral physiology is limited by heat diffusion and thermal screening. Here, we introduce a tolerance-aware, full-profile framework for determining when different intratumoral perfusion patterns become observationally equivalent at the breast surface. The framework is applied to a steady-state modified Pennes model of a two-dimensional multilayer breast cross-section containing a finite tumor with uniform, rim-enhanced, necrotic-core, or anisotropic perfusion. The equation is discretized using a conservative finite-difference scheme and solved using a biconjugate-gradient stabilized (BiCGSTAB) solver with an incomplete-LU (IncompleteLUT) preconditioner, with mesh refinement and an analytical benchmark used for numerical verification. For the representative configuration, the equivalent-pair fractions are $2/6$, $5/6$, and $6/6$ at observational tolerances of $0.020$, $0.050$, and $0.100~^\circ\mathrm{C}$, respectively. Noise tests further show that the classification depends on how profile-level uncertainty is modeled. After matching the tumor-averaged perfusion, the rim-enhanced--uniform and necrotic-core--uniform distances decrease by $97.7\%$ and $99.0\%$, respectively, whereas anisotropic perfusion remains more distinguishable. This shows that spatial organization can leave a surface signature even when the mean perfusion is fixed. Increasing tumor depth promotes thermal-signature equivalence, whereas increasing tumor diameter enhances surface distinguishability. Sensitivity analysis identifies heat generation as the strongest positive influence on mean profile separation among the four parameter groups examined. For the fixed representative geometry and perfusion-pattern shapes, a 300-point Latin-hypercube analysis using independent uniform $\pm20\%$ perturbations of these parameter groups shows that, at $\varepsilon_T=0.020~^\circ\mathrm{C}$, approximately $79\%$ of the sampled configurations retain at least one thermally equivalent pair. Our results show that a detectable static surface anomaly does not uniquely determine its internal perfusion structure.
\end{abstract}




\maketitle

\section{Introduction}
\label{sec:introduction}

Infrared breast thermography offers a contact-free and
radiation-free way to measure skin-surface temperature, and it has therefore
been explored as a complementary tool for breast assessment
\cite{ng2009review,mashekova2022early}. Several studies have examined its
potential as an adjunct modality for detecting or characterizing breast
abnormalities \cite{goni2024breast,d2024review}. In parallel, thermographic
classification and diagnostic-support methods have been developed using
statistical analysis, machine learning, and image-based features
\cite{nicandro2013evaluation,rassiwala2014evaluation,saniei2016parameter,ryan2025breast}.
Despite this progress, the temperature pattern measured at the
skin surface remains an indirect signal. It can be influenced by
tumor-associated metabolism and blood perfusion, but also by heat diffusion
through the tissue, heat exchange at the boundary, and measurement conditions
\cite{ng2009review,kandlikar2017infrared}.

At the modeling level, the Pennes bioheat equation remains a widely used
baseline model for biological heat-transfer problems
\cite{pennes1948analysis}. It combines heat conduction, blood-mediated heat
exchange, and metabolic heat generation in a continuum formulation that is
simple enough for forward simulations and inverse analyses
\cite{charny1992mathematical,shrivastava2009generic}. The
perfusion term in this equation, however, is an effective volumetric
description of blood flow and does not resolve the detailed vascular network
\cite{wissler1998pennes}. Alternative bioheat models have been proposed to
describe vascular heat transport beyond the classical Pennes approximation
\cite{weinbaum1985new,duck2013physical}. For this reason, we
use a Pennes-type model as a practical continuum model for studying heat
transfer at the tissue scale, while recognizing that it does not describe
microvascular transport explicitly.

{\color{black}
Developments in heat-transfer modeling have also addressed variable
thermal properties, generalized heat-conduction laws, and coupling
between thermal and mechanical fields. Tiwari et
al.~\cite{Tiwari2022VariableProperties} investigated thermoelastic waves
induced by a sinusoidal heat source in a half-space medium with variable
thermal properties. Das et al.~\cite{Das2026MemoryMGT} considered
hydro-thermo-electromechanical responses in a size-dependent porous
piezoelectric medium using a memory-dependent Moore--Gibson--Thompson
formulation, while Seema and
Singhal~\cite{SeemaSinghal2025Rayleigh} examined Rayleigh-type wave
propagation using Green--Naghdi type III, Moore--Gibson--Thompson,
and three-phase-lag theories. These studies provide methodological
context for representing property variation, thermal memory, and
coupled material responses when such effects are relevant to the
material system and thermal excitation. The choice of a generalized
heat-flux law concerns the constitutive description of conduction
and is distinct from the representation of blood-mediated heat
exchange in a Pennes-type heat balance. The present study uses
Fourier heat conduction with prescribed tissue-dependent properties,
metabolic heat generation, and spatially heterogeneous perfusion
to examine stationary surface signatures under controlled conditions.
This formulation supports systematic comparisons of intratumoral
perfusion organization, while constitutive temperature dependence
and thermoelastic coupling would require additional material
relations and parameters.

Hybrid approaches have also connected generalized thermal modeling
with machine learning in biological applications. Singhal et
al.~\cite{Singhal2026SkinFractionalML} investigated a skin-tissue
framework incorporating Atangana--Baleanu and Cattaneo--Vernotte
formulations together with k-nearest-neighbor (KNN) and support
vector machine (SVM) classifiers. Seema et
al.~\cite{Huang2026HybridFractionalML} used simulation-derived
datasets to train KNN, SVM, and convolutional neural network (CNN)
classifiers within a fractional thermoelastic framework. Such
approaches connect physics-based thermal calculations with
data-driven classification. Their predictive performance must be
interpreted in relation to the modeled conditions, training data,
and classification targets; it does not, by itself, establish
whether a measured surface-temperature pattern uniquely determines
the underlying intratumoral perfusion organization. The present
study addresses this complementary question through direct
comparisons of complete surface temperature-rise profiles at a
prescribed observational tolerance. This perspective distinguishes
the prediction or classification of a thermal response from the
identifiability of the internal physiological configuration that
produced it.
}

In breast thermography, the modeling problem is not only to compute the
temperature field inside the tissue. The more relevant question
is how much information about an internal lesion can still be seen in the
temperature profile measured at the skin surface. Numerical bioheat models
have been combined with thermography to study how tumor size, depth, heat
generation, blood perfusion, and tissue properties influence surface
temperature patterns \cite{sudharsan1999surface,ng2001effect}. Related
computational studies have also used inverse heat-transfer methods to estimate
tumor parameters from surface-temperature data
\cite{hatwar2017inverse,gonzalez2020inverse}. More recent work has moved toward
more realistic settings, including three-dimensional breast geometries, surface
scanning, magnetic resonance imaging, and patient-specific thermal models
\cite{lozano2020determining,mukhmetov2021thermal,mukhmetov2021inverse,mukhmetov2021patient}.
These studies show that surface thermograms can be linked to
internal heat-transfer processes, but they also underline the difficulty of
inferring internal tumor properties from surface data alone.

A common simplifying assumption in many bioheat models is that tumor perfusion
is spatially uniform. This assumption is useful for controlled simulations, but
it does not fully reflect the biological complexity of tumor vascularization.
Tumors may exhibit heterogeneous vascular organization, including viable
peripheral regions, hypoxic zones, necrotic cores, and spatially irregular
perfusion patterns
\cite{jain2005normalization,goel2011normalization,li2021hypoxia,chang2017delineation}.
Spatially varying perfusion has therefore been introduced in modified
Pennes-type formulations to represent intratumoral thermal heterogeneity more
realistically \cite{singh2024modified}. This raises a natural
question for infrared thermography: when different perfusion patterns change the
temperature field inside a tumor, are those differences still visible at the
skin surface?

The answer is not obvious, because a surface thermogram is a
filtered projection of the internal temperature field rather than a direct image
of the tumor. Heat generated or redistributed within a lesion must pass through
the surrounding tissue before reaching the surface. Diffusion, perfusion, tissue
layering, and convective heat exchange can smooth internal thermal variations
and reduce their surface contrast
\cite{sudharsan1999surface,ng2001effect,lozano2020determining}. Consequently,
different intratumoral perfusion structures may produce surface temperature
profiles that are very close to one another within a finite observational
tolerance. This non-uniqueness is related to the general difficulty of inferring
internal tumor properties from static surface-temperature data
\cite{gonzalez2020inverse,mukhmetov2021inverse,mukhmetov2021patient,gutierrez2024breast,gutierrez2025detectability}.
Thus, detecting a thermal anomaly and identifying the internal
perfusion structure that produced it are two different problems.

Motivated by this issue, we study a modified Pennes bioheat model for
multilayer breast tissue containing a finite-sized tumor with spatially
heterogeneous intratumoral perfusion. The model is formulated in a
two-dimensional cross-sectional geometry with either an idealized or weakly
deformed outer surface. We compare four idealized perfusion
patterns: uniform, rim-enhanced, necrotic-core, and anisotropic perfusion.
These patterns are motivated by reported features of tumor vascular
heterogeneity, but they are used here as simplified model
cases, not as patient-specific perfusion maps
\cite{jain2005normalization,goel2011normalization,li2021hypoxia,chang2017delineation,singh2024modified}. Our aim is to quantify how differences in intratumoral
perfusion are transmitted to the surface temperature profile, rather than to
propose a clinical diagnostic rule. The potential clinical role of the proposed framework is
therefore interpretive rather than diagnostic. It provides a model-based
means of identifying conditions under which a detectable surface thermal
anomaly may be compatible with more than one internal perfusion
organization. Subject to validation in three-dimensional, patient-specific,
and clinical data, this information could complement adjunct thermographic
screening by indicating when a surface pattern should not be assigned a
unique physiological interpretation and when corroboration by other imaging
modalities is warranted.

To make this comparison quantitative, we introduce a
surface-profile distance for thermal-signature equivalence. Two tumor
configurations are considered thermally equivalent at the surface when the
difference between their surface temperature-rise profiles is smaller than a
prescribed observational tolerance. This tolerance is treated
as a profile-level uncertainty scale, which may include the effects of camera
thermal sensitivity, calibration, environmental variation, preprocessing, and
physiological repeatability
\cite{kandlikar2017infrared,d2024review}. With this definition,
we can ask when different internal perfusion patterns become indistinguishable
from the viewpoint of static infrared thermography.

Using this framework, we examine how tumor depth, tumor size, fat-layer
thickness, and outer-surface deformation affect the distinguishability of
surface thermal signatures. The results are interpreted in terms of thermal
screening by the overlying tissue. The numerical
implementation is verified through mesh-convergence tests and a
one-dimensional analytical transfer-matrix benchmark. The novelty of the
present work lies not in proposing a new governing bioheat equation, but in
introducing a tolerance-aware, full-profile framework for quantifying when
heterogeneous intratumoral perfusion patterns become observationally
equivalent at the breast surface. Unlike conventional Pennes-based studies
that primarily predict surface temperature or estimate tumor parameters, the
present framework distinguishes anomaly detectability from perfusion-pattern
identifiability, separates the effects of tumor-averaged perfusion from those
of its spatial organization through mean-matched controls, and evaluates the
robustness of the inferred similarity across alternative distance metrics,
independent and spatially correlated noise models, parameter sensitivity, and
uncertainty. The resulting framework provides a forward-modeling assessment
of how intratumoral perfusion heterogeneity is attenuated before reaching the
skin surface and explains why a detectable thermal anomaly does not
necessarily identify a unique internal perfusion structure.

\begin{figure*}[pos = tb]
\centering
\includegraphics[width=\textwidth]{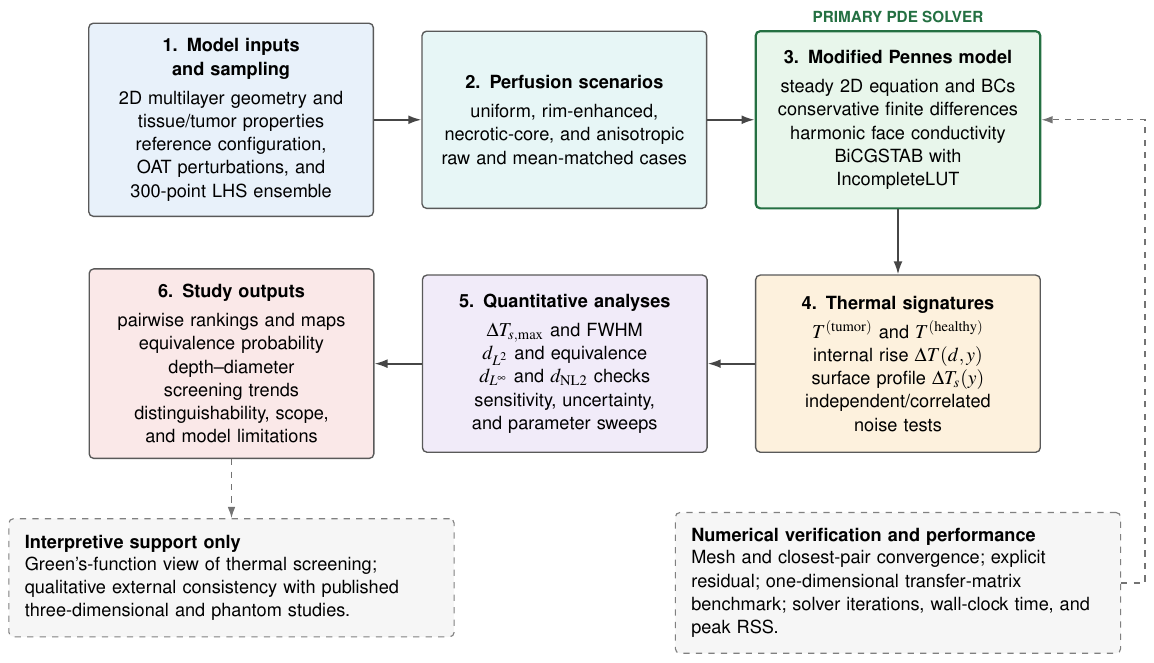}
\caption{Overview of the computational and analytical workflow. The modified Pennes equation is solved using conservative finite differences and the biconjugate-gradient stabilized (BiCGSTAB) method with an incomplete lower--upper factorization using a dual-threshold strategy (IncompleteLUT) as the preconditioner, followed by thermal-signature, robustness, sensitivity, and uncertainty analyses. Dashed boxes denote supporting numerical verification and physical interpretation; the
Green's-function analysis is not used as the primary numerical solver.
Abbreviations used in the workflow are: 2D, two-dimensional; PDE, partial
differential equation; BCs, boundary conditions; OAT, one-at-a-time; LHS,
Latin-hypercube sampling; FWHM, full width at half maximum; NL2,
normalized-$L^2$ distance; and RSS, resident set size.}
\label{fig:fig01_new}
\end{figure*}

The overall computational and analytical workflow is summarized in
Figure~\ref{fig:fig01_new}. The main numerical results are generated
by solving the steady two-dimensional modified Pennes equation using
a conservative finite-difference discretization. The accompanying
analytical developments are used for numerical verification and
physical interpretation rather than as substitutes for the primary
two-dimensional solver.

\section{Mathematical model}
\label{sec:model}

We model the breast as a multilayer tissue domain containing a
finite-sized tumor whose perfusion may vary inside the tumor region. The model
is based on the Pennes bioheat equation, which provides a practical continuum
description of heat conduction, blood-mediated heat exchange, and metabolic heat
generation in biological tissue \cite{pennes1948analysis}. This framework has
been widely used in breast thermal modeling and inverse heat-transfer studies
because it retains the main physical mechanisms relevant to tissue heat transfer
while remaining computationally tractable
\cite{charny1992mathematical,shrivastava2009generic}.

{\color{black}
The Pennes formulation was selected to provide a controlled comparison
of how heterogeneous intratumoral perfusion influences stationary
surface-temperature signatures. It retains heat conduction,
blood-mediated heat exchange, and metabolic heat generation while
allowing the prescribed perfusion patterns and their mean-matched
counterparts to be compared under common geometrical and boundary
conditions. Its parameter requirements and computational cost are
compatible with the repeated forward solutions needed for the
parameter sweeps and Latin-hypercube uncertainty analysis.
The perfusion term represents effective volumetric heat exchange
rather than individual vascular pathways~\cite{wissler1998pennes}.
More anatomically resolved vascular descriptions would require
additional information about vessel geometry, orientation, and
blood flow~\cite{weinbaum1985new}.

The adopted heat balance uses Fourier heat conduction. The choice
of a heat-flux constitutive law is distinct from the representation
of perfusion: non-Fourier formulations can retain Pennes-type
blood-mediated heat exchange while introducing additional
constitutive time scales. Dual-phase-lag bioheat models have,
for example, been applied to tissue heating and hyperthermia
problems~\cite{Chaudhary2023DPL,Ansari2025DPL}.
For the standard Cattaneo--Vernotte relation,
$\mathbf{q}+\tau_q\partial_t\mathbf{q}=-k\nabla T$,
where $\mathbf{q}$ is the conductive heat flux and $\tau_q$ is
its relaxation time, a stationary heat flux recovers the Fourier
relation. The stationary behavior of other generalized formulations
depends on their constitutive structure and couplings. The present
formulation therefore provides a Fourier-based stationary reference
for examining perfusion-induced surface-profile differences;
assessment of constitutive memory effects would require a suitable
thermal protocol and independently justified material parameters.

Generalized thermoelastic and porous piezoelectric formulations
introduce additional coupled fields and constitutive parameters
appropriate to their material systems and loading
conditions~\cite{Das2026MemoryMGT,SeemaSinghal2025Rayleigh}.
Here, the geometry, including the outer-surface deformation,
is prescribed; mechanical displacement, stress, and electrical
fields are not solved. Variable-property
studies~\cite{Tiwari2022VariableProperties} also motivate distinguishing
spatial heterogeneity from constitutive temperature dependence.
The present model assigns fixed thermal conductivities to the
tissue regions within each forward calculation. A relation such
as $k(T)$ could modify even a stationary temperature field and
would require an appropriate material law. The parameter sweeps
and uncertainty analysis vary prescribed coefficients between
calculations and do not represent such temperature-dependent
constitutive behavior.

The computational strategy is likewise distinct from the choice
of physical model. Hybrid fractional thermoelastic--machine-learning
studies combine thermal formulations with classification
methods~\cite{Singhal2026SkinFractionalML,Huang2026HybridFractionalML},
whereas the surface profiles reported here are obtained directly
from conservative finite-difference discretization and a BiCGSTAB
solver with an IncompleteLUT preconditioner. The equivalence
framework compares the resulting profiles at a prescribed
observational tolerance. It could also be applied to profiles
generated by an extended physical model or an independently
assessed surrogate, but the profile distances and equivalence
classifications would need to be evaluated for that implementation.
This separation between the forward model and the observational
comparison preserves the central objective of quantifying which
differences in internal perfusion remain distinguishable at the
surface.
}

\subsection{Geometry and governing bioheat equation}

The computational domain is a two-dimensional cross-section of the breast. We
use a Cartesian coordinate system in which the posterior, or chest-wall, side is
located at $x=0$, while the outer breast surface is represented by
$x=H_{\eta}(y)$. The case $\eta=0$ corresponds to the flat reference geometry,
for which $H_0(y)=L_x$, whereas $\eta>0$ describes a weakly deformed outer
surface. The computational domain is written as
\begin{equation}
\Omega_{\eta}
=
\left\{(x,y): 0\le x\le H_{\eta}(y),\ 0\le y\le L_y\right\}.
\label{eq:domain}
\end{equation}

Although the global coordinate $x$ is measured from the chest-wall side toward
the outer surface, the tissue layers are assigned using the local depth measured
inward from the outer surface, $d(x,y)=H_{\eta}(y)-x$. Thus, the skin occupies
$0\le d<\delta_s$, the fat layer occupies
$\delta_s\le d<\delta_s+\delta_f$, the glandular layer occupies
$\delta_s+\delta_f\le d<\delta_s+\delta_f+\delta_g$, and the remaining
posterior region is treated as muscle. Such multilayer representations are
commonly used in breast bioheat modeling because tissue thickness and
thermophysical properties influence the transmission of heat from internal
sources to the surface \cite{sudharsan1999surface,ng2001effect}. Recent studies
have also emphasized the role of breast geometry and tissue-specific properties
in surface-temperature modeling
\cite{lozano2020determining,mukhmetov2021thermal}.

The tumor is modeled as a finite circular inclusion,
\begin{equation}
\Omega_t
=
\left\{(x,y)\in\Omega_{\eta}:
(x-x_t)^2+(y-y_t)^2\le R_t^2
\right\},
\label{eq:tumor_domain}
\end{equation}
where $(x_t,y_t)$ is the tumor center and $R_t$ is the tumor radius. The
tumor-center depth measured from the local outer surface is
$d_t=H_{\eta}(y_t)-x_t$. In the ideal flat geometry, $H_0(y)=L_x$, so that
$d_t=L_x-x_t$. {\color{black}
The Cartesian two-dimensional geometry provides a controlled
cross-sectional setting for comparing the surface signatures of
different intratumoral perfusion patterns. The formulation resolves
heat conduction in the $x$ and $y$ directions and assumes invariance
in the out-of-plane direction; the circular inclusion has finite
extent within the modeled plane.
}

The temperature field $T(\mathbf{r})$, with $\mathbf{r}=(x,y)$, satisfies the
steady-state modified Pennes equation
\begin{equation}
\nabla\!\cdot\!\bigl(k(\mathbf{r})\nabla T(\mathbf{r})\bigr)
-
\rho_b c_b\,\omega(\mathbf{r})
\bigl(T(\mathbf{r})-T_a\bigr)
+
Q_m(\mathbf{r})
=
0,
\quad
\mathbf{r}\in\Omega_{\eta}.
\label{eq:pennes}
\end{equation}
Here, $k(\mathbf{r})$ is the thermal conductivity, $\omega(\mathbf{r})$ is the
blood perfusion rate, and $Q_m(\mathbf{r})$ is the metabolic heat-generation
rate. The parameters $\rho_b$, $c_b$, and $T_a$ denote the blood density, blood
specific heat, and arterial blood temperature, respectively. In the healthy
background, the material parameters are piecewise constant in each layer:
$(k_s,\omega_s,Q_s)$ for skin, $(k_f,\omega_f,Q_f)$ for fat,
$(k_g,\omega_g,Q_g)$ for glandular tissue, and
$(k_m,\omega_m,Q_m^{(h)})$ for muscle, where the superscript $(h)$ denotes the
healthy-muscle value and avoids confusion with the field $Q_m(\mathbf{r})$.

Inside the tumor, the thermal conductivity and metabolic heat generation are
taken as $k_t$ and $Q_t$, while the perfusion field is allowed to vary
spatially. Thus, in $\Omega_t$ we set
$k(\mathbf{r})=k_t$, $Q_m(\mathbf{r})=Q_t$, and
$\omega(\mathbf{r})=\omega_t(\mathbf{r})$. The tumor region therefore satisfies
\begin{equation}
\nabla\!\cdot\!\bigl(k_t\nabla T(\mathbf{r})\bigr)
-
\rho_b c_b\,\omega_t(\mathbf{r})
\bigl(T(\mathbf{r})-T_a\bigr)
+
Q_t
=
0,
\quad
\mathbf{r}\in\Omega_t.
\label{eq:tumor_pennes}
\end{equation}
Here the word ``modified'' simply refers to the spatially
varying tumor perfusion field within the Pennes-type equation. The governing
equation itself remains Pennes-based; the modification is introduced to test how
different intratumoral perfusion patterns are transmitted to the surface
temperature profile.

\subsection{\label{sec2.2}Idealized intratumoral perfusion scenarios}

Tumor vascularization can be spatially heterogeneous, with viable peripheral
regions, hypoxic zones, necrotic cores, and irregular vascular organization
\cite{jain2005normalization,goel2011normalization,li2021hypoxia,chang2017delineation}.
Spatially varying perfusion has also been introduced in modified Pennes-type
formulations to relax the assumption of spatially uniform blood perfusion
\cite{singh2024modified}. Based on these features, we compare
four idealized perfusion cases. The purpose is to represent broad forms of
intratumoral heterogeneity and to test whether their thermal effects remain
visible at the skin surface.

The reference case is uniform perfusion,
$\omega_t(\mathbf{r})=\omega_0$, where $\omega_0$ is taken from the
corresponding representative tumor parameter set. To describe spatial
heterogeneity, we define the distance from the tumor center as
$\rho=\sqrt{(x-x_t)^2+(y-y_t)^2}$, with $0\le \rho\le R_t$. Here,
$\mathbf{r}=(x,y)$ denotes the position vector and $(x_t,y_t)$ is the tumor
center. Rim-enhanced perfusion is represented by
\begin{equation}
\omega_t(\rho) = 
\omega_0
\left[
1+\alpha
\left(\frac{\rho}{R_t}\right)^m
\right],
\label{eq:omega_radial}
\end{equation}
where $\alpha$ controls the amplitude of peripheral enhancement and $m$ controls
the radial sharpness. We use $\alpha=1.3$ and $m=2$, which produce a smooth
increase from the tumor center toward the boundary without introducing an abrupt
discontinuity.

A necrotic-core scenario is represented by a piecewise profile with reduced
perfusion in the central region and higher perfusion in the outer region:
$\omega_t=\omega_c$ for $0\le \rho<r_n$ and
$\omega_t=\omega_r$ for $r_n\le \rho\le R_t$. We set $r_n/R_t=0.45$, so that the
low-perfusion core occupies a finite but not dominant fraction of the tumor
cross-section. Directional asymmetry is represented by
\begin{equation}
\omega_t(\rho,\theta) = 
\omega_0(1+\epsilon\cos\theta),
\label{eq:omega_anisotropic}
\end{equation}
where $\theta$ is the polar angle measured around the tumor center. With
$\epsilon=0.65$, the perfusion ratio satisfies
$\omega_t/\omega_0\in[0.35,1.65]$, introducing a moderate angular contrast while
keeping the perfusion positive throughout the tumor.

The profiles defined above are used as raw, or absolute-perfusion, scenarios.
Thus, both the spatial organization and the tumor-averaged perfusion may differ
among cases. For example, in a two-dimensional circular tumor, the raw
rim-enhanced and necrotic-core profiles have tumor-averaged perfusions
\begin{align}
\left\langle \omega_t^{\mathrm{rim}} \right\rangle_{\Omega_t}
&=
\omega_0
\left(
1+\frac{2\alpha}{m+2}
\right), \\
\left\langle \omega_t^{\mathrm{nec}} \right\rangle_{\Omega_t}
&=
\left(\frac{r_n}{R_t}\right)^2\omega_c
+
\left[
1-\left(\frac{r_n}{R_t}\right)^2
\right]\omega_r .
\label{eq:raw_mean_perfusion_examples}
\end{align}
For the anisotropic profile in
Eq.~\eqref{eq:omega_anisotropic}, the angular average of $\cos\theta$
vanishes over the circular tumor domain, giving
$\langle\omega_t^{\mathrm{ani}}\rangle_{\Omega_t}=\omega_0$.
Thus, the anisotropic profile already satisfies the mean-matching
condition analytically and has the same tumor-averaged perfusion as the
uniform case before rescaling.

By contrast, the raw rim-enhanced and necrotic-core
profiles generally have tumor-averaged perfusion values different from
$\omega_0$. Consequently, differences among the raw surface signatures
can reflect both the spatial organization of perfusion and differences
in its tumor-averaged magnitude. To separate these contributions, we
performed a mean-matched control analysis. For each heterogeneous
perfusion field, the raw perfusion profile
$\omega_t^{\mathrm{raw}}(\mathbf{r})$ was rescaled as
\begin{equation}
\omega_t^{\mathrm{mm}}(\mathbf{r})
=
\omega_0
\frac{
\omega_t^{\mathrm{raw}}(\mathbf{r})
}{
\left\langle\omega_t^{\mathrm{raw}}\right\rangle_{\Omega_t}
},
\label{eq:mean_matched_perfusion}
\end{equation}
where
\begin{equation}
\left\langle\omega_t^{\mathrm{raw}}\right\rangle_{\Omega_t}
=
\frac{1}{|\Omega_t|}
\int_{\Omega_t}
\omega_t^{\mathrm{raw}}(\mathbf{r})\,dA.
\label{eq:raw_mean_perfusion}
\end{equation}
This normalization enforces
$\langle\omega_t^{\mathrm{mm}}\rangle_{\Omega_t}=\omega_0$ for every
perfusion class while preserving the spatial form of each profile.
For the anisotropic case, the numerical scaling factor differs from
unity only through negligible discretization effects associated with
representing the circular tumor on a finite grid. The mean-matched
cases therefore provide a control that isolates the influence of
spatial perfusion organization from that of the tumor-averaged
perfusion magnitude. Their effects on the surface thermal signatures
are examined in Figure~\ref{fig:fig06_new}.

The numerical values used for the raw perfusion scenarios are summarized in
Table~\ref{tab:perfusion_profiles}. The baseline value $\omega_0$ is taken from the corresponding representative parameter set, whereas $\alpha$, $m$,
$r_n/R_t$, and $\epsilon$ are dimensionless shape-control parameters. Together, these choices represent four simplified classes of
intratumoral perfusion: nearly uniform perfusion, peripheral enhancement,
central hypoperfusion, and directional asymmetry. They should be read as model
parameters for sensitivity analysis, not as direct measurements for a particular tumor. A fully data-constrained version of the model would require independent perfusion information, for example from dynamic contrast-enhanced magnetic resonance imaging (DCE-MRI), histology, or patient-specific vascular imaging.

\begin{table*}[pos = tbp]
\centering
\caption{Idealized intratumoral perfusion scenarios used in the simulations.}
\label{tab:perfusion_profiles}

\begin{threeparttable}
\small
\setlength{\tabcolsep}{5pt}
\renewcommand{\arraystretch}{1.18}

\begin{tabularx}{\linewidth}{
@{}
p{0.18\linewidth}
p{0.30\linewidth}
>{\raggedright\arraybackslash}X
@{}
}
\toprule
Perfusion class & Functional form & Parameter choice and interpretation \\
\midrule

Uniform &
$\omega_t=\omega_0$ &
Reference case with spatially uniform tumor perfusion. The baseline value is
$\omega_0=0.01600~\mathrm{s}^{-1}$ for the Khomsi-based set and
$\omega_0=1.114\times10^{-2}~\mathrm{s}^{-1}$ for the Lozano-inspired set.
\\

Rim-enhanced &
$\omega_t(\rho)=\omega_0\left[1+\alpha(\rho/R_t)^m\right]$ &
$\alpha=1.3$ and $m=2$. These values generate a smooth increase toward the
tumor boundary, representing peripheral perfusion enhancement without an
abrupt discontinuity.
\\

Necrotic-core &
$\omega_t=\omega_c$ for $\rho<r_n$; 
$\omega_t=\omega_r$ for $r_n\leq \rho\leq R_t$ &
$r_n/R_t=0.45$. For the Khomsi-based set,
$\omega_c=0.0020~\mathrm{s}^{-1}$ and
$\omega_r=0.0250~\mathrm{s}^{-1}$. For the Lozano-inspired set,
$\omega_c=7.356\times10^{-3}~\mathrm{s}^{-1}$ and
$\omega_r=1.114\times10^{-2}~\mathrm{s}^{-1}$. This case represents central
hypoperfusion with a more perfused outer region.
\\

Anisotropic &
$\omega_t(\rho,\theta)=\omega_0(1+\epsilon\cos\theta)$ &
$\epsilon=0.65$, giving $\omega_t/\omega_0\in[0.35,1.65]$. This case introduces
directional perfusion asymmetry while keeping $\omega_t>0$ throughout the tumor.
\\

\bottomrule
\end{tabularx}

\begin{tablenotes}[flushleft]
\footnotesize
\item \textit{Note.} The listed profiles define the raw, or absolute-perfusion,
cases. They are idealized perfusion patterns used for model comparison, not
patient-specific perfusion maps. The baseline value $\omega_0$ is taken from
the corresponding representative parameter set, whereas $\alpha$, $m$,
$r_n/R_t$, and $\epsilon$ are dimensionless shape-control parameters. The
mean-matched control cases in Figure~\ref{fig:fig06_new} are obtained
by rescaling these raw fields according to
Eq.~\eqref{eq:mean_matched_perfusion}, so that all perfusion classes have the
same tumor-averaged perfusion,
$\langle\omega_t\rangle_{\Omega_t}=\omega_0$.
\end{tablenotes}

\end{threeparttable}
\end{table*}

\subsection{Boundary conditions and surface thermal signature}

The boundary conditions represent heat exchange with the environment at the
outer surface and thermal coupling to the body core at the posterior side. On
the outer surface
$\Gamma_{\mathrm{surf}}=\{(H_{\eta}(y),y)\}$, we impose the Robin condition
\begin{equation}
-k(\mathbf{r})
\frac{\partial T}{\partial n}
=
h(T-T_\infty),
\qquad
\mathbf{r}\in\Gamma_{\mathrm{surf}},
\label{eq:bc_surface}
\end{equation}
where $h$ is the convective heat-transfer coefficient, $T_\infty$ is the
ambient temperature, and $\mathbf{n}$ is the outward unit normal vector. On the
posterior, or chest-wall, side
$\Gamma_{\mathrm{cw}}=\{(0,y)\}$, we impose the prescribed core temperature,
$T=T_{\mathrm{core}}$. The lateral boundaries are treated as insulated,
$\partial T/\partial n=0$ on $\Gamma_{\mathrm{lat}}$. Across tissue interfaces
and at the tumor boundary, temperature and normal heat flux are assumed to be
continuous.

Infrared thermography measures the temperature at the outer
surface. In the present model, this surface temperature is
$T_s(y)=T(H_{\eta}(y),y)$. To isolate the tumor-induced contribution from the
healthy background, we define
\begin{equation}
\Delta T_s(y)
=
T_s^{(\mathrm{tumor})}(y)
-
T_s^{(\mathrm{healthy})}(y).
\label{eq:deltaTs}
\end{equation}
The profile $\Delta T_s(y)$ is used as the surface thermal signature. It allows
tumor configurations with the same size and location, but different
intratumoral perfusion patterns, to be compared directly at the surface. This
focus on the skin-temperature profile is consistent with thermography-based and
inverse thermal modeling studies, where surface temperature provides the primary
observable \cite{hatwar2017inverse,lozano2020determining,mukhmetov2021inverse}.
In this study, the profile is used to compare surface
distinguishability among perfusion patterns, not to assign a unique clinical
interpretation to a measured thermogram.

\begin{figure}[pos =tb]
    \centering
    \includegraphics[width=\linewidth]{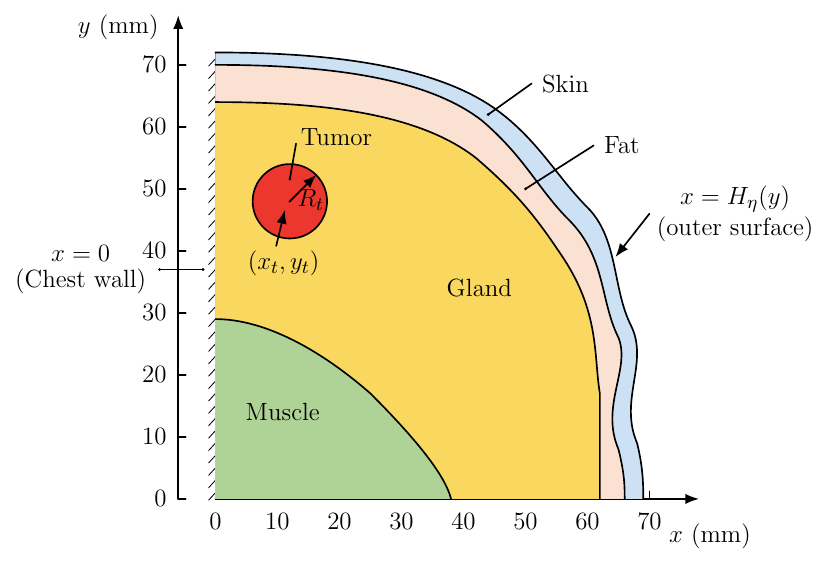}
    \caption{
    Schematic geometry of the multilayer breast model used in the modified
    Pennes bioheat formulation. The global coordinate $x$ is measured from the
    posterior, or chest-wall, side toward the outer surface. Thus, the chest wall
    is located at $x=0$, while the outer surface is represented by
    $x=H_{\eta}(y)$. The case $\eta=0$ corresponds to the flat geometry,
    $H_0(y)=L_x$, whereas $\eta>0$ represents a smooth surface deformation. The
    healthy tissue is divided into skin, fat, glandular, and muscle layers
    according to the local depth measured inward from the outer surface,
    $d(x,y)=H_{\eta}(y)-x$. The tumor is modeled as a circular inclusion with
    center $(x_t,y_t)$ and radius $R_t$. The tumor-center depth is
    $d_t=H_{\eta}(y_t)-x_t$.
    }
    \label{fig:fig02_new}
\end{figure}

The governing bioheat equation is solved numerically on the
two-dimensional domain using the prescribed material properties,
perfusion fields, heat-generation rates, and boundary conditions. The
discretization is chosen to resolve tissue interfaces, the tumor boundary, and
the surface-temperature profile. A mesh-convergence test is performed because
the central quantities in this study are small differences between surface
profiles. A schematic illustration of the model geometry, tumor location, and
boundary conditions is shown in Figure~\ref{fig:fig02_new}.

\section{Surface thermal signatures and numerical setup}
\label{sec:signatures_numerics}
Using the stationary surface temperature-rise profile
$\Delta T_s(y)$ defined in Eq.~\eqref{eq:deltaTs}, we compare
tumor configurations with the same geometry, size, and location
but different intratumoral perfusion patterns. The following
subsections define the surface descriptors, profile-distance
metrics, and observational equivalence criterion, then describe
the numerical implementation, verification procedures, and
parameter-sensitivity and uncertainty analyses.

\subsection{Surface descriptors and thermal-signature equivalence}

To describe the surface anomaly, we first use several simple thermal
descriptors: the maximum temperature rise
$\Delta T_{s,\max}=\max_y \Delta T_s(y)$, the hotspot position
$y_{\max}=\arg\max_y \Delta T_s(y)$, the full width at half maximum (FWHM), and
the hotspot centroid
\begin{equation}
y_c =
\dfrac{
\displaystyle \int_{\Gamma_{\mathrm{surf}}} y\,\Delta T_s(y)\,dy
}{
\displaystyle \int_{\Gamma_{\mathrm{surf}}} \Delta T_s(y)\,dy
}.
\label{eq:centroid_main}
\end{equation}
These quantities characterize the amplitude, position, lateral spread, and
effective center of the surface thermal anomaly. Similar surface-based
descriptors have been used in bioheat and inverse thermography studies to
relate skin-temperature patterns to internal heat-source properties such as
tumor size, depth, and location
\cite{sudharsan1999surface,ng2001effect,das2015simultaneous}.

Although these scalar descriptors are useful, they do not fully represent the
shape of the surface thermal signature. Therefore, the main comparison in this
work is based on the full profile $\Delta T_s(y)$. For two tumor configurations
$\mathcal{C}_1$ and $\mathcal{C}_2$, we define the profile distance
\begin{equation}
d_{L^2}(\mathcal{C}_1,\mathcal{C}_2)
=
\left[
\frac{1}{L_\Gamma}
\int_{\Gamma_{\mathrm{surf}}}
\left\{
\Delta T_s(y;\mathcal{C}_1)
-
\Delta T_s(y;\mathcal{C}_2)
\right\}^2
dy
\right]^{1/2},
\label{eq:L2_main}
\end{equation}
where $L_\Gamma$ is the length of the surface interval used for comparison. Smaller values of $d_{L^2}$ indicate more similar surface
temperature-rise profiles. Because the metric uses the full profile rather than
only the peak temperature, it captures differences in both amplitude and shape.

To examine whether the inferred similarity ordering depends on the
choice of metric, we additionally evaluate the maximum pointwise
distance
\begin{equation}
d_{L^\infty}(\mathcal{C}_i,\mathcal{C}_j)
=
\max_{y\in\Gamma_{\mathrm{surf}}}
\left|
\Delta T_s(y;\mathcal{C}_i)
-
\Delta T_s(y;\mathcal{C}_j)
\right|,
\label{eq:dLinf}
\end{equation}
and the normalized $L^2$ distance
\begin{equation}
d_{\mathrm{NL2}}(\mathcal{C}_i,\mathcal{C}_j)
=
\frac{
\left\|
\Delta T_s^{(i)}-\Delta T_s^{(j)}
\right\|_2
}{
\frac{1}{2}
\left(
\left\|\Delta T_s^{(i)}\right\|_2+
\left\|\Delta T_s^{(j)}\right\|_2
\right)
}.
\label{eq:dNL2}
\end{equation}
Here, $d_{L^\infty}$ measures the largest local difference between
the two surface profiles and has units of temperature, whereas
$d_{\mathrm{NL2}}$ is dimensionless and measures their difference
relative to their average profile magnitude. These two measures are
used as robustness checks for the pairwise similarity rankings in
Table~\ref{tab:metric_robustness}. Unless otherwise stated, the
thermal-equivalence classification in the remainder of this study is
based on $d_{L^2}$.

We define two tumor configurations as thermally equivalent at
the surface when their profile distance is smaller than a prescribed
observational tolerance,
\begin{equation}
d_{L^2}(\mathcal{C}_1,\mathcal{C}_2) \le \varepsilon_T,
\label{eq:equiv_main}
\end{equation}
where $\varepsilon_T$ is the tolerance for the difference between two
surface-temperature-rise profiles. This tolerance is not a
universal clinical threshold and is not the noise-equivalent temperature
difference of a single detector pixel. Instead, it is used as an effective
profile-level uncertainty for the processed surface-temperature-rise profile.
It represents, in a lumped way, the combined influence of detector thermal
sensitivity, calibration uncertainty, surface-emissivity assumptions, camera
distance and viewing conditions, environmental fluctuations, preprocessing
choices, and physiological repeatability.

Modern medical infrared systems may reach thermal sensitivity scales of order
$0.01$--$0.05~^\circ\mathrm{C}$ depending on whether cooled or uncooled
detectors are used \cite{lahiri2012medical}, whereas a more conservative
practical scale of order $0.1~^\circ\mathrm{C}$ has also been discussed in the
context of breast thermography \cite{kandlikar2017infrared}. Therefore, in this
work we use $\varepsilon_T=0.020$, $0.050$, and
$0.100~^\circ\mathrm{C}$ as representative bracketing tolerances, corresponding
respectively to an optimistic laboratory-scale tolerance, a typical
high-sensitivity infrared-camera scale, and a conservative practical tolerance.

Since $d_{L^2}$ compares the difference between two profiles rather than the
absolute temperature at a single point, $\varepsilon_T$ is interpreted as an
effective root-mean-square tolerance for profile differences. If each processed
profile has an effective RMS uncertainty $\sigma_T$ and the two profile
uncertainties are approximately independent, then the uncertainty in their
difference scales as $\sigma_\Delta\simeq\sqrt{2}\sigma_T$. Thus, the three
values used above span single-profile uncertainty scales of approximately
$0.014$, $0.035$, and $0.071~^\circ\mathrm{C}$. The
classification is therefore evaluated across a range of plausible measurement
scales, rather than being tied to a single fixed threshold.

To make the observational meaning of $\varepsilon_T$ explicit, we also consider
a noise-perturbed surface profile of the form
\begin{equation}
\Delta T_s^{\mathrm{obs}}(y) = \Delta T_s(y)+\xi(y),
\label{eq:observed_profile_noise}
\end{equation}
where $\xi(y)$ represents the combined uncertainty in the processed
temperature-rise profile. Two limiting noise models are used to test the
robustness of the equivalence classification. The first is an independent
Gaussian noise model,
\begin{equation}
\xi(y_j) \sim \mathcal{N}(0,\sigma_T^2),
\label{eq:independent_noise}
\end{equation}
where $y_j$ denotes a discrete surface point. This model approximates
uncorrelated pointwise temperature uncertainty in the processed profile. The
second is a spatially correlated Gaussian noise model with covariance
\begin{equation}
\left\langle \xi(y)\xi(y') \right\rangle =
\sigma_T^2
\exp\left[
-\frac{(y-y')^2}{2\ell_c^2}
\right],
\label{eq:correlated_noise}
\end{equation}
where $\ell_c$ is the correlation length. This model represents slowly varying
profile-level perturbations caused, for example, by calibration drift, residual
background variation, imperfect surface correction, physiological variability,
or preprocessing-induced smoothing. The two models therefore
represent two limiting forms of profile uncertainty: pointwise random
fluctuations and spatially coherent distortions.

For each noise model, the equivalence classification can be repeated over
$N_{\mathrm{MC}}$ Monte Carlo realizations. If
$d_{L^2}^{(k),\mathrm{obs}}(\mathcal{C}_i,\mathcal{C}_j)$ denotes the profile
distance obtained from the $k$th noisy realization, the corresponding
probability of equivalence is defined as
\begin{equation}
P_{\mathrm{eq}}(\mathcal{C}_i,\mathcal{C}_j)
=
\frac{1}{N_{\mathrm{MC}}}
\sum_{k=1}^{N_{\mathrm{MC}}}
I\left[
d_{L^2}^{(k),\mathrm{obs}}(\mathcal{C}_i,\mathcal{C}_j)
\le
\varepsilon_T
\right],
\label{eq:prob_equivalence_noise}
\end{equation}
where $I[\cdot]$ is an indicator function. For each selected value of $\varepsilon_T$, the uncertainty assigned
to an individual processed profile is set to
$\sigma_T=\varepsilon_T/\sqrt{2}$. Consequently, when the
perturbations applied to two profiles are statistically independent,
the standard-deviation scale of their difference is
$\sqrt{2}\sigma_T=\varepsilon_T$. The equivalence probability is
estimated using $N_{\mathrm{MC}}=1000$ Monte Carlo realizations for
each profile pair and tolerance. For the spatially correlated model
in Eq.~\eqref{eq:correlated_noise}, the Gaussian correlation length
is fixed at $\ell_c=5~\mathrm{mm}$. The independent and correlated
noise ensembles use random seeds 20260529 and 20261529,
respectively. Wilson 95\% confidence intervals for the corresponding
binomial equivalence probabilities are reported in Figure~\ref{fig:fig05_new}. This noisy-profile
test complements the deterministic distance in Eq.~\eqref{eq:L2_main}. It shows
how stable the equivalent or distinguishable classification remains when
representative measurement uncertainties are added to the surface profiles.

This definition separates two questions. The first is whether
a tumor produces a detectable thermal anomaly at the surface. The second is
whether the perfusion structure inside the tumor can be identified from that
surface anomaly. These two questions are not equivalent in inverse thermal
imaging, because different internal configurations may lead to similar surface
temperature fields \cite{hatwar2017inverse,gonzalez2020inverse}. Thus, two
tumors may have different intratumoral perfusion organizations but still be
thermographically indistinguishable if their surface profiles differ by less
than $\varepsilon_T$. {\color{black}
The criterion in Eq.~\eqref{eq:equiv_main} quantifies observational equivalence
of stationary surface profiles under the prescribed geometry,
physiological inputs, and boundary conditions, using the
profile-distance metric in Eq.~\eqref{eq:L2_main}.
}

For an ensemble containing $N_C$ tumor configurations, the pairwise profile
distances are collected in the matrix
\begin{equation}
D_{ij} =
d_{L^2}(\mathcal{C}_i,\mathcal{C}_j),
\label{eq:distance_matrix_main}
\end{equation}
where $i,j=1,\ldots,N_C$. This matrix is used to construct distance maps and equivalence maps. Distance
maps show the magnitude of separation between surface signatures, whereas
equivalence maps classify each pair as distinguishable or thermally equivalent
for a chosen value of $\varepsilon_T$. In this way, the
analysis focuses on differences that remain visible at the surface, rather than
on differences inside the tissue alone.

For an ensemble of $N_C$ perfusion configurations, we define the
equivalent-pair fraction at tolerance $\varepsilon_T$ as
\begin{equation}
f_{\mathrm{eq}}(\varepsilon_T)
=
\frac{2}{N_C(N_C-1)}
\sum_{i<j}
I\!\left[D_{ij}\leq\varepsilon_T\right],
\label{eq:equivalent_pair_fraction}
\end{equation}
where the summation includes the $N_C(N_C-1)/2$ unique off-diagonal
pairs and excludes self-comparisons. Thus, $f_{\mathrm{eq}}=0$
indicates that all profile pairs are distinguishable at the selected
tolerance, whereas $f_{\mathrm{eq}}=1$ indicates that all pairs are
classified as thermally equivalent. For the four perfusion patterns
considered here, six unique pairs are evaluated, so that
$f_{\mathrm{eq}}$ changes in increments of $1/6$.

\subsection{\label{sec:numerical_imp}Numerical implementation, verification, and computational performance}

Numerical computations are performed on the two-dimensional multilayer breast
domain described in Sec.~\ref{sec:model}. The tumor is placed within the
glandular layer and represented as a finite circular inclusion. The healthy
tissue properties are taken to be piecewise constant within each layer, while
the tumor has its own thermal conductivity, metabolic heat generation, and
spatially dependent perfusion field. The numerical ensemble is constructed by
varying the tumor radius $R_t$, tumor-center depth $d_t$, fat-layer thickness
$\delta_f$, outer-surface geometry, and intratumoral perfusion class.

Most parameter variations are carried out as OAT
sweeps around a representative reference configuration. This makes it possible
to see how each factor changes the surface profile. The depth sweep examines
thermal screening by the overlying tissue, the diameter sweep probes the effect of tumor size on the total thermal perturbation, the fat-layer sweep tests the role of the adipose layer as an overlying thermal filter, and the
outer-geometry sweep examines sensitivity to the surface through which heat is
observed. In addition, because depth and diameter are expected to play
competing roles, we include a limited two-parameter depth--diameter map.
This map is used to check whether the trends from the separate depth and diameter sweeps persist when both parameters are varied together.

The geometrical variables are examined primarily through separate OAT sweeps around the representative configuration, together with a two-parameter depth--diameter map. In addition, the joint effects of four thermophysical and physiological parameter groups are examined
through the Latin-hypercube uncertainty-screening analysis described in
Sec.~\ref{sec:sensitivity_uncertainty}.

The governing equations are solved on a structured Cartesian grid with spatially
varying material coefficients. The conductive term $\nabla\cdot(k\nabla T)$ is
discretized in conservative flux form. Thermal conductivities at cell faces are
computed using harmonic averaging, which helps preserve normal heat-flux
continuity across tissue interfaces and at the tumor boundary. At the outer
surface, the Robin boundary condition is incorporated directly into the matrix
stencil. In the flat reference geometry, with the coordinate measured inward from
the surface, the discrete Robin condition at the first grid point can be written
as
\begin{equation}
\frac{k_{1/2}}{\Delta x}(T_1-T_0) = h(T_0-T_\infty),
\label{eq:robin_discrete_flux}
\end{equation}
or equivalently,
\begin{equation}
\left(\frac{k_{1/2}}{\Delta x}+h\right)T_0
-
\frac{k_{1/2}}{\Delta x}T_1
=
hT_\infty .
\label{eq:robin_discrete_matrix}
\end{equation}
Here, $T_0$ is the surface-node temperature, $T_1$ is the adjacent interior-node
temperature, and $k_{1/2}$ is the face conductivity between them. The posterior
wall is imposed as a Dirichlet boundary, $T=T_{\mathrm{core}}$, and the lateral
boundaries are treated as insulated using a zero-normal-gradient condition. For
each parameter set, the healthy-background problem is solved first. The same
geometry is then solved with the tumor included, and the difference between the
two surface temperature profiles gives $\Delta T_s(y)$.

The resulting sparse linear systems are solved using an iterative BiCGSTAB
solver with an incomplete-LU preconditioner. Solver convergence is monitored
using the normalized residual
\begin{equation}
r_{\mathrm{rel}} =
\frac{\|A\mathbf{T}-\mathbf{b}\|_2}{\|\mathbf{b}\|_2},
\label{eq:normalized_residual}
\end{equation}
where $A\mathbf{T}=\mathbf{b}$ is the discretized linear system. \textcolor{black}{The solver tolerance is set to $10^{-10}$. The normalized residual is
used to assess algebraic convergence, while the mesh-refinement tests
reported below assess the discretization sensitivity of the monitored
surface-temperature quantities and pairwise profile distances.}

Because the quantities compared in this study are small differences between
surface temperature profiles, grid convergence must be checked carefully. We
therefore performed a mesh-convergence test using the Khomsi-based
representative parameter set \cite{khomsi2024deep}. The monitored quantities
were the maximum surface temperature rise $\Delta T_{s,\max}$, the FWHM of the surface profile, and the profile distance
$d_{L^2}^{(\mathrm{N-U})}$ between the necrotic-core and uniform-perfusion
cases. Relative errors were computed with respect to the finest mesh included
in the test, $\Delta x=\Delta y=0.10~\mathrm{mm}$. The results are summarized in
Table~\ref{tab:mesh_convergence}. The amplitude and width descriptors show
rapid convergence. For the mesh used in the subsequent simulations,
$h_{\mathrm{target}}=0.15~\mathrm{mm}$, the relative deviations are $0.232\%$ for
$\Delta T_{s,\max}$ and $0.098\%$ for FWHM. {\color{black}
The profile-distance measure shows greater relative sensitivity to grid
refinement than the surface-amplitude and width descriptors in this
test, as it compares two close surface profiles. At the nominal target
grid spacing $h_{\mathrm{target}}=0.15~\mathrm{mm}$, the relative
deviation of $d_{L^2}^{(N-U)}$ from its value on the
$0.10~\mathrm{mm}$ reference mesh is $0.504\%$, calculated from the
unrounded numerical outputs. This target spacing was adopted as a
practical compromise between the discretization sensitivity of the
monitored quantities and computational cost.}

\begin{table*}[pos = tbp]
\centering
\caption{Mesh-convergence test for the Khomsi-based representative case.}
\label{tab:mesh_convergence}

\begin{threeparttable}
\scriptsize
\setlength{\tabcolsep}{4.5pt}
\renewcommand{\arraystretch}{1.15}

\begin{tabular*}{\textwidth}{@{\extracolsep{\fill}}llcccccc@{}}
\toprule
\multirow{2}{*}{$h_{\mathrm{target}}$}
& \multirow{2}{*}{Grid}
& \multicolumn{3}{c}{Monitored quantity}
& \multicolumn{3}{c}{Relative error} \\
\cmidrule(lr){3-5}
\cmidrule(lr){6-8}
&
& $\Delta T_{s,\mathrm{max}}$
& FWHM
& $d_{L^2}^{(\mathrm{N-U})}$
& $\Delta T_{s,\mathrm{max}}$
& FWHM
& $d_{L^2}^{(\mathrm{N-U})}$ \\
(mm)
&
& ($^\circ$C)
& (mm)
& ($^\circ$C)
& (\%)
& (\%)
& (\%) \\
\midrule
1.000 & $66\times121$ & 0.343856 & 37.731834 & 0.019036 & 2.506 & 2.333 & 5.844 \\
0.750 & $88\times161$ & 0.342611 & 38.261715 & 0.016309 & 2.135 & 0.962 & 9.314 \\
0.500 & $131\times241$ & 0.340523 & 38.442720 & 0.016165 & 1.513 & 0.493 & 10.117 \\
0.400 & $164\times301$ & 0.340692 & 38.528238 & 0.018240 & 1.563 & 0.272 & 1.419 \\
0.300 & $218\times401$ & 0.338209 & 38.517365 & 0.017832 & 0.823 & 0.300 & 0.846 \\
0.250 & $261\times481$ & 0.336743 & 38.562355 & 0.017824 & 0.386 & 0.183 & 0.894 \\
0.200 & $326\times601$ & 0.336336 & 38.518738 & 0.017786 & 0.264 & 0.296 & 1.102 \\
0.150 & $434\times801$ & 0.336226 & 38.595200 & 0.018075 & 0.232 & 0.098 & 0.504 \\
0.125 & $521\times961$ & 0.335869 & 38.622486 & 0.017844 & 0.125 & 0.028 & 0.781 \\
0.100 & $651\times1201$ & 0.335449 & 38.633243 & 0.017985 & 0.000 & 0.000 & 0.000 \\
\bottomrule
\end{tabular*}

\begin{tablenotes}[flushleft]
\scriptsize
\item \textit{Note.} The representative Khomsi-based case uses the
ideal outer geometry, a tumor-center depth of $22~\mathrm{mm}$,
and a tumor diameter of $12~\mathrm{mm}$. Here,
$h_{\mathrm{target}}$ denotes the nominal grid spacing; the actual
Cartesian spacings are determined by the fixed domain dimensions
and integer numbers of grid intervals. The quantities
$\Delta T_{s,\max}$ and FWHM are evaluated from the necrotic-core
surface temperature-rise profile, whereas
$d_{L^2}^{(\mathrm{N-U})}$ denotes the profile distance between
the necrotic-core and uniform-perfusion cases. Relative errors
are calculated from the unrounded numerical outputs using the
$h_{\mathrm{target}}=0.100~\mathrm{mm}$ mesh as the reference;
consequently, values recalculated from the rounded entries may
differ slightly. At $h_{\mathrm{target}}=0.150~\mathrm{mm}$,
the relative deviations are $0.232\%$ for $\Delta T_{s,\max}$,
$0.098\%$ for FWHM, and $0.504\%$ for
$d_{L^2}^{(\mathrm{N-U})}$. The small non-monotonic variations
in the profile-distance error are consistent with changes in
the grid representation of the circular tumor boundary and
tissue interfaces. The $h_{\mathrm{target}}=0.150~\mathrm{mm}$
mesh is therefore adopted for the main simulations as a
practical compromise between numerical accuracy and computational
cost. An additional convergence test for the closest profile
pair, rim-enhanced--necrotic-core, is reported in
Appendix~\ref{app:mesh_closest_pair}.
\end{tablenotes}

\end{threeparttable}
\end{table*}

{\color{black}
Because the smallest off-diagonal distance in the representative
raw-perfusion case occurs for the rim-enhanced--necrotic-core pair,
we also performed an additional mesh-refinement test for
$d_{L^2}^{(R-N)}$, as reported in Appendix~\ref{app:mesh_closest_pair}. At
$h_{\mathrm{target}}=0.15~\mathrm{mm}$, the distance is
$0.016529~^\circ\mathrm{C}$, compared with
$0.016603~^\circ\mathrm{C}$ on the finest reference mesh included
in this test, $h_{\mathrm{target}}=0.10~\mathrm{mm}$.
The absolute intermesh deviation is approximately
$7.4\times10^{-5}~^\circ\mathrm{C}$, corresponding to a relative
deviation of $0.449\%$ calculated from the unrounded outputs.
Both distances remain below the smallest observational tolerance,
$\varepsilon_T=0.020~^\circ\mathrm{C}$. Moreover, the intermesh
deviation is much smaller than the approximately
$3.47\times10^{-3}~^\circ\mathrm{C}$ margin between the distance
on the main simulation mesh and this threshold. These results
support the stability of the equivalent classification for this
pair under the tested mesh refinement. The maximum normalized
residual in the additional test was below $2\times10^{-11}$.
}

To quantify the computational cost associated with grid refinement,
the representative necrotic-core case was benchmarked at nominal mesh
sizes of 0.30, 0.15, and 0.10 mm. All computations were performed in
C++17 using GCC 12.2.0 and Eigen 3.5.0 on a computing node equipped
with a 13th Gen Intel Core i9-13900K processor and approximately
64 GB of system memory. The
code was compiled using the \texttt{-O3} and \texttt{-DNDEBUG}
optimization flags. The resulting computational-performance
measurements are summarized in Table~\ref{tab:computational_performance}.

\begin{table*}[t]
\centering
\caption{Computational performance for the representative necrotic-core case. Reported wall-clock times are medians of 5 runs after 1 warm-up run. All calculations used one CPU thread. The solver time includes IncompleteLUT setup and BiCGSTAB iterations.}
\label{tab:computational_performance}
\small
\renewcommand{\arraystretch}{1.15}
\begin{tabular*}{\textwidth}{@{\extracolsep{\fill}}cccccccc@{}}
\hline
$h_{\mathrm{target}}$ & Grid & $N_{\mathrm{unk}}$ & NNZ & Iterations & Assembly & Solver & $r_{\mathrm{rel}}$\\
(mm) & & & & & (s) & (s) &\\
\hline
0.30 & $218\times401$ & 87\,418 & 432\,987 & 59 & 0.0054 & 0.3887 & 7.22e-13 \\
0.15 & $434\times801$ & 347\,634 & 1\,729\,971 & 127 & 0.0187 & 3.3949 & 7.76e-12 \\
0.10 & $651\times1201$ & 781\,851 & 3\,896\,954 & 194 & 0.0340 & 11.3478 & 1.13e-11 \\
\hline
\end{tabular*}
\begin{minipage}{\textwidth}
\footnotesize \textit{Note.} Here, $h_{\mathrm{target}}$ denotes the nominal
grid spacing; the actual Cartesian spacings are determined by
the fixed domain dimensions and integer numbers of grid intervals.
$N_{\mathrm{unk}}$ denotes the number of unknowns, and NNZ denotes
the number of nonzero entries in the discretized matrix. The
assembly time includes construction of the sparse matrix and
right-hand-side vector, together with compression of the matrix.
The solver time is the sum of the preconditioner setup time and
the iterative solution time. The explicitly evaluated relative
residual is
$r_{\mathrm{rel}}=
\lVert A\mathbf{T}-\mathbf{b}\rVert_2/
\lVert\mathbf{b}\rVert_2$.
\end{minipage}
\end{table*}

As the nominal mesh size was reduced from 0.30 to 0.10 mm, the
number of unknowns increased from 87,418 to 781,851, while the
number of nonzero matrix entries remained approximately five times
the number of unknowns, consistent with the adopted five-point
finite-difference stencil. The median BiCGSTAB iteration count
increased from 59 to 194, and the corresponding solver time increased
from 0.389 to 11.348 s. Peak memory consumption increased from
55.5 to 444.6 MiB. For the 0.15-mm mesh employed in the main
simulations, a representative solution required 127 iterations,
3.395 s of solver time, and approximately 184 MiB of peak memory.
The explicit relative residual remained below
$1.2\times10^{-11}$ for all meshes. Together with the mesh-independence
results, these measurements support the 0.15-mm mesh as a practical
compromise between numerical accuracy and computational cost.

As an additional numerical-verification step, the laterally uniform
healthy-tissue problem was compared with the analytical transfer-matrix
solution for the multilayer slab, as detailed in Appendix~\ref{app:benchmark_1d}. At the mesh used
in the main simulations, the absolute surface-temperature error was
$5.49\times10^{-4}~^\circ\mathrm{C}$, the depth-averaged $L^2$ error was
$2.22\times10^{-3}~^\circ\mathrm{C}$, and the normalized residual was
$4.78\times10^{-13}$. Together with the mesh-convergence and closest-pair
tests, these results verify the numerical implementation within the
investigated discretization range. These numerical checks do not constitute
validation against independent experimental or clinical thermographic data.

\subsection{Parameter sensitivity and uncertainty analysis}
\label{sec:sensitivity_uncertainty}

A parameter-sensitivity and uncertainty analysis was conducted for the
representative Khomsi-based configuration to examine whether the
surface-profile distinguishability remains stable under variations in
the principal thermophysical and physiological parameters. The tumor
diameter, tumor-center depth, tissue-layer geometry, and perfusion-pattern
shapes were fixed at their representative values. Four dimensionless
multipliers were introduced for blood perfusion, volumetric heat
generation, thermal conductivity, and the outer-surface convection
coefficient:
\begin{equation}
\begin{aligned}
\omega(\mathbf{r}) &\longrightarrow
m_{\omega}\,\omega(\mathbf{r}),\\
Q_m(\mathbf{r}) &\longrightarrow
m_Q\,Q_m(\mathbf{r}), \qquad
Q_t \longrightarrow m_Q\,Q_t,\\
k(\mathbf{r}) &\longrightarrow
m_k\,k(\mathbf{r}),\\
h &\longrightarrow m_h\,h.
\end{aligned}
\label{eq:uncertainty_multipliers}
\end{equation}
The same multiplier was applied globally to the corresponding healthy-
tissue and tumor parameters. Consequently, the relative contrasts among
the tissue layers and the spatial forms of the four intratumoral
perfusion patterns were preserved.

First, a one-at-a-time (OAT) sensitivity analysis was performed. For
each parameter group, the corresponding multiplier was varied over
\begin{equation}
m_p\in\{0.8,\,0.9,\,1.0,\,1.1,\,1.2\},
\quad
p\in\{\omega,Q,k,h\},
\label{eq:oat_multipliers}
\end{equation}
while the other three multipliers were held at unity. For each
simulation, the four raw perfusion scenarios were evaluated and the
following outputs were recorded: the maximum surface-temperature rise
for each perfusion pattern, the six pairwise $d_{L^2}$ distances, their
mean
\begin{equation}
\overline{d}_{L^2}
=
\frac{1}{6}
\sum_{i<j}d_{L^2}(\mathcal{C}_i,\mathcal{C}_j),
\label{eq:mean_pairwise_distance}
\end{equation}
and the equivalent-pair fraction $f_{\mathrm{eq}}$ at
$\varepsilon_T=0.020~^\circ\mathrm{C}$.

The relative response of the mean pairwise distance was calculated as
\begin{equation}
\Delta\overline{d}_{L^2}(m_p)
=
100
\left[
\frac{\overline{d}_{L^2}(m_p)}
     {\overline{d}_{L^2}(1)}
-1
\right]\%.
\label{eq:relative_mean_distance}
\end{equation}
To compare the pair-specific effects of the four parameter groups, a
dimensionless endpoint sensitivity was estimated using the central
secant
\begin{equation}
S_p^{(ij)}
=
\frac{
d_{L^2}^{(ij)}(1.2)-d_{L^2}^{(ij)}(0.8)
}{
(1.2-0.8)\,d_{L^2}^{(ij)}(1.0)
}.
\label{eq:normalized_sensitivity}
\end{equation}
Positive values of $S_p^{(ij)}$ indicate that increasing the parameter
increases the separation between the corresponding surface profiles,
whereas negative values indicate increasing similarity.

Second, the joint effects of the four parameter groups were examined
using Latin-hypercube sampling (LHS). A total of 300 parameter
combinations were generated by independently sampling
\begin{equation}
m_{\omega},m_Q,m_k,m_h
\sim \mathcal{U}(0.8,1.2),
\label{eq:lhs_distribution}
\end{equation}
using the random seed 20260716. For every sample, the healthy-background
problem and the four tumor-bearing perfusion cases were recomputed using
the same sampled parameter multipliers. Uncertainty in each pairwise
distance was summarized using its median and 2.5th--97.5th percentile
interval. The distribution of $f_{\mathrm{eq}}$ was also evaluated at
$\varepsilon_T=0.020~^\circ\mathrm{C}$.

Spearman rank correlations were calculated between each input
multiplier and the mean pairwise distance across the 300 LHS samples.
The Spearman coefficient was computed as the Pearson correlation between
the corresponding ranked variables and was used to identify the
parameters most strongly associated with variations in
$\overline{d}_{L^2}$. {\color{black}
The common $\pm20\%$ multiplier range provides a consistent relative
perturbation scale for comparing the four input groups. This range
and the uniform distributions in Eq.~(32) are prescribed choices
for controlled uncertainty screening around the representative
Khomsi-based configuration; they were not inferred from clinical
measurements. Independence is assumed between the four multipliers.
Within the perfusion, heat-generation, and conductivity groups,
however, global rescaling imposes proportional changes across
the corresponding healthy-tissue and tumor parameters, preserving
their relative contrasts. The LHS ensemble therefore explores
the joint effects of these four prescribed multipliers at fixed
geometry and fixed perfusion-pattern shapes. The percentile summaries of the profile distances are conditional
on the prescribed input model and comparison metric. The
equivalent-pair frequencies additionally depend on the specified
observational tolerance.
The Spearman coefficients characterize input--output rank
associations within the ensemble; they neither estimate
physiological correlations among the inputs nor provide a
variance-based sensitivity decomposition. The resulting analysis
quantifies model responses within the prescribed parameter domain,
without assigning population-level clinical probabilities.
}

\section{Results and discussion}
\label{sec:results}

We first compare how the prescribed heterogeneous
intratumoral perfusion patterns modify the internal temperature field and how
much of this variation remains observable in the surface profile. Unless
otherwise stated, all forward simulations reported below used a conservative
finite-difference discretization with a target grid spacing of
$0.15~\mathrm{mm}$. The resulting sparse linear systems were solved using
BiCGSTAB with an IncompleteLUT preconditioner. The corresponding
mesh-convergence, solver-convergence, and analytical-verification results are
reported in Sec.~\ref{sec:numerical_imp} and
Appendices~\ref{app:benchmark_1d} and \ref{app:mesh_closest_pair}.

We first consider a representative Khomsi-based multilayer case. The
tissue-layer thicknesses, temperature conditions, and thermophysical parameters
are summarized in Tables~\ref{tab:geom_khomsi} and \ref{tab:param_khomsi}.
Unless otherwise stated, the tumor-center depth is $22~\mathrm{mm}$ and the
tumor diameter is $12~\mathrm{mm}$. In this reference
comparison, the geometry, tumor size, and tumor position are kept fixed, while
only the intratumoral perfusion pattern is changed.

\begin{table}[pos = tbp]
\centering
\caption{Geometry and temperature conditions for the Khomsi-based representative case.}
\label{tab:geom_khomsi}

\small
\setlength{\tabcolsep}{6pt}
\renewcommand{\arraystretch}{1.15}

\begin{tabular*}{\linewidth}{@{\extracolsep{\fill}}lc@{}}
\toprule
Quantity & Value \\
\midrule
\multicolumn{2}{@{}l}{\textit{Tissue-layer geometry}} \\
Skin thickness, $\delta_s$       & $1.6~\mathrm{mm}$ \\
Fat thickness, $\delta_f$        & $5.0~\mathrm{mm}$ \\
Glandular thickness, $\delta_g$  & $43.4~\mathrm{mm}$ \\
Muscle thickness                 & $15.0~\mathrm{mm}$ \\
\midrule
\multicolumn{2}{@{}l}{\textit{Temperature conditions}} \\
Arterial blood temperature, $T_a$        & $37^\circ\mathrm{C}$ \\
Chest-wall temperature, $T_{\mathrm{core}}$ & $37^\circ\mathrm{C}$ \\
Ambient temperature, $T_\infty$          & $25^\circ\mathrm{C}$ \\
\midrule
\multicolumn{2}{@{}l}{\textit{Tumor geometry}} \\
Tumor-center depth, $d_t$ & $22~\mathrm{mm}$ \\
Tumor diameter, $D$     & $12~\mathrm{mm}$ \\
\bottomrule
\end{tabular*}
\end{table}

\begin{table}[pos = tbp]
\centering
\caption{Thermal and physiological parameters for the
Khomsi-based representative case.}
\label{tab:param_khomsi}

\begin{threeparttable}
\small
\setlength{\tabcolsep}{4pt}
\renewcommand{\arraystretch}{1.15}

\begin{tabular*}{\linewidth}{
@{\extracolsep{\fill}}lccc@{}
}
\toprule
Tissue
& Heat source
& Conductivity
& Perfusion \\
&
$(\mathrm{W\,m^{-3}})$
&
$(\mathrm{W\,m^{-1}\,K^{-1}})$
&
$(\mathrm{s^{-1}})$ \\
\midrule
Skin      & $368.1$          & $0.45$ & $0.00018$ \\
Fat       & $400$            & $0.21$ & $0.00022$ \\
Glandular & $700$            & $0.48$ & $0.00054$ \\
Muscle    & $700$            & $0.48$ & $0.00270$ \\
Tumor     & $7.0\times 10^4$ & $0.62$ & $0.01600$ \\
\bottomrule
\end{tabular*}

\begin{tablenotes}[flushleft]
\footnotesize
\item \textit{Note.} For healthy tissues, the listed heat-generation
and perfusion values are the layerwise constants defining
$Q_m(\mathbf r)$ and $\omega(\mathbf r)$, respectively.
For the tumor, the listed values are the prescribed heat-generation
rate $Q_t$ and the perfusion $\omega_0$ of the uniform reference
case. The spatially heterogeneous tumor-perfusion fields are
specified in Sec.~\ref{sec2.2}.
\end{tablenotes}

\end{threeparttable}
\end{table}

\begin{figure*}[pos = tb]
\centering
\includegraphics[width=\linewidth]{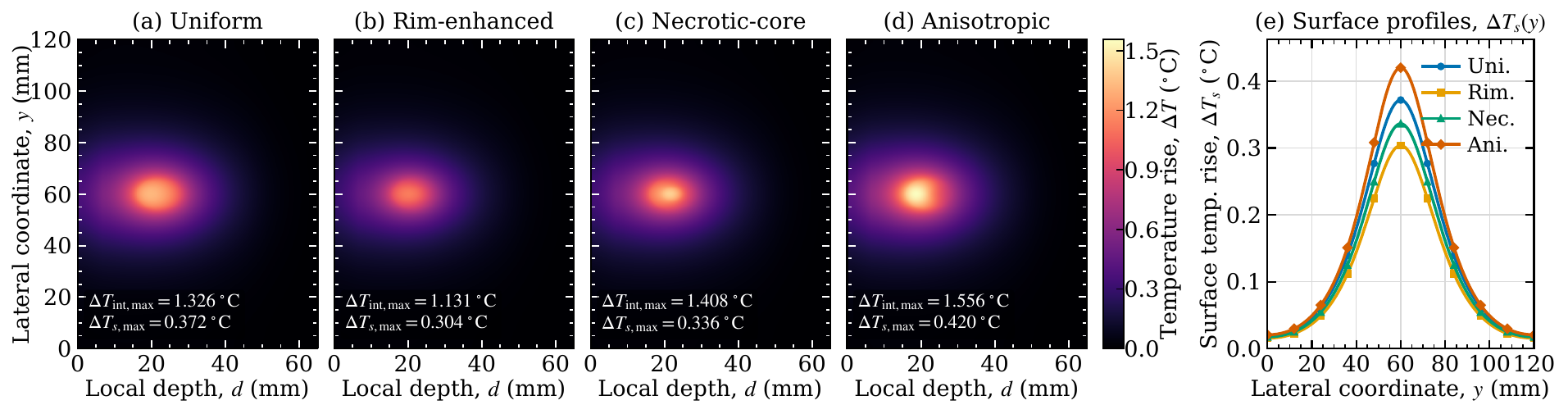}
\caption{Internal and surface thermal responses of the same
representative tumor geometry under four raw (not mean-matched) intratumoral
perfusion scenarios. Panels (a)--(d) show the tumor-induced internal
temperature rise $\Delta T(d,y)$ in the local-depth coordinate $d$ measured
inward from the outer surface, for uniform, rim-enhanced, necrotic-core, and
anisotropic perfusion, respectively. The annotations report
the maximum internal temperature rise $\Delta T_{\mathrm{int},\max}$ and the
maximum surface temperature rise $\Delta T_{s,\max}$ for each scenario.
Panel (e) shows the corresponding surface thermal signatures,
$\Delta T_s(y)$. A common color scale is used in panels
(a)--(d) to enable direct comparison of the internal fields. Relative to the
internal responses, the surface profiles are spatially smoother and attenuated,
although scenario-dependent differences in their peak magnitudes remain
visible, illustrating the filtering effect of the overlying tissue.
}

\label{fig:fig03_new}
\end{figure*}

Figure~\ref{fig:fig03_new} compares four raw, or absolute-perfusion, scenarios: uniform, rim-enhanced, necrotic-core, and anisotropic perfusion. Panels (a)--(d) show the tumor-induced internal temperature rise plotted in the local-depth coordinate $d$ measured inward from the outer surface. The corresponding peak internal temperature rises are
$1.326$, $1.131$, $1.408$, and $1.556~^\circ\mathrm{C}$, respectively.
Thus, the internal maxima span $0.425~^\circ\mathrm{C}$, or approximately
$37.6\%$ relative to the smallest value, demonstrating a substantial
dependence on the intratumoral perfusion scenario. This variation arises
because perfusion changes the local balance between metabolic heat generation
and blood-mediated heat exchange. Regions with stronger perfusion exchange
heat more efficiently with blood, whereas low-perfusion regions exchange heat
less effectively and may retain heat. Hence, even for the same tumor size and
location, the internal thermal field can be reshaped by the intratumoral
perfusion scenario.

By contrast, the surface profiles in Figure~\ref{fig:fig03_new}(e) are much closer to one another than the corresponding internal fields. This
difference between the internal and surface responses is the main physical
mechanism behind thermal-signature equivalence. As heat travels from the tumor to the surface, conduction, perfusion, and boundary heat exchange smooth the spatial details of intratumoral heterogeneity. The overlying
tissue therefore acts as a thermal low-pass filter: clear internal differences
can appear only as weakly separated surface profiles. This interpretation is
consistent with the Green's-function representation in Appendix~\ref{app:green}, where the tumor-induced perturbation reaches the surface through a smoothing propagation kernel.

\begin{figure*}[pos = tb]
\centering
\includegraphics[width=\linewidth]{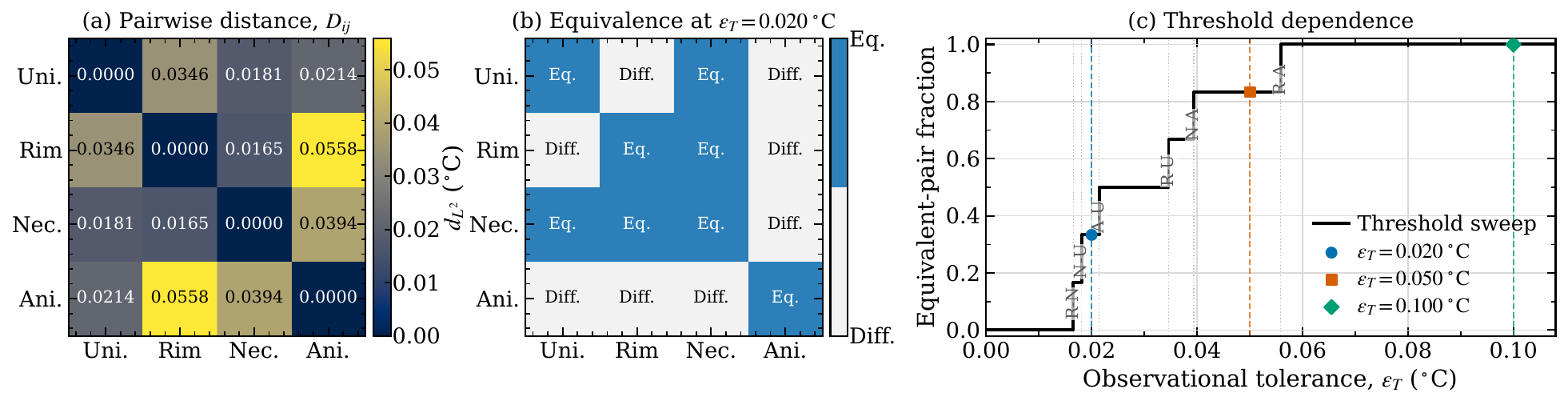}
\caption{
Thermal-signature equivalence analysis for the representative Khomsi-based
case using the raw (absolute-perfusion) scenarios.
(a) Pairwise surface-profile distance matrix $D_{ij}$ for uniform (Uni.),
rim-enhanced (Rim), necrotic-core (Nec.), and anisotropic (Ani.) perfusion
patterns. (b) Binary equivalence map for
$\varepsilon_T=0.020~^\circ\mathrm{C}$, where pairs with
$D_{ij}\leq\varepsilon_T$ are classified as equivalent (Eq.) and the others as
distinguishable (Diff.). At this tolerance, the Necrotic-core--Uniform and Rim-enhanced--Necrotic-core pairs are equivalent;
the diagonal entries represent self-comparisons. (c)
Stepwise equivalent-pair fraction computed over the six
unique off-diagonal pairs. The gray dotted vertical lines mark the individual
pair distances, whereas the colored dashed lines and symbols indicate the
representative tolerances $0.020$, $0.050$, and
$0.100~^\circ\mathrm{C}$. The corresponding equivalent-pair fractions are
$2/6$, $5/6$, and $6/6$, respectively. Increasing $\varepsilon_T$ progressively increases the fraction
of profile pairs classified as thermally equivalent, illustrating
the tolerance-dependent nature of thermographic distinguishability.
}

\label{fig:fig04_new}
\end{figure*}

The similarity among the surface profiles is quantified in
Figure~\ref{fig:fig04_new}. Panel (a) shows the pairwise distance matrix $D_{ij}$
between the four surface signatures. The smallest
off-diagonal distance is obtained for the rim-enhanced--necrotic-core pair,
$D_{ij}=0.0165~^\circ\mathrm{C}$, whereas the largest is obtained for the
rim-enhanced--anisotropic pair, $D_{ij}=0.0558~^\circ\mathrm{C}$, approximately
$3.38$ times the smallest value. Panel (b) converts the distance matrix into a
binary equivalence map using the optimistic observational tolerance
$\varepsilon_T=0.020~^\circ\mathrm{C}$. At this tolerance,
the uniform--necrotic-core and rim-enhanced--necrotic-core pairs are classified
as equivalent, giving an equivalent-pair fraction of $2/6$ ($33.3\%$) when
self-comparisons are excluded. Panel (c) shows the
stepwise equivalent-pair fraction over a continuous range of
$\varepsilon_T$, with representative values $0.020$, $0.050$, and
$0.100~^\circ\mathrm{C}$ marked explicitly. The corresponding
equivalent-pair fractions are $2/6$ ($33.3\%$), $5/6$ ($83.3\%$), and $6/6$
($100\%$), respectively. Thus, thermographic distinguishability is not an
absolute property of the tumor model alone, but depends quantitatively on the
observational tolerance used to compare the surface profiles.

The deterministic equivalence maps in Figure~\ref{fig:fig04_new} are based on clean
surface profiles. In practice, however, the measured surface-temperature profile
is affected by detector noise, calibration uncertainty, emissivity assumptions,
environmental fluctuations, preprocessing choices, and physiological
repeatability. We therefore tested how robust the equivalence classification is
when the clean surface profiles are perturbed by representative measurement
uncertainties. The noise-perturbed equivalence probability $P_{\mathrm{eq}}$ was
computed from Monte Carlo realizations using the observational model described
in Eqs.~\eqref{eq:observed_profile_noise}--\eqref{eq:prob_equivalence_noise}.

\begin{figure*}[pos = tb]
\centering
\includegraphics[width=0.9\linewidth]{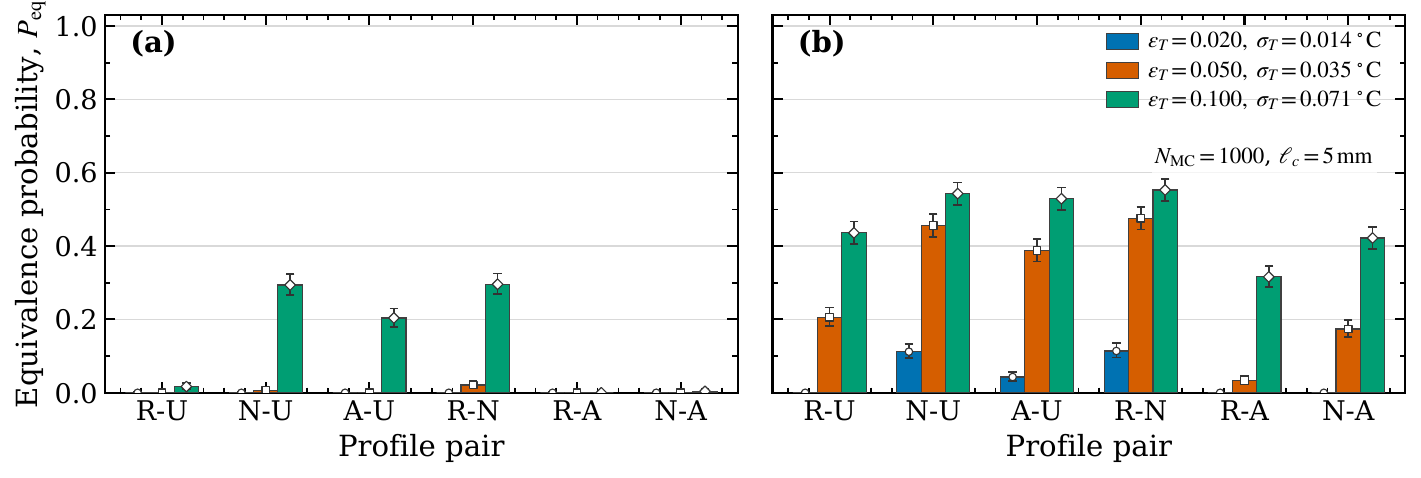}
\caption{
Noise-perturbed robustness of thermal-signature equivalence. The equivalence
probability $P_{\mathrm{eq}}$ is estimated from
$N_{\mathrm{MC}}=1000$ noisy realizations of the surface profiles using
$\sigma_T=\varepsilon_T/\sqrt{2}$.
The bars correspond to
$\varepsilon_T=0.020$, $0.050$, and $0.100~^\circ\mathrm{C}$, and the error
bars show Wilson $95\%$ confidence intervals for the estimated binomial
probabilities. (a) Independent Gaussian noise.
Equivalence probabilities remain close to zero at the two
lower tolerances and increase only moderately for selected pairs at
$\varepsilon_T=0.100~^\circ\mathrm{C}$. (b) Spatially correlated Gaussian
noise with $\ell_c=5~\mathrm{mm}$.
Spatial correlation produces substantially higher
equivalence probabilities for most profile pairs, particularly the N--U and R--N pairs, demonstrating that the robustness of the equivalence
classification depends strongly on the assumed spatial structure of profile
uncertainty. U, R, N, and A denote uniform, rim-enhanced, necrotic-core, and
anisotropic perfusion, respectively.
}
\label{fig:fig05_new}
\end{figure*}

The results are shown in Figure~\ref{fig:fig05_new}. For independent Gaussian
noise, all equivalence probabilities remain at or below
approximately $0.02$ for $\varepsilon_T=0.020$ and
$0.050~^\circ\mathrm{C}$. Even at $\varepsilon_T=0.100~^\circ\mathrm{C}$,
the largest probabilities are only approximately $0.29$--$0.30$, obtained for
the uniform--necrotic-core and rim-enhanced--necrotic-core pairs. This
behavior reflects the accumulation of uncorrelated pointwise fluctuations in
the RMS profile distance. In contrast, spatially correlated noise gives higher
$P_{\mathrm{eq}}$ values for several borderline pairs.
For example, the equivalence probabilities of the
uniform--necrotic-core and rim-enhanced--necrotic-core pairs are approximately
$0.11$ at $\varepsilon_T=0.020~^\circ\mathrm{C}$, increase to approximately
$0.46$--$0.48$ at $0.050~^\circ\mathrm{C}$, and reach approximately
$0.54$--$0.55$ at $0.100~^\circ\mathrm{C}$. Such correlated perturbations
represent slowly varying profile-level uncertainties, such as calibration
drift, residual background variation, imperfect surface correction, or
preprocessing-induced smoothing. These results show
that the equivalence classification depends not only on the selected threshold,
but also quantitatively on the assumed spatial structure of uncertainty in the
processed surface profile. This supports the interpretation of
$\varepsilon_T$ as an effective profile-level tolerance rather than as the
noise-equivalent temperature difference (NETD) of a single detector pixel.

The distances in Figure~\ref{fig:fig04_new} correspond to the raw perfusion scenarios
defined in Table~\ref{tab:perfusion_profiles}. They therefore reflect the
combined influence of spatial perfusion organization and differences in
tumor-averaged perfusion. Thus, a large surface-profile
distance in the raw comparison does not by itself prove that the surface
thermogram is sensitive only to the spatial pattern of perfusion. It may also
come from a difference in the mean perfusion level. To separate these two
contributions, we next perform a mean-matched control analysis.

\begin{figure*}[pos = tb]
\centering
\includegraphics[width=0.85\linewidth]{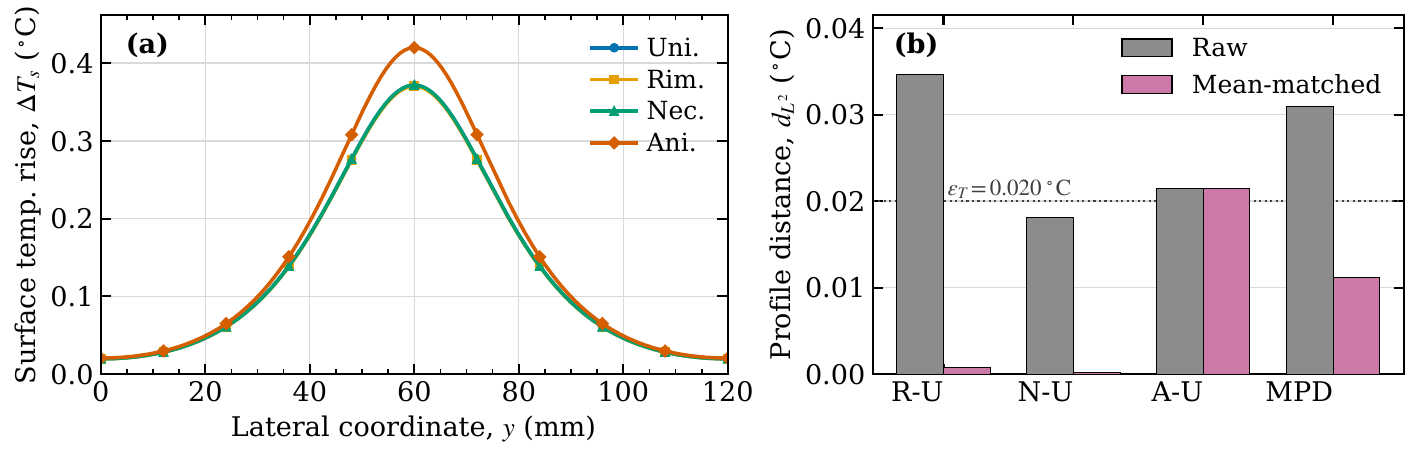}
\caption{
Mean-matched perfusion control for the representative Khomsi-based case.
Each heterogeneous perfusion field is multiplicatively
rescaled so that
$\langle\omega_t^{\mathrm{mm}}\rangle_{\Omega_t}=\omega_0$, while its spatial
pattern is preserved.
(a) Mean-matched surface temperature-rise profiles.
The uniform, rim-enhanced, and necrotic-core profiles nearly
overlap, whereas the anisotropic profile remains visibly separated.
(b) Profile distances before and after mean matching.
R--U, N--U, and A--U denote the rim-enhanced--uniform,
necrotic-core--uniform, and anisotropic--uniform comparisons, respectively,
whereas MPD denotes the mean pairwise distance $\bar d_{L^2}$ over all six
unique profile pairs. The horizontal dotted line marks the observational
tolerance $\varepsilon_T=0.020~^\circ\mathrm{C}$. Mean matching reduces the
R--U and N--U distances to well below this tolerance, indicating that much of
their separation in the raw cases originates from differences in
tumor-averaged perfusion. By contrast, the A--U distance remains nearly
unchanged and slightly above the tolerance, indicating a residual surface
signature associated with directional perfusion organization.
}

\label{fig:fig06_new}
\end{figure*}

In the mean-matched control, each heterogeneous perfusion field is rescaled
according to Eq.~\eqref{eq:mean_matched_perfusion}, so that its tumor-averaged
perfusion is equal to that of the uniform case,
$\langle\omega_t\rangle_{\Omega_t}=\omega_0$, while its spatial pattern is
preserved. The resulting surface profiles are shown in
Figure~\ref{fig:fig06_new}. After mean matching, the uniform,
rim-enhanced, and necrotic-core profiles become nearly indistinguishable at the
surface. This result indicates that much of the raw separation
among the radially organized profiles comes from differences in average
perfusion, not from radial heterogeneity alone. By contrast, the anisotropic
profile remains separated from the uniform case, suggesting that directional
perfusion organization can leave a surface-observable signature even when the
mean perfusion level is fixed.

Panel (b) of Figure~\ref{fig:fig06_new} confirms this interpretation
quantitatively. After mean matching, the Rim--Uniform distance
decreases from $0.0346$ to $0.000801~^\circ\mathrm{C}$, corresponding to a
$97.7\%$ reduction, while the Necrotic--Uniform distance decreases from
$0.0181$ to $0.000173~^\circ\mathrm{C}$, a $99.0\%$ reduction. Both values
therefore fall well below $\varepsilon_T=0.020~^\circ\mathrm{C}$. In contrast,
the Anisotropic--Uniform distance changes by less than $0.02\%$, from
$0.021432$ to $0.021435~^\circ\mathrm{C}$, and remains slightly above the
observational tolerance. The mean distance over all six profile pairs decreases
from $0.03097$ to $0.01113~^\circ\mathrm{C}$, corresponding to an overall
reduction of $64.0\%$. The mean-matched control therefore
shows quantitatively that surface thermal signatures depend on both the
tumor-averaged perfusion magnitude and the spatial organization of perfusion.
In the present representative case, radially heterogeneous patterns become
almost thermally equivalent to the uniform case when the mean perfusion is
fixed, while directional anisotropy remains more robustly distinguishable.

To determine whether the inferred pairwise similarity depends
on the selected distance metric, the surface-temperature-rise profiles for the
original and mean-matched perfusion cases were re-evaluated using the
$L^\infty$ and normalized-$L^2$ distances in addition to the baseline $L^2$
metric. The resulting distances and similarity rankings are reported in
Table~\ref{tab:metric_robustness}. For the original absolute-perfusion cases,
the $L^2$ and $L^\infty$ metrics produce identical rankings for all six
profile pairs. The normalized-$L^2$ metric only reverses the order of the two
most similar pairs, while the rim-enhanced--anisotropic pair remains the most
dissimilar under all three metrics. For the mean-matched cases, all metrics
identify the uniform--necrotic-core pair as the most similar and the
rim-enhanced--anisotropic pair as the most dissimilar, although some
intermediate rankings differ. Thus, the principal conclusions concerning the
closest and most separated thermal signatures are robust to the choice of
metric, while the precise ordering of borderline pairs retains some
metric dependence. The equivalence classifications reported elsewhere remain
based on the dimensional $L^2$ distance and the prescribed observational
tolerance $\varepsilon_T$.

\begin{table*}[tb]
\centering
\caption{Pairwise distances and similarity rankings of the surface
temperature-rise profiles for (a) the original absolute-perfusion cases
and (b) the mean-matched perfusion cases. Rank 1 denotes the most similar
pair.}
\label{tab:metric_robustness}
\small

\begin{tabular*}{\textwidth}{
@{\extracolsep{\fill}}
l
r r
r r
r r
@{}
}
\toprule
& \multicolumn{2}{c}{$d_{L^2}$}
& \multicolumn{2}{c}{$d_{L^\infty}$}
& \multicolumn{2}{c}{$d_{\mathrm{NL2}}$} \\
\cmidrule(lr){2-3}
\cmidrule(lr){4-5}
\cmidrule(lr){6-7}
Pair
& Value ($^\circ$C) & Rank
& Value ($^\circ$C) & Rank
& Value & Rank \\
\midrule

\multicolumn{7}{l}{
\textit{(a) Original absolute-perfusion cases}}\\

Rim-enhanced--Uniform
& 0.0346035 & 4
& 0.0680060 & 4
& 0.207300 & 4 \\

Necrotic-core--Uniform
& 0.0180751 & 2
& 0.0354073 & 2
& 0.103177 & 1 \\

Anisotropic--Uniform
& 0.0214322 & 3
& 0.0483338 & 3
& 0.110002 & 3 \\

Rim-enhanced--Necrotic-core
& 0.0165286 & 1
& 0.0325987 & 1
& 0.104683 & 2 \\

Rim-enhanced--Anisotropic
& 0.0558400 & 6
& 0.116340 & 6
& 0.314518 & 6 \\

Necrotic-core--Anisotropic
& 0.0393565 & 5
& 0.0837411 & 5
& 0.211819 & 5 \\

\midrule

\multicolumn{7}{l}{
\textit{(b) Mean-matched perfusion cases}}\\

Rim-enhanced--Uniform
& 0.000801150 & 3
& 0.00135149 & 2
& 0.00435827 & 3 \\

Necrotic-core--Uniform
& 0.000172707 & 1
& 0.000274344 & 1
& 0.000937707 & 1 \\

Anisotropic--Uniform
& 0.0214347 & 4
& 0.0483389 & 5
& 0.110014 & 4 \\

Rim-enhanced--Necrotic-core
& 0.000715054 & 2
& 0.00148243 & 3
& 0.00389070 & 2 \\

Rim-enhanced--Anisotropic
& 0.0221985 & 6
& 0.0496904 & 6
& 0.114166 & 6 \\

Necrotic-core--Anisotropic
& 0.0214874 & 5
& 0.0482080 & 4
& 0.110305 & 5 \\
\bottomrule
\end{tabular*}
\end{table*}

We next examine how geometrical and anatomical factors affect
the surface distinguishability of the perfusion patterns.
Tumor-center depth, fat-layer thickness, outer-surface
deformation, and tumor diameter are first examined through OAT
sweeps around the reference configuration, with the remaining geometrical and
model parameters held fixed. This makes it easier to identify the role of each
factor. Because depth and diameter exhibit the strongest and
opposing geometrical effects, their combined influence is subsequently
examined using a two-parameter depth--diameter map. The joint effects of the
principal thermophysical and physiological parameter groups are treated
separately through the sensitivity and Latin-hypercube uncertainty-screening
analyses described in Sec.~\ref{sec:sensitivity_uncertainty} and discussed in
connection with Figure~\ref{fig:fig12_new}.

\begin{figure*}[pos = tb]
\centering
\includegraphics[width=0.95\linewidth]{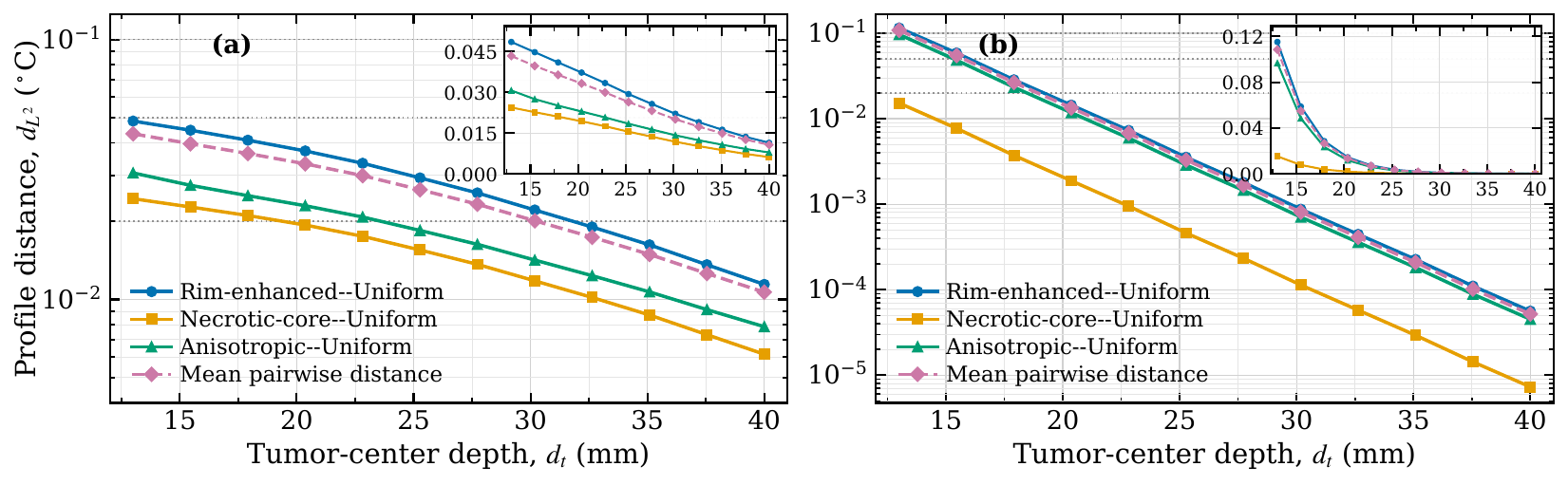}
\caption{
Depth dependence of surface-profile distinguishability. The profile distance
$d_{L^2}$ is shown as a function of tumor-center depth for (a) the
Khomsi-based parameter set and (b) the Lozano-inspired parameter set.
The distances compare each heterogeneous perfusion pattern
with the uniform-perfusion case, while the mean pairwise distance is averaged
over all six pairs of perfusion patterns. The main axes use a logarithmic
scale for $d_{L^2}$, while the insets show the same data on a linear scale.
The horizontal dotted lines indicate the representative
observational tolerances
$\varepsilon_T=0.020$, $0.050$, and $0.100~^\circ\mathrm{C}$.
All profile distances decrease with increasing depth, with a substantially steeper decay for the Lozano-inspired parameter set,
demonstrating that thermal screening by the overlying tissue progressively reduces the surface distinguishability of intratumoral perfusion patterns.
}
\label{fig:fig07_new}
\end{figure*}

Figure~\ref{fig:fig07_new} shows the surface-profile distance $d_{L^2}$ as a
function of tumor-center depth for the Khomsi-based parameter set in panel (a)
and the Lozano-inspired parameter set in panel (b). The main axes use a
logarithmic scale for $d_{L^2}$, while the insets show the same data on a
linear scale. For the Khomsi-based set, all monitored profile
distances decrease with increasing tumor-center depth. The
rim-enhanced--uniform comparison remains the largest, whereas the
necrotic-core--uniform comparison is the smallest. The mean pairwise
distance decreases from approximately $0.0434~^\circ\mathrm{C}$ at
$d_t=13.0~\mathrm{mm}$ to $0.0107~^\circ\mathrm{C}$ at $d_t=40.0~\mathrm{mm}$, corresponding to a reduction of approximately $75.3\%$. It first falls below $\varepsilon_T=0.020~^\circ\mathrm{C}$ at the sampled depth of
$32.64~\mathrm{mm}$. These results quantify the progressive loss of surface
distinguishability as the thermal propagation path through the overlying
tissue increases.

The Lozano-inspired case shows a much stronger depth dependence. The
rim-enhanced--uniform, anisotropic--uniform, and mean pairwise distances
decrease rapidly with depth, while the necrotic-core--uniform distance is much
smaller than the other comparisons. At larger depths, several distances fall
well below the representative observational tolerances used in
Figure~\ref{fig:fig04_new}. Increasing depth therefore strengthens
thermal screening: the heat perturbation travels through a longer tissue path,
and less information about the internal perfusion structure reaches the
surface.

\begin{figure*}[pos = tb]
\centering
\includegraphics[width=0.95\linewidth]{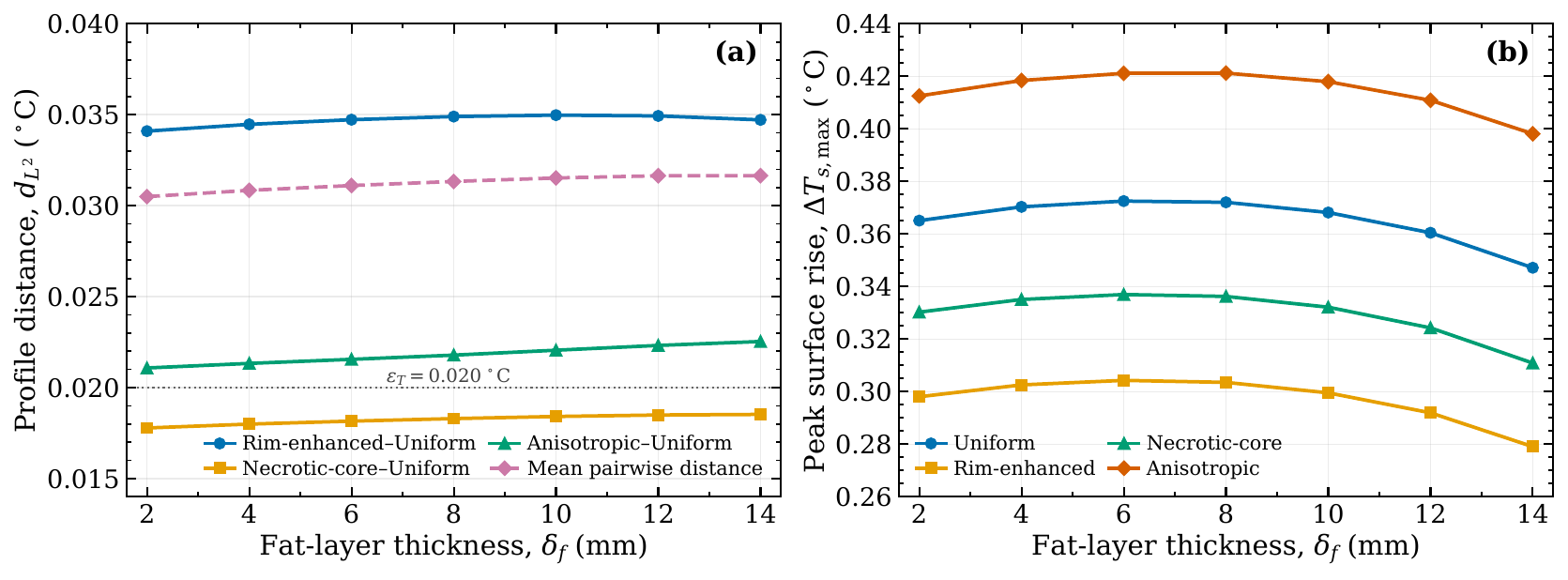}
\caption{
Effect of fat-layer thickness on surface thermal signatures in
the Khomsi-based representative case. (a) Profile distance $d_{L^2}$
for the rim-enhanced, necrotic-core, and anisotropic cases
relative to the uniform-perfusion case, together with the mean pairwise
distance. The horizontal dotted line denotes the observational tolerance
$\varepsilon_T=0.020~^\circ\mathrm{C}$. (b) Peak surface temperature rise
$\Delta T_{s,\max}$ for the four perfusion patterns.
Across the investigated range, the profile distances vary
only modestly, whereas the peak temperature rises reach maxima at fat-layer
thicknesses of approximately $6$--$8~\mathrm{mm}$ and decrease thereafter.
}
\label{fig:fig08_new}
\end{figure*}

In addition to tumor depth, the fat-layer thickness can modulate heat
transmission from the tumor to the surface. Figure~\ref{fig:fig08_new} shows
this effect for the Khomsi-based representative case.
Panel (a) shows that the profile distances vary only modestly
over the investigated fat-layer thickness range of $2$--$14$ mm. The
rim-enhanced--uniform distance remains within approximately
$0.0341$--$0.0350~^\circ\mathrm{C}$, the necrotic-core--uniform distance
within $0.0178$--$0.0187~^\circ\mathrm{C}$, and the
anisotropic--uniform distance within $0.0210$--$0.0226~^\circ\mathrm{C}$.
The mean pairwise distance changes by only approximately $4\%$, from about
$0.0305$ to $0.0317~^\circ\mathrm{C}$. The dotted horizontal line marks the
optimistic observational tolerance
$\varepsilon_T=0.020~^\circ\mathrm{C}$.
The necrotic-core--uniform distance remains below this
tolerance throughout the investigated range, whereas the
rim-enhanced--uniform, anisotropic--uniform, and mean pairwise distances remain
above it. Thus, varying the fat-layer thickness does not change the equivalence
classification of these comparisons at the selected tolerance.

Panel (b) shows the peak surface temperature rise for the four perfusion
patterns. The response is weakly nonmonotonic, reaching a
maximum at a fat-layer thickness of approximately $6$--$8$ mm and decreasing
thereafter. Across the complete thickness range, the peak surface rises are
approximately $0.347$--$0.372~^\circ\mathrm{C}$ for uniform,
$0.279$--$0.305~^\circ\mathrm{C}$ for rim-enhanced,
$0.311$--$0.337~^\circ\mathrm{C}$ for necrotic-core, and
$0.398$--$0.421~^\circ\mathrm{C}$ for anisotropic perfusion. The anisotropic
case gives the largest peak response, while the rim-enhanced case gives the
smallest. Therefore, within the investigated range, the fat
layer does not behave as a simple monotonic attenuator. Its influence reflects
a balance between low thermal conductivity, weak perfusion in adipose tissue,
heat redistribution within the multilayer structure, and the fixed tumor depth
used in this test. Compared with the $75.3\%$ depth-induced
reduction and the much stronger diameter dependence found for the same
Khomsi-based parameter set, the approximately $4\%$ variation in the mean
pairwise distance confirms that fat-layer thickness acts mainly as a secondary
modulator of surface-profile distinguishability over the tested range.

\begin{figure*}[pos = tb]
\centering
\includegraphics[width=0.95\linewidth]{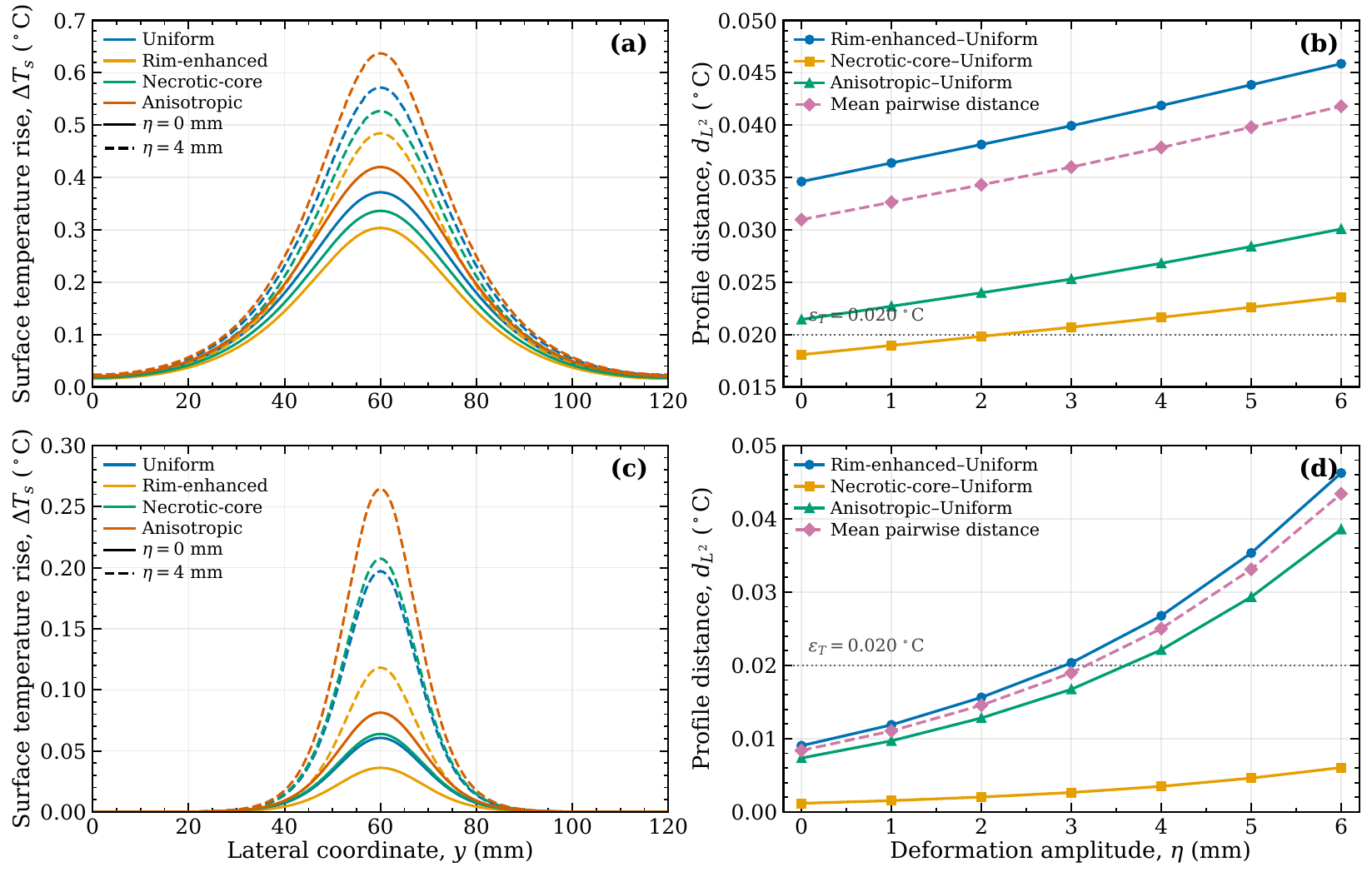}
\caption{
Influence of outer-surface geometry on thermal-signature
distinguishability.
Panels (a,b) correspond to the Khomsi-based parameter set, and panels
(c,d) to the Lozano-inspired parameter set. Panels (a,c) compare surface
temperature-rise profiles $\Delta T_s(y)$ for flat
($\eta=0$~mm, solid lines) and deformed
($\eta=4$~mm, dashed lines) geometries. Panels (b,d) show
three profile distances relative to the uniform case
and the mean distance over all six profile pairs as functions of
deformation amplitude $\eta$; the dotted horizontal line marks
$\varepsilon_T=0.020~^\circ\mathrm{C}$.
Because the tumor coordinate is fixed, its local depth
varies as $d_t=22-\eta$~mm; therefore, the trends reflect both surface
deformation and the associated change in overlying tissue thickness.
}
\label{fig:fig09_new}
\end{figure*}

Another anatomical factor that can affect the surface signature is the
outer-surface geometry. Figure~\ref{fig:fig09_new} shows the effect of the
deformation amplitude $\eta$ for the Khomsi-based and Lozano-inspired parameter
sets. Panels (a) and (c) compare the surface profiles for the flat geometry,
$\eta=0~\mathrm{mm}$, and a deformed geometry,
$\eta=4~\mathrm{mm}$. In both parameter sets, increasing
$\eta$ changes the amplitude and width of $\Delta T_s(y)$. However, because
the tumor coordinate is held fixed as the outer surface is deformed, these
changes reflect both the modified surface shape and the accompanying reduction
in the thickness of the tissue overlying the tumor.

Panels (b) and (d) show the corresponding profile distances as functions of
$\eta$. The local tumor-center depth varies according to
$d_t=22~\mathrm{mm}-\eta$; consequently, increasing $\eta$ from $0$ to
$6~\mathrm{mm}$ reduces the overlying tissue thickness from $22$ to
$16~\mathrm{mm}$. For the Khomsi-based case, the
rim-enhanced--uniform distance is the largest over the whole range, followed by
the mean pairwise distance, anisotropic--uniform distance, and
necrotic-core--uniform distance. The mean pairwise distance
increases from approximately $0.0310~^\circ\mathrm{C}$ at $\eta=0$ to
$0.0418~^\circ\mathrm{C}$ at $\eta=6~\mathrm{mm}$, corresponding to an
increase of approximately $35\%$. For the Lozano-inspired case, the
rim-enhanced--uniform, anisotropic--uniform, and mean pairwise distances
increase more noticeably, whereas the necrotic-core--uniform distance remains
small. The mean pairwise distance increases from approximately
$0.0084$ to $0.0435~^\circ\mathrm{C}$, representing an increase by a factor of
approximately $5.2$. Several Lozano-inspired comparisons consequently cross
the $\varepsilon_T=0.020~^\circ\mathrm{C}$ tolerance between approximately
$\eta=3$ and $4~\mathrm{mm}$, whereas the necrotic-core--uniform comparison
remains below the tolerance throughout the investigated range. These trends
should therefore be interpreted as a coupled geometry--depth effect rather
than as an isolated effect of outer-surface deformation.

\begin{figure*}[pos = tb]
\centering
\includegraphics[width=0.95\linewidth]{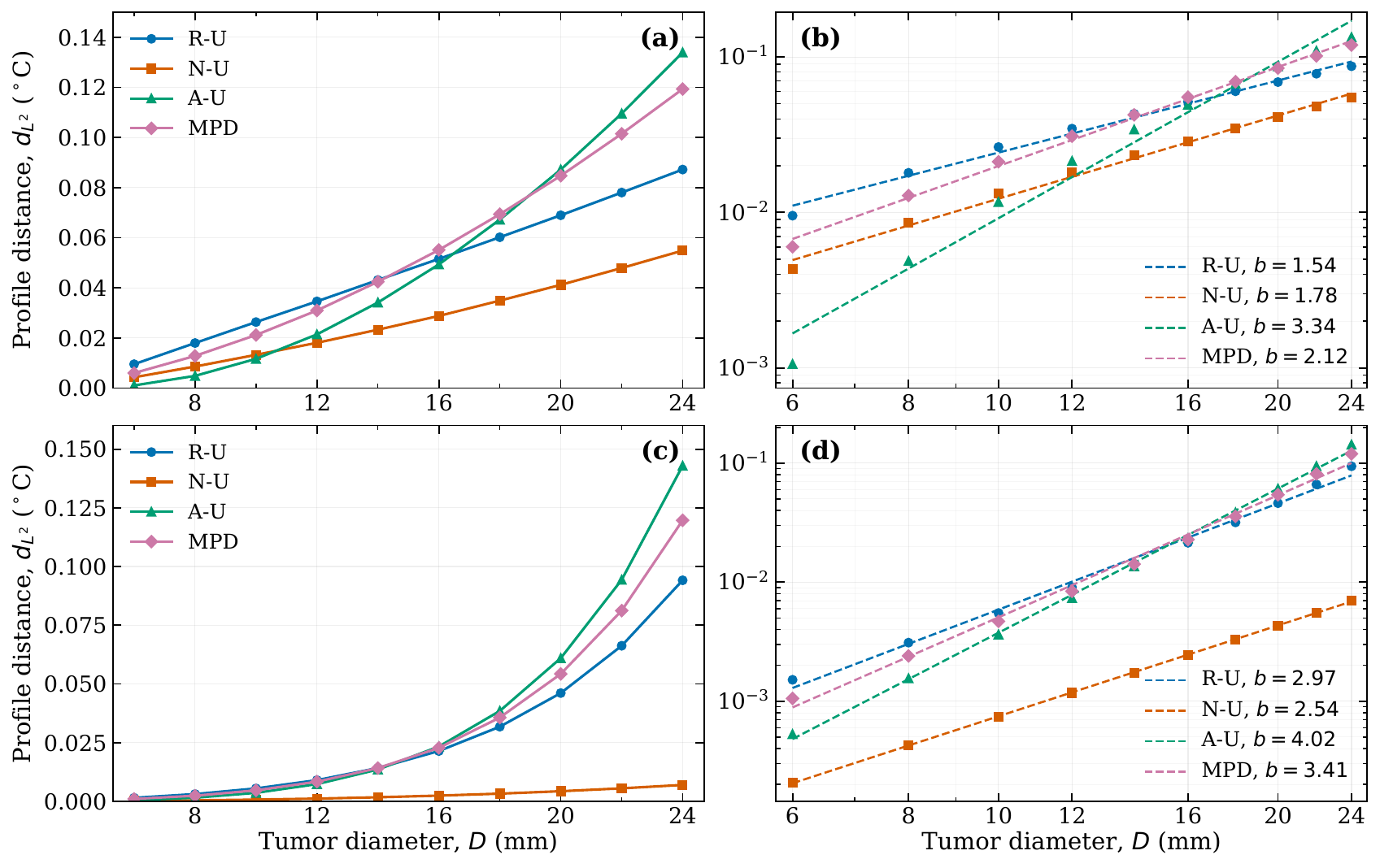}
\caption{
Tumor-diameter dependence of thermal-signature distinguishability.
Panels (a) and (b) correspond to the Khomsi-based
parameter set, whereas panels (c) and (d) correspond to the
Lozano-inspired parameter set.
Panels (a,c) show the surface-profile distance $d_{L^2}$ as a function
of tumor diameter $D$ for each heterogeneous perfusion pattern relative
to the uniform-perfusion case, together with the mean pairwise distance.
Here, R--U, N--U, and A--U denote
rim-enhanced--uniform, necrotic-core--uniform, and
anisotropic--uniform comparisons, respectively, while MPD denotes the
mean distance over all six pairwise comparisons.
Panels (b,d) show the same data on logarithmic  scales, with dashed lines
denoting least-squares finite-range fits in logarithmic
space according to $d_{L^2}(D)=A D^b$.
Fit statistics are reported in Table~\ref{tab:size_power_law_fit} in Appendix~\ref{app:power_law_fit}.
Within the investigated diameter range and the fixed
two-dimensional geometry, increasing tumor diameter generally enhances
surface-signature distinguishability, whereas smaller tumors are more
prone to thermal-signature equivalence.
}
\label{fig:fig10_new}
\end{figure*}

After examining parameters that weaken or modulate thermal observability, we
next consider the role of tumor size at fixed tumor-center depth. In contrast
to depth, which mainly controls the propagation path through the overlying
tissue, the tumor diameter $D$ controls the spatial extent of the thermal
perturbation generated inside the domain. Figure~\ref{fig:fig10_new} shows the
dependence of $d_{L^2}$ on $D$ for the Khomsi-based and Lozano-inspired
parameter sets. Panels (a) and (c) present the data on a linear scale, while
panels (b) and (d) show the corresponding log--log representations.

For both parameter sets, $d_{L^2}$ increases with tumor diameter for all
perfusion comparisons, indicating that larger tumors produce
greater separation among the modeled surface thermal signatures. Over the
investigated range $D=6$--$24~\mathrm{mm}$, the mean pairwise distance for the
Khomsi-based set increases from $0.006003$ to
$0.119343~^\circ\mathrm{C}$, corresponding to a factor of approximately
$19.9$. For the Lozano-inspired set, it increases from $0.001055$ to
$0.119699~^\circ\mathrm{C}$, corresponding to a substantially larger factor
of approximately $113.5$. In the Khomsi-based case, the
anisotropic--uniform distance grows most rapidly and becomes the largest
distance for large tumors, followed by the mean pairwise distance, the
rim-enhanced--uniform distance, and the necrotic-core--uniform distance. The
Lozano-inspired case shows the same qualitative behavior, but with a sharper
increase in the anisotropic--uniform and mean pairwise distances, while the
necrotic-core--uniform distance remains comparatively small.
Thus, although increasing diameter consistently enhances
surface-profile distinguishability, the magnitude of this enhancement depends
strongly on both the thermophysical parameter set and the spatial structure of
the perfusion contrast.

The log--log panels show that the simulated trends can be summarized
empirically by $d_{L^2}(D)=A D^b$ over the diameter range considered here. For
the Khomsi-based set, the fitted exponents are $b=1.54$ for
rim-enhanced--uniform, $b=1.78$ for necrotic-core--uniform, $b=3.34$ for
anisotropic--uniform, and $b=2.12$ for the mean pairwise distance. For the
Lozano-inspired set, the corresponding values are $b=2.97$, $2.54$, $4.02$,
and $3.41$. The anisotropic comparison therefore shows the strongest diameter
dependence in both parameter sets. These fits are used only as
finite-range empirical summaries of the simulated data, not as universal
asymptotic scaling laws. The corresponding regression statistics are reported
in Table~\ref{tab:size_power_law_fit}.

The fitted exponents should not be compared directly with the leading-order
quadratic size estimate in Appendix~\ref{app:scaling_depth_size}. That estimate
assumes a small source with size-independent effective contrast and a slowly
varying propagation kernel. In contrast, the numerical quantity $d_{L^2}$ is a
distance between two surface profiles, so it depends on the spatial structure of
the perfusion contrast, finite-depth screening, tissue layering, and boundary
effects. The exponent is therefore expected to vary with the parameter set and
the perfusion comparison.

The separate depth and diameter sweeps indicate opposite effects: increasing
depth suppresses surface distinguishability, whereas increasing diameter
enhances it. To check that this conclusion is not only a consequence of separate
OAT sweeps, we computed a limited two-parameter depth--diameter map
for the Khomsi-based parameter set. For each point in the $(d_t,D)$ plane, the
four perfusion scenarios were simulated and the six pairwise surface-profile
distances were used to compute the mean pairwise distance $\bar d_{L^2}$ and
the equivalent-pair fraction $f_{\mathrm{eq}}$.

\begin{figure*}[pos = tb]
\centering
\includegraphics[width=0.9\linewidth]{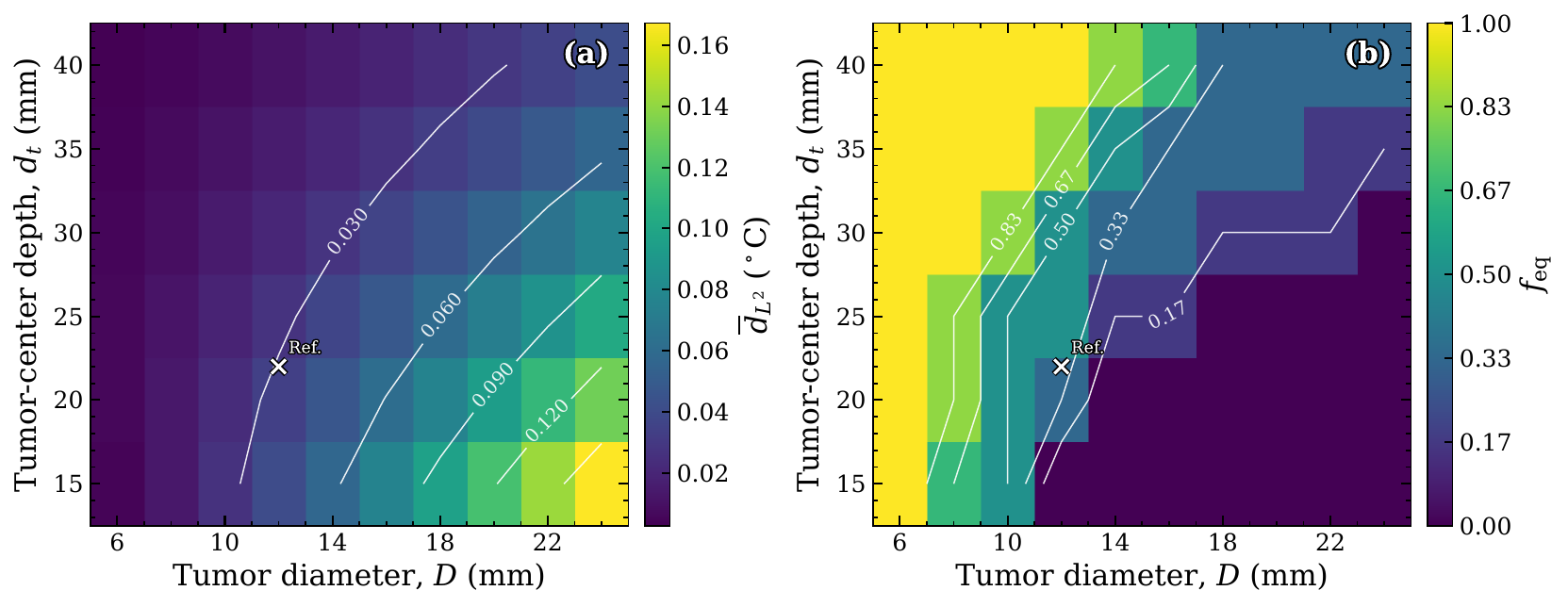}
\caption{
Two-parameter depth--diameter map of thermal-signature distinguishability
for the raw absolute-perfusion scenarios of the
Khomsi-based parameter set. (a) Mean pairwise surface-profile distance
$\overline{d}_{L^2}$ calculated over the six unique
comparisons among the four perfusion patterns. (b) Equivalent-pair
fraction $f_{\mathrm{eq}}$, defined as the fraction of
these six comparisons satisfying
$d_{L^2}\leq\varepsilon_T$, with
$\varepsilon_T=0.020~^\circ\mathrm{C}$;
consequently, $f_{\mathrm{eq}}$ varies in increments
of $1/6$. White contours indicate the labeled metric levels.
Within the investigated parameter ranges, increasing tumor-center depth
strengthens thermal screening, decreasing
$\overline{d}_{L^2}$ and promoting equivalence, whereas increasing
tumor diameter generally enhances surface-profile
distinguishability. The cross labeled ``Ref.'' marks the representative
case ($D=12$~mm, tumor-center depth $=22$~mm)
used in the preceding analyses.
}
\label{fig:fig11_new}
\end{figure*}

\begin{figure*}[pos = tb!]
\centering
\includegraphics[width=0.9\linewidth]{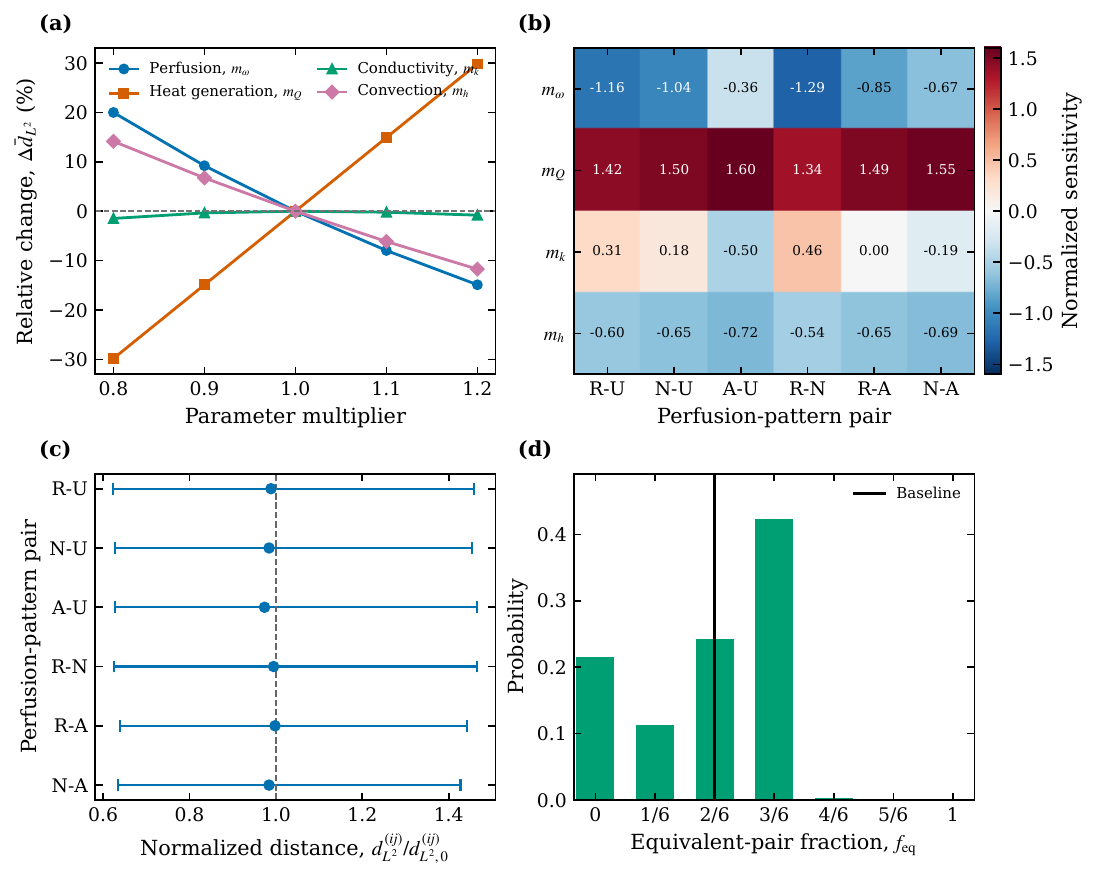}
\caption{
Sensitivity and uncertainty of surface thermal-signature
distinguishability to blood perfusion, volumetric heat generation,
thermal conductivity, and surface convection for the representative
Khomsi-based configuration. The geometry and shapes of the four
raw perfusion patterns are fixed.
(a) Percentage change in the mean pairwise surface-profile distance,
$\overline{d}_{L^2}$, relative to its baseline value when each
parameter-group multiplier is varied individually from $0.8$ to
$1.2$, with the other multipliers held at unity.
(b) Pair-specific normalized sensitivities estimated from the
endpoints of the same OAT analysis. Positive values indicate
increasing profile separation, whereas negative values indicate
increasing similarity. U, R, N, and A denote uniform, rim-enhanced,
necrotic-core, and anisotropic perfusion, respectively; paired
letters identify the corresponding comparisons.
(c) Medians and central 95\% empirical ranges
(2.5th--97.5th percentiles) of the six normalized pairwise
distances, $d_{L^2}^{(ij)}/d_{L^2,0}^{(ij)}$, across 300 LHS
configurations. Here, $d_{L^2,0}^{(ij)}$ is the baseline distance
with all four multipliers set to unity. The dashed vertical
line marks a normalized distance of one.
(d) Empirical relative frequencies of the equivalent-pair
fraction $f_{\mathrm{eq}}$ across the same LHS ensemble.
For each configuration, $f_{\mathrm{eq}}$ is the fraction of
the six unique profile pairs satisfying
$d_{L^2}^{(ij)}\leq\varepsilon_T$, with
$\varepsilon_T=0.020~^\circ\mathrm{C}$.
The solid vertical line marks the baseline value of
$f_{\mathrm{eq}}$.
For panels (c,d), the four group multipliers were sampled
independently from uniform distributions over $[0.8,1.2]$.
Global rescaling of the tissue parameters preserves their
relative contrasts. The percentile ranges and relative
frequencies are conditional summaries of this prescribed
computational ensemble.
}
\label{fig:fig12_new}
\end{figure*}

The resulting maps are shown in Figure~\ref{fig:fig11_new}. Panel (a) shows that
$\bar d_{L^2}$ increases toward shallow and large tumors,
indicating greater separation among the modeled surface
thermal signatures. It decreases toward deeper and smaller tumors, where
thermal screening dominates. Panel (b), computed using
$\varepsilon_T=0.020~^\circ\mathrm{C}$, shows the complementary trend:
the equivalent-pair fraction reaches values close to unity for
small, deep tumors and decreases to zero for sufficiently large, shallow
tumors. Because only six unique profile pairs are compared, $f_{\mathrm{eq}}$
changes discretely in increments of $1/6$ as individual pairs cross the
prescribed tolerance. Thus, rather than constituting a continuous transition,
the equivalence boundaries shift systematically across the combined
depth--diameter parameter space. At the representative
configuration, $D=12~\mathrm{mm}$ and $d_t=22~\mathrm{mm}$, the mean pairwise
distance is approximately $0.03097~^\circ\mathrm{C}$ and
$f_{\mathrm{eq}}=2/6$. The representative case therefore lies in an
intermediate region, making it suitable for illustrating both distinguishable
and thermally equivalent profile pairs.

Figures~\ref{fig:fig03_new}--\ref{fig:fig11_new} show that intratumoral
perfusion patterns that remain clearly distinct inside the tissue can become
substantially less separable at the surface. Increasing tumor-center depth
strengthens thermal screening, whereas increasing tumor diameter enlarges the
thermal perturbation and makes perfusion-induced differences more visible. The
depth--diameter map confirms that these opposing effects persist when both
parameters are varied simultaneously. Consequently, larger, shallower tumors
are more likely to retain surface-observable information about perfusion
heterogeneity, whereas smaller or deeper tumors may still produce a detectable
thermal anomaly whose internal perfusion organization is difficult to identify
from the surface profile alone
\citep{mukhmetov2021inverse,gutierrez2025detectability,
sritharan2024breast}. Fat-layer thickness and outer-surface geometry also
modify the surface response but act mainly as secondary modulators over the
parameter ranges considered here. Thermal-signature equivalence in the present
model therefore arises when physically distinct intratumoral perfusion
organizations are mapped to weakly distinguishable, or effectively
indistinguishable, profiles in static surface thermography.

From a clinical perspective, the proposed framework addresses
a complementary problem to disease classification by quantifying when
different intratumoral perfusion organizations can produce observationally
similar surface thermal anomalies. Subject to appropriate validation, it could
support thermography-based screening as a physics-informed confidence layer,
identifying cases in which surface measurements alone do not justify a unique
physiological interpretation and corroboration by other imaging modalities may
be warranted. Evaluation alongside artificial-intelligence classifiers or
inverse thermal models, including the estimation of sensitivity, specificity,
and clinical decision thresholds, lies beyond the present forward-modeling
scope and would require independent labeled thermographic data.

{\color{black}
The present interpretation is restricted to stationary surface-temperature
profiles obtained with time-independent perfusion, metabolic heat
generation, and boundary conditions. This approximation describes the
response after thermal relaxation under fixed inputs. Its applicability to an acquisition would require residual thermal
drift and surface-temperature changes induced by physiological or environmental variability during the measurement interval to remain sufficiently small relative to the prescribed observational tolerance. A quasi-steady
interpretation may be appropriate when these inputs vary much more slowly than the relevant tissue thermal-relaxation times. Otherwise, heat storage and the history of the thermal forcing must be considered. The stationary solutions reported here should therefore not generally be interpreted as time averages of a physiologically varying system.

A transient extension can be formulated within the Pennes framework by retaining tissue heat storage. For fixed geometry, thermal conductivity, and arterial blood temperature, an illustrative extension of Eq.~\eqref{eq:pennes} is
\begin{equation}
 \begin{aligned}
C_v(\mathbf{r})\frac{\partial T}{\partial t}
={}&
\nabla\cdot\!\left[k(\mathbf{r})\nabla T\right]
\\
&-\rho_b c_b\,\omega(\mathbf{r},t)\left(T-T_a\right)
+Q_m(\mathbf{r},t),
\end{aligned}   
\end{equation}
where $C_v(\mathbf{r})$ is the tissue volumetric heat capacity,
$\omega(\mathbf{r},t)$ is the time-dependent perfusion field, and
$Q_m(\mathbf{r},t)$ is the time-dependent metabolic heat-generation
rate. Such a model would additionally require specified initial
conditions and the boundary-condition histories associated with the
measurement protocol. A changing surface heating or cooling stimulus
can produce transient temperature fields even when perfusion and
metabolic heat generation remain constant. Retaining this time
dependence does not, by itself, require a non-Fourier heat-conduction
model. Numerical optimization of cooling protocols for breast
thermography illustrates the importance of the imposed thermal
stimulus~\cite{ZhouHerman2018Cooling}.

Time-dependent perfusion would modify both blood-mediated heat exchange
and the rate at which the tissue temperature responds to a perturbation.
The direction of this exchange depends on the local temperature
relative to $T_a$: increased perfusion removes heat where $T>T_a$ but
supplies heat where $T<T_a$, including tissue cooled below arterial
blood temperature. Consequently, the steady-state sensitivity trends
reported here cannot be assumed to apply throughout a cooling and
recovery sequence. Active breast thermography has investigated vascular
responses to a temperature stimulus, including vasomodulation,
supporting the relevance of physiological responses during dynamic
acquisition~\cite{Gershenson2023Dynamic}. Time-dependent metabolic heat
generation would likewise alter the internal thermal forcing, while
heat storage and transport through the overlying tissue would smooth
and delay its manifestation at the surface. The magnitude and time
scale of these physiological changes would require justification for
the selected protocol. Changes in perfusion and heat generation may
also produce partially compensating surface responses, so additional
temporal information does not automatically provide a unique
physiological interpretation.

Dynamic thermography would compare temperature histories rather than
only stationary profiles. Two perfusion configurations with similar
stationary signatures could exhibit different cooling rates, recovery
rates, or evolving spatial profile shapes. Their separation could
therefore exceed the observational tolerance during part of a transient,
even if their final profiles satisfy Eq.~\eqref{eq:equiv_main}. Conversely, differences
may remain below the measurement resolution, or uncertainty in the
physiological inputs and thermal stimulus may obscure them. The present
OAT sweeps and Latin-hypercube analysis compare stationary solutions
under different parameter choices, while the noisy-profile tests
perturb the resulting spatial observations. These analyses do not
resolve physiological time courses, thermal response delays, or
correlations between successive thermographic frames.

One possible extension of the equivalence framework would apply
Eq.~\eqref{eq:L2_main} at each time to the profiles
$\Delta T_s^{(i)}(y,t)$ and $\Delta T_s^{(j)}(y,t)$, giving an
instantaneous spatial distance $d_{L_2}^{(ij)}(t)$. For consistency with Eq.~\eqref{eq:deltaTs}, the simulated temperature-rise history
would be defined as
$\Delta T_s^{(i)}(y,t)
=T_s^{(i)}(y,t)-T_s^{(\mathrm{healthy})}(y,t)$,
with the healthy-background response evaluated under the same
preparation protocol and boundary-condition history. A temperature
change measured relative to each configuration's own pre-stimulus
baseline would constitute a different observable and would require
a corresponding definition of the comparison metric. For a common
observation interval $[t_0,t_1]$, a distance incorporating the complete
recorded histories could then be defined as
\begin{equation}
 d_{\mathrm{dyn}}^{(ij)}
=
\left[
\frac{1}{t_1-t_0}
\int_{t_0}^{t_1}
\left[d_{L_2}^{(ij)}(t)\right]^2\,\mathrm{d}t
\right]^{1/2}.   
\end{equation}
This quantity has units of temperature and reduces to the existing
spatial distance when both profiles are time independent. Dynamic
equivalence could be assessed using
$d_{\mathrm{dyn}}^{(ij)}\leq\varepsilon_{\mathrm{dyn}}$, where the
tolerance is specified for the chosen acquisition and analysis
protocol. Because averaging over time can reduce the contribution of
brief differences, the observation window and temporal weighting
would need to be specified in advance; a maximum-over-time distance
could instead be considered when resolving short-lived differences is
the objective. These alternatives express different observational
requirements and would require separate evaluation.

A meaningful dynamic comparison would require a common preparation and
stimulation protocol, consistently defined initial conditions,
synchronized observations, spatial registration, and a justified
temperature-reference strategy. Its uncertainty analysis would need to
account for temporal sampling, camera response, measurement drift,
physiological repeatability, and correlations across both space and
time. Accordingly, $\varepsilon_{\mathrm{dyn}}$ cannot be assigned
automatically from the static tolerances used here. The transient
formulation and temporal distance above are proposed extensions and
have not been evaluated in the present study. The reported pairwise
distances, equivalent-pair fractions, and threshold crossings therefore
remain conditional on stationary observations; whether dynamic
thermography resolves any of the identified equivalent pairs remains
an open question requiring transient modeling and matched
time-resolved measurements.
}

The geometrical studies were conducted primarily through OAT
sweeps of tumor depth, tumor diameter, fat-layer thickness, and
outer-surface deformation, while the depth--diameter map provided a
limited two-parameter assessment of the dominant competing geometrical
effects. Separately, the sensitivity and uncertainty analyses described
in Sec.~\ref{sec:sensitivity_uncertainty} were used to assess whether the
preceding distinguishability trends remain stable under controlled
variations in the principal thermophysical and physiological parameter
groups.

The OAT results are shown in Figure~\ref{fig:fig12_new}(a, b). Volumetric heat generation
produces the strongest positive change in the mean pairwise surface-profile
distance $\overline{d}_{L^2}$, which decreases by approximately $30\%$
when $m_Q=0.8$ and increases by approximately $30\%$ when $m_Q=1.2$.
Perfusion and surface convection have the opposite effect: increasing either
parameter reduces the separation among the surface signatures and increases
the tendency toward thermal equivalence. Thermal conductivity has only a weak
net effect on $\overline{d}_{L^2}$, although its influence depends on the
specific pair of perfusion patterns. The normalized pairwise sensitivities in
Figure~\ref{fig:fig12_new}(b) quantify this ordering. Heat
generation produces consistently positive sensitivities ranging from $1.34$ to
$1.60$, whereas perfusion produces negative sensitivities ranging from
$-1.29$ to $-0.36$. Surface convection also produces consistently negative
sensitivities, within the narrower range $-0.72$ to $-0.54$. By contrast,
thermal conductivity gives smaller, pair-dependent values ranging from
$-0.50$ to $0.46$, indicating that its effect can change sign depending on
the perfusion-pattern pair.

{\color{black}
The joint uncertainty results are summarized in Figure~\ref{fig:fig12_new}(c,d)
for the 300-point LHS ensemble defined in Sec.~3.3. The central
95\% empirical range, corresponding to the 2.5th--97.5th
percentiles, of the mean pairwise distance $\overline{d}_{L_2}$
is approximately $0.019$--$0.044~^\circ\mathrm{C}$. The medians
of the six normalized pairwise distances in Figure~\ref{fig:fig12_new}(c) remain
close to unity, indicating that their median values are close
to the corresponding baseline distances. Within this ensemble,
the heat-generation multiplier has the strongest positive
Spearman rank association with $\overline{d}_{L_2}$,
with $\rho\simeq0.82$. The convection and perfusion multipliers
have negative associations of approximately $-0.46$ and $-0.43$,
respectively, whereas the conductivity multiplier has a weaker
overall association, $\rho\simeq-0.07$. These coefficients summarize the direction and strength of the
monotonic associations between the sampled parameter multipliers
and the mean pairwise surface-profile distance within the prescribed
LHS ensemble.

At $\varepsilon_T=0.020~^\circ\mathrm{C}$, approximately $79\%$
of the 300 sampled parameter combinations yield at least one
thermally equivalent pair, corresponding to
$f_{\mathrm{eq}}\geq 1/6$. Thus, the occurrence of thermal-signature
equivalence persists across many configurations under simultaneous
perturbations of the four parameter groups within the prescribed
ensemble.
}

{\color{black}
The uncertainty analysis provides a controlled assessment of robustness
around the representative Khomsi-based configuration. The independent
uniform multipliers over $[0.8,1.2]$ define a prescribed ensemble of
$\pm20\%$ parameter perturbations; their ranges and probability
distributions were not estimated from a clinical cohort. Independence
is assumed between the four input groups, whereas the common
multiplier within each group imposes proportional changes across
the corresponding healthy-tissue and tumor parameters. Consequently,
relative tissue contrasts are preserved, and tissue-specific
properties cannot vary independently within a group. The LHS
ensemble also retains the representative tumor size and depth,
tissue-layer geometry, and shapes of the four raw perfusion
patterns. These choices isolate the response to specified parameter
changes while restricting the anatomical and physiological
variability represented by the ensemble.

The reported Spearman coefficients do not decompose output variance
into individual parameter contributions and interaction effects.
As input--output association measures, they also do not estimate
physiological correlations among tissue properties. Their magnitudes
and relative ordering remain conditional on the prescribed parameter
ranges and sampling assumptions. Likewise, the spatially correlated
noisy-profile tests describe correlation in the processed surface
observations rather than a joint distribution of physiological
inputs. Although geometrical effects were investigated separately,
their joint variation with tissue properties was not sampled in
the present LHS analysis. Correlations among inputs could reinforce
or compensate their effects on the surface profiles, potentially
broadening or narrowing the resulting distance distributions.
The direction and magnitude of these changes cannot be inferred
from the independent-multiplier ensemble alone.

The equivalent-pair frequencies and output percentile ranges are
conditional on the prescribed independent uniform parameter
multipliers, fixed geometry, and selected perfusion-pattern families.
They characterize variability across the computational ensemble and
should not be interpreted as patient-population probabilities or
clinical confidence intervals.

A patient-oriented extension would distinguish physiological
variation between patients from uncertainty in estimating the
parameters of an individual patient. Joint input distributions
would need to be informed by anatomical and physiological
measurements and their uncertainties, allowing tissue-specific
properties and their relationships with anatomy and acquisition
conditions to be represented. These distributions could then be
propagated through an appropriately matched forward model to assess
surface profiles, pairwise distances, and equivalence classifications
under a prespecified observation protocol. Clinical generalizability
would require evaluation on subjects not used to estimate the
population-level distributions, with thermographic predictions
assessed using data withheld from model calibration. The present
screening provides a controlled baseline for determining how such
data-informed uncertainty models alter perfusion-pattern
distinguishability.
}

\begingroup
\color{black}

The mesh-refinement tests, explicit residual evaluations, closest-pair
convergence test, and analytical transfer-matrix comparison reported in
Sec.~\ref{sec:numerical_imp} and Appendices~\ref{app:benchmark_1d} and~\ref{app:mesh_closest_pair} provide numerical verification of the
computational implementation. These checks assess algebraic convergence
and the discretization sensitivity of the monitored temperature and
surface-profile quantities, with analytical agreement established for a
reduced, laterally uniform healthy-tissue problem. They do not establish
the accuracy of the physical model against independent measurements,
consistent with the distinction between verification and validation in
computational modeling~\cite{RoyOberkampf2011}. No experimental or clinical
thermographic dataset matched to the modeled geometry, material
properties, heat sources, perfusion representation, and boundary
conditions was analyzed in the present study. Experimental validation of
the predicted surface profiles and their equivalence classifications
therefore remains outstanding.

Published studies provide external context for the predictions.
Mukhmetov et al.~\cite{mukhmetov2021thermal} combined three-dimensional breast thermal modeling with phantom experiments and demonstrated the importance of breast geometry and heat-source depth. In the computational part of that study, the maximum
surface-temperature differences between breast models with and
without tumors ranged from $0.34$ to $1.37~^\circ\mathrm{C}$.
The maximum surface-temperature rise of
$0.336~^\circ\mathrm{C}$ predicted for the present representative
necrotic-core case is comparable to the lower end of this range.
Bezerra et al.~\cite{bezerra2020empirical} investigated the inverse
estimation of breast and nodule thermal properties from
thermographic measurements. These studies illustrate the importance of geometry and thermophysical inputs, but their configurations and
parameter-estimation procedures differ from the present controlled comparison. Agreement in the magnitude of the peak temperature rise therefore provides only a physical-consistency reference; it does not quantify the model-form error of the present formulation or validate its small inter-pattern profile distances.

An initial validation study could use a multilayer tissue-mimicking
phantom with a finite heated inclusion and controllable regional heat
exchange. A direct test of the present two-dimensional formulation would
require approximately uniform conditions in the out-of-plane direction,
with residual end effects assessed; otherwise, the forward model should
reproduce the three-dimensional phantom geometry. Layer thicknesses,
inclusion size and depth, thermal properties, heating power, and boundary
conditions would be characterized independently of the thermograms used
for validation. Controlled fluid channels or heat-exchange elements
could approximate the four prescribed perfusion scenarios, provided
that their effective volumetric heat exchange is characterized over the
relevant temperature range and related to the Pennes perfusion term.
Heat generation and perfusion-mediated exchange would be controlled
separately, including mean-matched perfusion configurations at fixed heat
generation to test the contribution of spatial organization.

After thermal drift becomes negligible relative to the measurement
uncertainty, repeated calibrated infrared measurements would be acquired
for each configuration. A healthy-reference measurement would reproduce
the modeled background thermal properties, heat generation, and
heat-exchange conditions when evaluating individual temperature-rise
profiles. Numerical predictions would be mapped to the camera sampling
and effective spatial resolution, and registered profiles would be
compared over a common observation interval using the normalization in
Eq.~\eqref{eq:L2_main}. Validation would assess the maximum temperature rise, profile
width, complete profile shape, and model--measurement discrepancies
together with numerical, input, and measurement uncertainties. The
measurement analysis would include calibration, emissivity,
environmental drift, registration, repeatability, and the joint error
structure of the compared profiles. In particular, noise contributions
to the squared profile distance and correlations introduced by shared
measurements would be accounted for. For configurations sharing the same
healthy reference, that reference cancels in the pairwise profile
difference, allowing the corresponding distance to be evaluated directly
from the two registered temperature profiles.

The equivalence criterion would be tested by comparing predicted and
experimental pairwise $d_{L_2}$ distances at a tolerance $\varepsilon_T$
specified before evaluating the results and justified by the measurement
capability and intended resolution. Tests would include both predicted
equivalent and distinguishable pairs over selected inclusion depths and
sizes. With uncertainty-interval construction and coverage specified in
advance, an upper distance bound at or below $\varepsilon_T$ would support
equivalence, whereas a lower bound above $\varepsilon_T$ would support
distinguishability. An interval crossing the tolerance would indicate an
inconclusive classification. These assessments would be accompanied by
the profile-level model--measurement comparisons, because matching a
binary equivalence classification alone would not establish accurate
temperature prediction. Persistent discrepancies beyond the quantified
uncertainties would motivate reassessment of the modeling assumptions.

A subsequent clinical assessment would require a geometrically matched
three-dimensional model, standardized repeated thermography, and
independent anatomical and perfusion-sensitive measurements. Surface
scanning and anatomical imaging could constrain breast geometry and
tumor morphology, while dynamic contrast-enhanced magnetic-resonance
imaging could provide additional physiological information. Its
tracer-kinetic parameters are model dependent, however, and their relation
to the effective Pennes perfusion coefficient would require explicit
justification and uncertainty assessment~\cite{SourbronBuckley2011}.
Because an individual patient's healthy counterfactual is not directly
available, clinical testing could begin with absolute
surface-temperature predictions; evaluating individual temperature-rise
profiles would additionally require a justified reference strategy.
Predictions would be compared with thermograms not used for parameter
calibration or model selection, using prespecified measures of temperature
bias, profile distance, and uncertainty. Such comparisons would test
forward predictions but would not, by themselves, establish the internal
perfusion ground truth needed for a direct test of inter-pattern
equivalence. Until matched experimental and clinical assessments are
performed, the present findings should be interpreted as conditional
predictions of a controlled forward model, without inferring clinical
sensitivity, specificity, or diagnostic performance.

\endgroup

{\color{black}
The Cartesian two-dimensional formulation provides a controlled setting
for quantifying the influence of intratumoral perfusion on complete
surface thermal signatures. Combined with the mean-matched controls,
it separates differences associated with mean perfusion from those
associated with spatial organization under the same geometrical and
boundary conditions. The model resolves heat conduction in the $x$
and $y$ directions and assumes invariance in the out-of-plane direction.
The circular tumor inclusion therefore has finite extent within the
modeled plane, while its out-of-plane extent is not resolved. A
three-dimensional extension would retain the steady-state Pennes heat
balance while allowing temperature gradients and heat redistribution
in the third direction. This additional spatial dependence can modify
the maximum surface-temperature rise, the extent and shape of the
thermal anomaly, and the separation between perfusion signatures.
Additional spreading can attenuate localized temperature differences,
but the direction and magnitude of the change relative to the present
predictions depend on the source representation, surrounding anatomy,
and boundary conditions.

Irregular tumor morphology introduces further geometrical effects.
Tumors with the same volume and center depth can have different minimum
distances from the skin, different orientations, and different spatial
distributions of heat generation and perfusion relative to the surface.
For example, an extension toward the skin shortens the local propagation
path and may increase the surface contribution of that part of the
tumor. Elongation or asymmetry may also shift the hotspot or change the
directional spread of the surface anomaly. The orientation of a
heterogeneous perfusion field relative to the skin can consequently
affect its distinguishability from uniform perfusion. Patient-specific
variations in breast curvature, tissue-layer thicknesses, and distance
from the chest wall additionally alter conduction paths and the geometry
through which boundary heat exchange occurs. Lozano et al.~\cite{lozano2020determining}
incorporated measured three-dimensional surface geometry and
imaging-derived tumor information into a subject-specific thermal model,
while Mukhmetov et al.~\cite{mukhmetov2021thermal} demonstrated the importance of breast
geometry and tumor depth through three-dimensional modeling and phantom
experiments. These studies provide physical context for extending the
present analysis to realistic anatomy. The surface-deformation sweep
in Figure~\ref{fig:fig09_new} already illustrates a coupled geometry--depth effect within
the modeled cross-section; a three-dimensional analysis would additionally
resolve out-of-plane curvature and anatomical variation.

A meaningful dimensional comparison must specify how the tumor source
is represented. Integrating a volumetric metabolic heat-generation rate
over a two-dimensional cross-section gives power per unit out-of-plane
length, whereas integration over a finite three-dimensional tumor gives
total power. Equal tumor radius and center depth therefore do not, by
themselves, define equivalent heat inputs. Comparisons should specify
the out-of-plane extent or an appropriate source-power normalization,
together with the material properties and boundary conditions. Within
three-dimensional morphology tests, equal-volume tumors could be compared
at fixed center depth, using a consistent depth definition and also
reporting their minimum distance from the skin. Comparisons among
perfusion patterns should retain a common tumor geometry and metabolic
heat-generation field. The mean-matched control in Eq.~\eqref{eq:mean_matched_perfusion} would then
use a volume average of perfusion, replacing the cross-sectional area
average in Eq.~\eqref{eq:raw_mean_perfusion}. This distinction matters because the relative
contributions of a low-perfusion core and a perfused rim depend on their
volume fractions, and directional patterns require a specified
orientation in three dimensions. These controls allow the influence of
spatial organization to be assessed at the same mean perfusion.

The consequences for thermal-signature equivalence follow from the
distance criterion in Eq.~\eqref{eq:equiv_main}. For a fixed observation protocol,
comparison metric, and prescribed tolerance $\varepsilon_T$, changes
in geometry modify the calculated pairwise distances and may therefore
change the equivalent-pair fraction, the ordering of profile similarities,
and the depth--diameter boundaries at which a distance crosses the
tolerance. Geometry does not itself prescribe $\varepsilon_T$, which
represents the observational resolution of the comparison. The relevant
question is therefore how geometrical changes move particular pairs
relative to a specified tolerance. Additional spreading may reduce some
profile differences, whereas variations in tumor shape, orientation,
or proximity to the surface may increase others. The resulting
classification boundaries would need to be evaluated for each
three-dimensional geometry rather than inferred directly from the
numerical boundaries obtained for the present cross-section.

In three dimensions, the observable can also be extended from a line
profile to a temperature-rise map on an observed skin-surface region
$S$. One possible generalization of Eq.~\eqref{eq:L2_main} is
\begin{equation}
 d_{L_2,S}^{(ij)}
=
\left[
\frac{1}{A_S}
\int_S
\left[
\Delta T_s^{(i)}(\mathbf{s})
-
\Delta T_s^{(j)}(\mathbf{s})
\right]^2
\,\mathrm{d}A
\right]^{1/2},   
\end{equation}
where $\mathbf{s}$ denotes a point on the surface,
$A_S=\int_S\mathrm{d}A$, and the two temperature-rise maps use a
consistently defined healthy reference on the same geometry. This
area-normalized distance has units of temperature and could be compared
with a prespecified tolerance $\varepsilon_{T,S}$ appropriate to the
surface-map analysis. The observation region, surface registration,
and area weighting would need to be specified, including the mapping
between camera pixels and the curved skin surface when experimental
images are used. Enlarging the comparison region to include areas with
negligible inter-pattern differences can reduce the normalized distance.
Consequently, the observation support is part of the criterion, and the
tolerance should reflect the sampling and uncertainty of the selected
surface-map comparison.

Surface maps may reveal spatial asymmetries that are absent from a
selected cross-section, allowing some configurations with similar line
profiles to become distinguishable. Heat diffusion smooths spatial
temperature variations, while perfusion and boundary heat exchange
modify their transmission to the surface; uncertainties in anatomy and
physiological inputs may also obscure the additional spatial information.
A useful evaluation would therefore first compare corresponding line
profiles from matched two- and three-dimensional calculations, and then
assess the additional information contained in the surface maps. This
would separate changes in the physical temperature field from changes
in the observable used for classification. The comparisons should
evaluate peak temperature rise, spatial extent and asymmetry, pairwise
distances, equivalent-pair fractions, and the locations of tolerance
crossings.

The surface-map formulation and three-dimensional comparisons
remain proposed extensions. The present results provide a controlled
two-dimensional reference for assessing how realistic anatomy changes
surface-profile distances and equivalence boundaries at prescribed
observational tolerances, with mean-matched controls distinguishing
mean-perfusion effects from spatial-organization effects.
}

\section{Summary and Conclusion}
\label{sec:conclusion}

This study introduced a tolerance-aware, full-profile framework
that distinguishes surface-anomaly detectability from the
identifiability of intratumoral perfusion organization. The framework
was applied to a steady-state modified Pennes model of a
two-dimensional multilayer breast cross-section containing a finite
circular tumor with four idealized perfusion patterns: uniform,
rim-enhanced, necrotic-core, and anisotropic. The governing equation
was solved using conservative finite differences and BiCGSTAB with
an IncompleteLUT preconditioner. Mesh-refinement tests and an
analytical transfer-matrix benchmark provided numerical verification.

Distinct internal temperature fields produced substantially smaller
differences at the surface because conduction, perfusion-mediated
exchange, tissue layering, and boundary heat transfer smooth and
attenuate the thermal perturbation. For the representative
raw-perfusion configuration, the fractions of unique profile pairs
satisfying $d_{L^2}\leq\varepsilon_T$ were $2/6$, $5/6$, and $6/6$
at $\varepsilon_T=0.020$, $0.050$, and
$0.100~^\circ\mathrm{C}$, respectively. These tolerances apply to
processed surface-profile differences. Noise tests further showed
that equivalence probabilities depend on the spatial structure of
the assumed measurement uncertainty.

Mean matching reduced the rim-enhanced--uniform and
necrotic-core--uniform distances by approximately $97.7\%$ and
$99.0\%$, respectively, placing both below
$\varepsilon_T=0.020~^\circ\mathrm{C}$. The anisotropic--uniform
distance changed by only approximately $0.012\%$ and remained
slightly above this tolerance. Thus, spatial organization can retain
a surface signature even when mean perfusion is fixed. Increasing
tumor depth reduced profile separation, whereas increasing diameter
enhanced it; these competing effects persisted in the joint
depth--diameter map. Fat-layer thickness and outer-surface
deformation acted mainly as secondary modulators over the
investigated ranges.

Among the four parameter groups examined, heat generation had the
strongest positive influence on mean profile separation, whereas
increasing perfusion or surface convection generally reduced it.
For fixed geometry and perfusion-pattern shapes, a 300-point
Latin-hypercube analysis using independent uniform multipliers over
$[0.8,1.2]$ yielded a central 95\% empirical range
(2.5th--97.5th percentiles) of approximately
$0.019$--$0.044~^\circ\mathrm{C}$ for the mean pairwise distance.
At $\varepsilon_T=0.020~^\circ\mathrm{C}$, approximately $79\%$
of sampled configurations retained at least one equivalent pair.
These results demonstrate persistence of equivalence within the
prescribed perturbation ensemble; population-level frequencies
would require clinically informed joint input distributions.

The findings remain conditional forward-model predictions without
matched experimental or clinical validation. Transient extensions
should retain tissue heat storage and time-dependent perfusion,
metabolic heat generation, and boundary conditions to test whether
temporal responses distinguish statically equivalent configurations.
Three-dimensional extensions should incorporate realistic anatomy,
tumor morphology, and volume-based mean-matched controls, with
distances and equivalence boundaries reevaluated for the selected
observation protocol. Validation should progress from independently
characterized phantoms to clinical comparisons using anatomical and
perfusion-sensitive information to constrain inputs and thermograms
withheld from calibration and model selection. These studies should
assess complete surface profiles, pairwise distances, and equivalence
classifications at experimentally justified tolerances.

Physics-informed surrogate and digital-twin
approaches~\cite{ren2025pielm,liu2026digitaltwin} could accelerate
repeated forward calculations and uncertainty analyses. Their
profile and distance errors should be evaluated against independent
reference PDE solutions relative to the margins from the equivalence
thresholds, with classification-sensitive cases checked using the
PDE solver. The present framework provides a controlled basis for
understanding why a detectable static surface anomaly can remain
compatible with distinct internal perfusion organizations.

\section*{Author Contributions}
\textbf{R.~Muslim:} Main contributor, Conceptualization,  Methodology, Writing, Software, Formal analysis, Validation, Visualisation, Review \& editing. \textbf{R.~Fardela:} Writing,  Formal analysis, Validation, Review \& editing. \textbf{T.~A.~I.~Kusuma:} Writing, Formal analysis, Validation, Review \& editing.  All authors read and reviewed the paper.
\section*{Declaration of Interests}
The contributors declare that they have no apparent competing business or personal connections that might have appeared to have influenced the reported work.
\section*{Acknowledgments}
\textbf{R.~Muslim} was supported by the YST Program of the Asia Pacific Center for Theoretical Physics (APCTP), funded by the Science and Technology Promotion Fund and Lottery Fund of the Korean Government, and by the Management Talent Program of the National Research and Innovation Agency of Indonesia (BRIN).

\appendix
\section{Additional Khomsi and Lozano data}

\subsection{\label{app:A1}Tumor depth}

Table~\ref{tab:depth_khomsi_lozano} lists the numerical
surface-profile distances obtained from the tumor-depth sweep for the
Khomsi-based and Lozano-inspired parameter sets. The table reports the
distances between each heterogeneous perfusion case and the uniform case,
together with the mean pairwise distance among the considered perfusion
patterns.

\begin{table*}[pos = tbp]
\centering
\caption{Surface-profile distance $d_{L^2}$ as a function of tumor-center
depth for the Khomsi-based and Lozano-inspired parameter sets, computed using
a target grid spacing of $0.15~\mathrm{mm}$.}
\label{tab:depth_khomsi_lozano}

\begin{threeparttable}
\scriptsize
\setlength{\tabcolsep}{3.8pt}
\renewcommand{\arraystretch}{1.12}

\begin{tabular*}{\textwidth}{@{\extracolsep{\fill}}crrrrrrrr@{}}
\toprule
\multirow{2}{*}{\shortstack{Tumor-center\\ depth, $d_t$ (mm)}}
& \multicolumn{4}{c}{Khomsi-based, $d_{L^2}$ ($^\circ\mathrm{C}$)}
& \multicolumn{4}{c}{Lozano-inspired, $d_{L^2}$ ($^\circ\mathrm{C}$)} \\
\cmidrule(lr){2-5}
\cmidrule(lr){6-9}
& R--U
& N--U
& A--U
& MPD
& R--U
& N--U
& A--U
& MPD \\
\midrule
13.00 & 0.0485 & 0.0244 & 0.0307 & 0.0434
      & 0.1151 & 0.0153 & 0.0967 & 0.1085 \\
15.45 & 0.0447 & 0.0227 & 0.0275 & 0.0397
      & 0.0588 & 0.0077 & 0.0485 & 0.0549 \\
17.91 & 0.0410 & 0.0210 & 0.0251 & 0.0364
      & 0.0285 & 0.0037 & 0.0234 & 0.0266 \\
20.36 & 0.0372 & 0.0193 & 0.0229 & 0.0332
      & 0.0144 & 0.0019 & 0.0118 & 0.0134 \\
22.82 & 0.0334 & 0.0175 & 0.0207 & 0.0299
      & 0.0073 & 0.0009 & 0.0059 & 0.0068 \\
25.27 & 0.0293 & 0.0155 & 0.0184 & 0.0264
      & 0.0035 & 0.0005 & 0.0029 & 0.0033 \\
27.73 & 0.0257 & 0.0136 & 0.0163 & 0.0232
      & 0.0018 & 0.0002 & 0.0015 & 0.0017 \\
30.18 & 0.0221 & 0.0118 & 0.0142 & 0.0201
      & 0.0009 & 0.0001 & 0.0007 & 0.0008 \\
32.64 & 0.0190 & 0.0102 & 0.0123 & 0.0174
      & 0.0004 & $5.77\times10^{-5}$ & 0.0004 & 0.0004 \\
35.09 & 0.0162 & 0.0087 & 0.0107 & 0.0149
      & 0.0002 & $2.93\times10^{-5}$ & 0.0002 & 0.0002 \\
37.55 & 0.0136 & 0.0073 & 0.0091 & 0.0126
      & 0.0001 & $1.43\times10^{-5}$
      & $8.87\times10^{-5}$ & 0.0001 \\
40.00 & 0.0114 & 0.0062 & 0.0079 & 0.0107
      & $5.62\times10^{-5}$ & $7.29\times10^{-6}$
      & $4.51\times10^{-5}$ & $5.19\times10^{-5}$ \\
\bottomrule
\end{tabular*}

\begin{tablenotes}[flushleft]
\footnotesize
\item \textit{Note.} R--U, N--U, and A--U denote the
rim-enhanced--uniform, necrotic-core--uniform, and anisotropic--uniform
comparisons, respectively. MPD denotes the arithmetic mean of all six unique
pairwise distances among the uniform, rim-enhanced, necrotic-core, and
anisotropic perfusion patterns. The reported depths are nominal values; the
corresponding grid-aligned depths differ by at most $0.075~\mathrm{mm}$.
\end{tablenotes}

\end{threeparttable}
\end{table*}

\subsection{\label{app:A2}Tumor diameter}

Table~\ref{tab:size_khomsi_lozano} gives the corresponding
surface-profile distances from the tumor-diameter sweep. The values are shown
for both parameter sets and for the same perfusion comparisons used in the main
text.

\begin{table*}[pos=tbp]
\centering
\caption{Surface-profile distance $d_{L^2}$ as a function of tumor
diameter for the Khomsi-based and Lozano-inspired parameter sets,
computed using a target grid spacing of $0.15$~mm.}
\label{tab:size_khomsi_lozano}

\begin{threeparttable}
\scriptsize
\setlength{\tabcolsep}{3.8pt}
\renewcommand{\arraystretch}{1.12}

\begin{tabular*}{\textwidth}{
@{\extracolsep{\fill}}
crrrrrrrr
@{}
}
\toprule
\multirow{2}{*}{\shortstack{Tumor diameter,\\ $D$ (mm)}}
& \multicolumn{4}{c}{Khomsi-based, $d_{L^2}$ ($^\circ$C)}
& \multicolumn{4}{c}{Lozano-inspired, $d_{L^2}$ ($^\circ$C)} \\
\cmidrule(lr){2-5}
\cmidrule(lr){6-9}
& R--U & N--U & A--U & MPD
& R--U & N--U & A--U & MPD \\
\midrule
 6 & 0.009532 & 0.004314 & 0.001059 & 0.006003
   & 0.001513 & 0.000206 & 0.000529 & 0.001055 \\
 8 & 0.017978 & 0.008600 & 0.004869 & 0.012836
   & 0.003107 & 0.000429 & 0.001556 & 0.002403 \\
10 & 0.026347 & 0.013269 & 0.011666 & 0.021181
   & 0.005509 & 0.000743 & 0.003630 & 0.004693 \\
12 & 0.034603 & 0.018075 & 0.021432 & 0.030973
   & 0.009037 & 0.001176 & 0.007354 & 0.008391 \\
14 & 0.043124 & 0.023308 & 0.034139 & 0.042431
   & 0.014202 & 0.001740 & 0.013569 & 0.014175 \\
16 & 0.051552 & 0.028788 & 0.049340 & 0.055124
   & 0.021435 & 0.002449 & 0.023333 & 0.022790 \\
18 & 0.060193 & 0.034900 & 0.067224 & 0.069361
   & 0.031798 & 0.003289 & 0.038508 & 0.035699 \\
20 & 0.068967 & 0.041200 & 0.087192 & 0.084727
   & 0.046116 & 0.004321 & 0.060988 & 0.054268 \\
22 & 0.078033 & 0.047913 & 0.109538 & 0.101483
   & 0.066272 & 0.005551 & 0.094378 & 0.081242 \\
24 & 0.087209 & 0.054867 & 0.133927 & 0.119343
   & 0.094155 & 0.006998 & 0.142934 & 0.119699 \\
\bottomrule
\end{tabular*}

\begin{tablenotes}[flushleft]
\footnotesize
\item \textit{Note.}
R--U, N--U, and A--U denote the rim-enhanced--uniform,
necrotic-core--uniform, and anisotropic--uniform comparisons,
respectively. MPD denotes the arithmetic mean of all six unique pairwise
distances among the uniform, rim-enhanced, necrotic-core, and anisotropic
perfusion patterns.
\end{tablenotes}

\end{threeparttable}
\end{table*}

\section{One-dimensional layered benchmark}
\label{app:benchmark_1d}

This appendix gives a one-dimensional analytical benchmark for
the steady-state temperature field in healthy multilayer tissue. The benchmark
is used to check the numerical solver in a laterally uniform setting and to show
how conduction and perfusion attenuate thermal perturbations before they reach
the outer surface. The benchmark is only a verification case;
the tumor-bearing simulations in the main text are solved with the full
two-dimensional model.

Because the global coordinate $x$ in the main text is measured from the
chest-wall side toward the outer surface, we introduce here a separate local
depth coordinate $s$ measured inward from the outer surface. Thus, $s=0$
corresponds to the surface and $s=L$ to the posterior wall. The healthy tissue
is represented as an $M$-layer slab, where the $j$th layer occupies
$s_{j-1}<s<s_j$, with $s_0=0$, $s_M=L$, and thickness
$\Delta_j=s_j-s_{j-1}$. In this layer, the steady-state modified Pennes equation
is
\begin{equation}
k_j \frac{d^2 T_j}{ds^2}
-
\beta_j \left(T_j-T_a\right)
+
Q_j
=0,
\qquad
s_{j-1}<s<s_j .
\label{eq:app_1d_pennes}
\end{equation}
Here, $k_j$ is the thermal conductivity, $Q_j$ is the metabolic heat-generation
rate, and $\beta_j=\rho_b c_b\omega_j$ is the perfusion coefficient. For
$\beta_j>0$, we define
$\lambda_j=\sqrt{\beta_j/k_j}$ and
$T_j^{(p)}=T_a+Q_j/\beta_j$. The general solution is
\begin{equation}
\begin{aligned}
T_j(s)
={}& T_j^{(p)}
+A_j\cosh\!\left[\lambda_j(s-s_{j-1})\right]
\\
&+B_j\sinh\!\left[\lambda_j(s-s_{j-1})\right].
\end{aligned}
\label{eq:app_1d_general_solution}
\end{equation}
The associated screening length is
$\ell_j=\lambda_j^{-1}=\sqrt{k_j/\beta_j}$, showing that stronger perfusion
shortens the penetration length, whereas larger thermal conductivity increases
it.

At the outer surface, the Robin condition becomes
\begin{equation}
k_1 \frac{dT_1}{ds}(0)
=
h\left[T_1(0)-T_\infty\right],
\label{eq:app_1d_robin}
\end{equation}
where $h$ is the convective heat-transfer coefficient and $T_\infty$ is the
ambient temperature. The sign follows from the inward coordinate convention:
the outward normal at the surface points opposite to increasing $s$. At the
posterior wall, the temperature is prescribed as $T_M(L)=T_P$. At each internal
interface $s=s_j$, temperature and heat flux are continuous:
\begin{align}
T_j(s_j^-)&=T_{j+1}(s_j^+), \\
k_j\frac{dT_j}{ds}(s_j^-)
&=
k_{j+1}\frac{dT_{j+1}}{ds}(s_j^+).
\label{eq:app_1d_interface}
\end{align}

For implementation, we write the solution in transfer-matrix
form. It is convenient to introduce the flux-like variable
$\Phi_j(s)=k_j\,dT_j/ds$. Let $T_j^L=T_j(s_{j-1}^+)$ and
$\Phi_j^L=\Phi_j(s_{j-1}^+)$. With
$C_j=\cosh(\lambda_j\Delta_j)$ and
$S_j=\sinh(\lambda_j\Delta_j)$, the temperature and flux-like variable at the
right side of the layer are
\begin{align}
T_j^R
&=
T_j^{(p)}
+
C_j\left(T_j^L-T_j^{(p)}\right)
+
\frac{S_j}{k_j\lambda_j}\Phi_j^L,
\label{eq:app_1d_transfer_T}
\\
\Phi_j^R
&=
k_j\lambda_j S_j\left(T_j^L-T_j^{(p)}\right)
+
C_j\Phi_j^L .
\label{eq:app_1d_transfer_phi}
\end{align}
Equivalently,
\begin{equation}
\begin{pmatrix}
T_j^R \\
\Phi_j^R \\
1
\end{pmatrix}
=
\mathbf{P}_j
\begin{pmatrix}
T_j^L \\
\Phi_j^L \\
1
\end{pmatrix},
\label{eq:app_1d_affine_matrix}
\end{equation}
where
\begin{equation}
\mathbf{P}_j
=
\begin{pmatrix}
C_j
&
\dfrac{S_j}{k_j\lambda_j}
&
\left(1-C_j\right)T_j^{(p)}
\\[2mm]
k_j\lambda_j S_j
&
C_j
&
-k_j\lambda_j S_j T_j^{(p)}
\\[2mm]
0 & 0 & 1
\end{pmatrix}.
\label{eq:app_1d_layer_matrix}
\end{equation}

The complete transfer matrix from the surface to the posterior wall is
\begin{equation}
\mathbf{P}
=
\mathbf{P}_M\mathbf{P}_{M-1}\cdots\mathbf{P}_1 .
\label{eq:app_1d_total_matrix}
\end{equation}
Writing the first row of $\mathbf{P}$ as
$(P_{11},P_{12},P_{13})$, the posterior temperature can be expressed as $T_P=P_{11}T_s+P_{12}\Phi_s+P_{13}$,
where $T_s=T_1(0)$ and $\Phi_s=\Phi_1(0)$. Using the surface condition
$\Phi_s=h(T_s-T_\infty)$ gives the explicit surface temperature
\begin{equation}
T_s
=
\frac{1
}{
P_{11}
+
P_{12}h
} \left[T_P
-
P_{13}
+
P_{12}hT_\infty\right]
\label{eq:app_1d_surface_temperature}
\end{equation}
Once $T_s$ is known, $\Phi_s$ follows from the Robin condition, and the solution
in each layer can be reconstructed recursively from
Eqs.~\eqref{eq:app_1d_transfer_T} and \eqref{eq:app_1d_transfer_phi}. At the
left side of layer $j$, the constants in
Eq.~\eqref{eq:app_1d_general_solution} are $A_j=T_j^L-T_j^{(p)}$, and $B_j=\Phi_j^L/{k_j\lambda_j}.$

For a homogeneous single-layer slab, the solution reduces to
\begin{equation}
T(s)
=
T^{(p)}
+
A\cosh(\lambda s)
+
B\sinh(\lambda s),
\label{eq:app_single_layer_solution}
\end{equation}
with $\lambda=\sqrt{\beta/k}$ and $T^{(p)}=T_a+Q/\beta$. Let
$C=\cosh(\lambda L)$, $S=\sinh(\lambda L)$, and
$\alpha=h/(k\lambda)$. The surface condition gives
$B=\alpha(T^{(p)}+A-T_\infty)$, while the posterior-wall condition gives
$T_P=T^{(p)}+AC+BS$. Hence
\begin{align}\label{eq:app_single_layer_coefficients}
A
=&
\frac{1
}{
C+\alpha S
} \left[T_P-T^{(p)}
-
\alpha S\left(T^{(p)}-T_\infty\right) \right], \\
B
=&
\alpha\left(T^{(p)}-T_\infty+A\right).
\end{align}

Equations~\eqref{eq:app_single_layer_solution} and
\eqref{eq:app_single_layer_coefficients} provide a compact analytical check for
the limiting case of a homogeneous perfused slab. If $\beta_j=0$, the screened solution is replaced by the conduction solution
\begin{equation}
T_j(s)
=
A_j
+
B_j\left(s-s_{j-1}\right)
-
\frac{Q_j}{2k_j}
\left(s-s_{j-1}\right)^2 .
\label{eq:app_beta_zero_solution}
\end{equation}
This case is not required for the parameter sets used in the present study,
where the healthy layers have nonzero perfusion, but it provides a useful
consistency check.

To verify the numerical implementation quantitatively, the
laterally uniform healthy-tissue problem was solved using the same
depth-direction discretization employed in the full two-dimensional
model. The resulting numerical temperature profile was compared
pointwise with the analytical transfer-matrix solution derived above.
The surface-temperature error, depth-averaged $L^2$ error, maximum
pointwise error, observed convergence order, and normalized solver
residual are summarized in Table~\ref{tab:analytical_verification}.

\begin{table*}[t]
\centering
\caption{Numerical verification against the analytical transfer-matrix
solution for the healthy multilayer slab. The laterally uniform reduction
uses the same depth-direction stencil as the two-dimensional solver.
Errors are evaluated over the complete depth profile.}
\label{tab:analytical_verification}

\begingroup
\small
\renewcommand{\arraystretch}{1.15}
\setlength{\tabcolsep}{3pt}

\begin{tabular*}{\textwidth}{
@{\extracolsep{\fill}}
ccccccccc
@{}
}
\hline
$\Delta x$
&
$N_x$
&
$T_{s,\mathrm{ana}}$
&
$T_{s,\mathrm{num}}$
&
$E_s$
&
$E_{L^2}$
&
$E_{L^\infty}$
&
$p_{L^2}$
&
$r_{\mathrm{rel}}$
\\

(mm)
&
&
($^\circ$C)
&
($^\circ$C)
&
($^\circ$C)
&
($^\circ$C)
&
($^\circ$C)
&
&
\\
\hline

0.5000
& 131
& 33.302215
& 33.273345
& 0.028869
& 0.016606
& 0.055937
& --
& $4.36\times10^{-14}$
\\

0.2500
& 261
& 33.302215
& 33.289109
& 0.013106
& 0.005236
& 0.016988
& --
& $1.56\times10^{-13}$
\\

$0.2000^{\dagger}$
& 326
& 33.302215
& 33.294824
& 0.007391
& 0.004461
& 0.017876
& --
& $2.54\times10^{-13}$
\\

$0.1501^{*}$
& 434
& 33.302215
& 33.301666
& 0.000549
& 0.002223
& 0.010134
& --
& $4.78\times10^{-13}$
\\

$0.1000^{\dagger}$
& 651
& 33.302215
& 33.298521
& 0.003694
& 0.002230
& 0.008960
& 1.000
& $1.06\times10^{-12}$
\\

$0.0500^{\dagger}$
& 1301
& 33.302215
& 33.300368
& 0.001847
& 0.001115
& 0.004486
& 1.000
& $4.26\times10^{-12}$
\\

\hline
\end{tabular*}

\vspace{3pt}

\begin{minipage}{\textwidth}
\footnotesize
\textit{Notes:}
$E_s=|T_{s,\mathrm{num}}-T_{s,\mathrm{ana}}|$.
The $L^2$ error is the depth-averaged root-mean-square temperature
error, whereas $E_{L^\infty}$ is the maximum pointwise error.
The superscript ${*}$ identifies the mesh used in the main simulations.
The superscript ${\dagger}$ identifies meshes for which all material
interfaces coincide with grid nodes. The observed order $p_{L^2}$ is
calculated only between successive interface-aligned meshes.
Non-aligned meshes are retained to assess the discretization used in
the main simulations.
\end{minipage}

\endgroup
\end{table*}

This benchmark makes the screening mechanism explicit.
Thermal perturbations generated at depth are attenuated by conduction,
perfusion, surface convection, and the thermal resistance of the overlying
layers. The same mechanism operates in the full
two-dimensional tumor model, where it reduces the surface differences between
distinct intratumoral perfusion patterns.

\section{Closest-pair mesh convergence}
\label{app:mesh_closest_pair}

This appendix reports an additional mesh-convergence test for
the closest pair of surface profiles in the representative raw-perfusion case.
This check is useful because the equivalence analysis depends on small
differences between surface temperature-rise profiles. The numerical values of
the profile distance, the errors relative to the finest mesh, and the maximum
solver residuals are summarized in Table~\ref{tab:mesh_convergence_rim_nec}.

\begin{table}[tbp]
\centering
\caption{Additional mesh-convergence test for the closest surface-profile pair.}
\label{tab:mesh_convergence_rim_nec}

\begin{threeparttable}
\scriptsize
\setlength{\tabcolsep}{4pt}
\renewcommand{\arraystretch}{1.12}

\begin{tabular*}{\linewidth}{@{\extracolsep{\fill}}lccccc@{}}
\toprule
$h_{\mathrm{target}}$
& Grid
& $d_{L^2}^{(\mathrm{R-N})}$
& Abs. error
& Rel. error
& Max. residual \\
$(\mathrm{mm})$
&
& $(^\circ\mathrm{C})$
& $(^\circ\mathrm{C})$
& $(\%)$
& \\
\midrule
1.000 & $66\times121$   & 0.015408 & 0.001195 & 7.199 & $7.13\times10^{-14}$ \\
0.750 & $88\times161$   & 0.017127 & 0.000524 & 3.157 & $1.31\times10^{-13}$ \\
0.500 & $131\times241$  & 0.017411 & 0.000808 & 4.864 & $3.55\times10^{-13}$ \\
0.400 & $164\times301$  & 0.016462 & 0.000141 & 0.852 & $6.03\times10^{-13}$ \\
0.300 & $218\times401$  & 0.016615 & 0.000012 & 0.070 & $1.31\times10^{-12}$ \\
0.250 & $261\times481$  & 0.016493 & 0.000110 & 0.665 & $8.08\times10^{-12}$ \\
0.200 & $326\times601$  & 0.016741 & 0.000138 & 0.830 & $3.49\times10^{-12}$ \\
0.150 & $434\times801$  & 0.016529 & 0.000074 & 0.449 & $6.66\times10^{-12}$ \\
0.125 & $521\times961$  & 0.016695 & 0.000092 & 0.551 & $1.07\times10^{-11}$ \\
0.100 & $651\times1201$ & 0.016603 & 0.000000 & 0.000 & $1.81\times10^{-11}$ \\
\bottomrule
\end{tabular*}

\begin{tablenotes}[flushleft]
\footnotesize
\item \textit{Note.} The test is performed for the representative Khomsi-based
case using the rim-enhanced--necrotic-core pair, which is the closest
off-diagonal pair in the raw equivalence analysis. The quantity
$d_{L^2}^{(\mathrm{R-N})}$ denotes the profile distance between the
rim-enhanced and necrotic-core surface temperature-rise profiles. Absolute
and relative errors are computed with respect to the finest reference mesh,
$h_{\mathrm{target}}=0.100~\mathrm{mm}$. For the mesh used in the main
simulations, $h_{\mathrm{target}}=0.150~\mathrm{mm}$, the distance is
$d_{L^2}^{(\mathrm{R-N})}=0.016529~^\circ\mathrm{C}$, with an absolute error of
$7.4\times10^{-5}~^\circ\mathrm{C}$ and a relative error of $0.449\%$. This
deviation is far smaller than the smallest observational tolerance considered
in the study, $\varepsilon_T=0.020~^\circ\mathrm{C}$.
\end{tablenotes}

\end{threeparttable}
\end{table}

\section{Green's-function view of thermal screening}
\label{app:green}

This appendix gives a Green's-function interpretation of how
a tumor-induced thermal perturbation is transmitted from the tissue interior to
the outer surface. The derivation is not meant to replace the full
two-dimensional numerical solution used in the main text. It
is included to explain why internal differences among intratumoral perfusion
patterns can be smoothed and attenuated before they appear in the surface
thermal signature.

We decompose the total temperature field as
\begin{equation}
T(\mathbf r)=T_{\mathrm{bg}}(\mathbf r)+u(\mathbf r),
\label{eq:app_temperature_decomposition}
\end{equation}
where $\mathbf r=(x,y)$, $T_{\mathrm{bg}}$ is the healthy-background solution,
and $u$ is the tumor-induced perturbation. In a laterally uniform flat geometry,
$T_{\mathrm{bg}}$ reduces to the one-dimensional layered solution derived in
Appendix~\ref{app:benchmark_1d}. The steady-state modified Pennes equation is
written as
\begin{equation}
\nabla\cdot\!\left(k\nabla T\right)
-
\beta\left(T-T_a\right)
+
Q
=
0 ,
\label{eq:app_full_steady_pennes}
\end{equation}
where $\beta(\mathbf r)=\rho_b c_b\,\omega(\mathbf r)$. In the healthy background, the coefficients
are denoted by $k_{\mathrm{bg}}$, $\beta_{\mathrm{bg}}$, and
$Q_{\mathrm{bg}}$. Inside the tumor, they are perturbed according to
$k=k_{\mathrm{bg}}+\delta k$,
$\beta=\beta_{\mathrm{bg}}+\delta\beta$, and
$Q=Q_{\mathrm{bg}}+\delta Q$, with the contrast terms nonzero only in the tumor
domain $\Omega_t$.

Substituting $T=T_{\mathrm{bg}}+u$ into
Eq.~\eqref{eq:app_full_steady_pennes} and subtracting the healthy-background
equation gives
\begin{equation}
\nabla\cdot\!\left(k_{\mathrm{bg}}\nabla u\right)
-
\beta_{\mathrm{bg}}u
=
-\mathcal{S}[u],
\label{eq:app_exact_perturbation}
\end{equation}
where
\begin{align}
\mathcal{S}[u]
=&
\nabla\cdot\!\left(\delta k\nabla T_{\mathrm{bg}}\right)
+
\nabla\cdot\!\left(\delta k\nabla u\right)
-
\delta\beta\left(T_{\mathrm{bg}}+u-T_a\right)
\nonumber \\
&+
\delta Q .
\end{align}
Thus, the tumor perturbs the thermal field through conductivity,
perfusion, and metabolic heat-generation contrasts.
Equation~\eqref{eq:app_exact_perturbation} is an exact reformulation
of the governing bioheat equation. Its effective source
$\mathcal{S}[u]$ remains implicit because it depends on the unknown
perturbation $u$. With the coefficient contrasts extended by zero
outside $\Omega_t$, the conductivity contrast $\delta k$ can jump
at the tumor boundary. The divergence terms in $\mathcal{S}[u]$
are therefore interpreted in the weak (distributional) sense,
including the interface contributions associated with this jump.
The same interpretation applies to the leading-order source
introduced below.

To expose the filtering structure of the solution, we retain only the
leading-order contrast terms and neglect the feedback terms proportional to $u$
inside the source. This gives
\begin{equation}
\mathcal{S}_0
=
\nabla\cdot\!\left(\delta k\nabla T_{\mathrm{bg}}\right)
-
\delta\beta\left(T_{\mathrm{bg}}-T_a\right)
+
\delta Q ,
\label{eq:app_leading_order_source}
\end{equation}
and the linearized perturbation equation
\begin{equation}
\nabla\cdot\!\left(k_{\mathrm{bg}}\nabla u\right)
-
\beta_{\mathrm{bg}}u
=
-\mathcal{S}_0 .
\label{eq:app_linearized_perturbation}
\end{equation}
This linearized form is used only for interpretation. In the
numerical simulations discussed in the main text, the full spatially varying
coefficients are retained.

Let
\begin{equation}
\mathcal{L}_{\mathrm{bg}}u
=
\nabla\cdot(k_{\mathrm{bg}}\nabla u)-\beta_{\mathrm{bg}}u
\label{eq:app_background_operator}
\end{equation}
be the background operator. The Green's function
$G(\mathbf r,\mathbf r')$ is defined by
\begin{equation}
\mathcal{L}_{\mathrm{bg}}G(\mathbf r,\mathbf r')
=
-\delta(\mathbf r-\mathbf r')
\label{eq:app_green_def}
\end{equation}
with homogeneous Robin, Dirichlet, and Neumann conditions at the
outer surface, posterior wall, and lateral boundaries, respectively.
The same external conditions hold for $u$. Across background tissue
interfaces, $G$ and its normal flux evaluated using $k_{\mathrm{bg}}$
are continuous. At the tumor boundary, continuity applies to total
temperature and total normal heat flux, with conductivity-contrast
interface contributions included in the weak source. The
leading-order perturbation can then be written in the following
weak Green's-function form:
\begin{equation}
u(\mathbf r)
=
\iint_{\Omega_t}
G(\mathbf r,\mathbf r')
\mathcal{S}_0(\mathbf r')
\,d\mathbf r' .
\label{eq:app_green_representation}
\end{equation}
This expression shows that the surface perturbation is not a
direct image of the internal source distribution. It is a weighted projection of
the tumor-induced source through the Green's function of the surrounding
tissue.

The filtering property is clearest in a homogeneous effective medium. For
constant effective parameters $k_{\mathrm{eff}}$ and $\beta_{\mathrm{eff}}$, the
screening length is
\begin{equation}
\ell_{\mathrm{eff}}
=
\sqrt{\frac{k_{\mathrm{eff}}}{\beta_{\mathrm{eff}}}} .
\label{eq:app_effective_screening_length}
\end{equation}
In an unbounded two-dimensional medium,
\begin{equation}
G_\infty(r)
=
\frac{1}{2\pi k_{\mathrm{eff}}}
K_0\!\left(\frac{r}{\ell_{\mathrm{eff}}}\right),
\label{eq:app_infinite_green}
\end{equation}
where $r=|\mathbf r-\mathbf r'|$ and $K_0$ is the modified Bessel function of
the second kind. This form illustrates the role of perfusion: increasing
$\beta_{\mathrm{eff}}$ decreases $\ell_{\mathrm{eff}}$ and strengthens thermal
screening, whereas increasing $k_{\mathrm{eff}}$ allows perturbations to spread
farther.

For a homogeneous slab with finite depth $0<s<L$ and unbounded
lateral extent $y\in\mathbb{R}$, let $s$ denote depth measured
inward from the outer surface. Thus, $s=0$ denotes the outer
surface and $s=L$ the posterior wall. This local coordinate is
distinct from the global coordinate $x$ used in the main text.
The laterally unbounded slab provides an interpretive approximation
to the finite-width computational domain. The homogeneous boundary
conditions for the perturbation are
\begin{equation}
k_{\mathrm{eff}}\frac{\partial u}{\partial s}=h u
\quad \text{at } s=0,
\qquad
u=0
\quad \text{at } s=L .
\label{eq:app_slab_boundary_conditions}
\end{equation}
Using separation of variables, the Green's function can be expressed as
\begin{equation}
G\bigl((s,y),(s',y')\bigr)
=
\sum_{n=1}^{\infty}
\frac{
\phi_n(s)\phi_n(s')
}{
2k_{\mathrm{eff}}\gamma_n N_n
}
\exp\!\left[-\gamma_n |y-y'|\right],
\label{eq:app_slab_green}
\end{equation}
where $\gamma_n=\sqrt{\mu_n^2+\ell_{\mathrm{eff}}^{-2}}$ and $N_n=\int_0^L \phi_n^2(s)\,ds $. A convenient choice of eigenfunctions is $\phi_n(s)=\sin[\mu_n(L-s)]$,
which satisfies the posterior Dirichlet condition. The Robin condition at the
outer surface gives
\begin{equation}
k_{\mathrm{eff}}\mu_n\cos(\mu_n L)
+
h\sin(\mu_n L)
=
0 .
\label{eq:app_eigenvalue_equation}
\end{equation}
The sign follows from the inward local-coordinate convention: the outward normal
at the surface points opposite to increasing $s$, so the homogeneous Robin
condition becomes
$k_{\mathrm{eff}}\partial u/\partial s=h u$ at $s=0$.

The modal representation above separates the dependence of the
Green's function on depth and lateral position. The factors
$\phi_n(s)\phi_n(s')$ describe the dependence on observation
and source depth, whereas
$\exp[-\gamma_n|y-y'|]$ describes attenuation with lateral
source--observation separation. Larger $\gamma_n$ therefore
correspond to stronger lateral attenuation of the associated
modes. The suppression of fine lateral features during
propagation toward the surface is made explicit by the
Fourier representation in
Appendix~\ref{app:scaling_depth_size}. In the homogeneous
half-space approximation, its depth-dependent factor
$\exp[-d\sqrt{\kappa^2+\ell_{\mathrm{eff}}^{-2}}]$
shows that higher lateral spatial frequencies are more
strongly attenuated as source depth increases.

This Green's-function view explains the thermal-signature
equivalence observed in the main text. Distinct perfusion patterns can
generate different internal source distributions $\mathcal{S}_0(\mathbf r')$,
but the measured surface profile is obtained only after these sources are
integrated against a smoothing and attenuating kernel. Increasing tumor depth
strengthens this filtering, whereas increasing tumor size increases the spatial
support of the effective source. Thus, small or deep tumors are more likely to
yield similar surface profiles, while larger or shallower tumors are more
likely to remain distinguishable. The same mechanism underlies the
profile-distance trends analyzed in
Appendix~\ref{app:scaling_depth_size}.

\section{Scaling with tumor depth and size}
\label{app:scaling_depth_size}
The Green's-function representation in Appendix~\ref{app:green}
provides an interpretation of the simulated dependence of the
surface-profile distance $d_{L^2}$ on tumor depth and size.
We use the linearized background problem to identify the roles
of source strength, spatial organization, and propagation through
the surrounding tissue. In the local flat-slab coordinates,
$\mathbf q=(s,y)$, the outer surface is at $s=0$ and the posterior
wall at $s=L$. A tumor of radius $R$ and diameter $D=2R$ is
centered at $\mathbf q_0=(d,y_0)$. The inward depth coordinate
$s$ is distinct from the global coordinate $x$ used in the main
text, which is measured from the chest-wall side.

To examine the leading size dependence, consider a localized
effective source of the form
\begin{equation}
\mathcal{S}_0(\mathbf q)
=
\bar S\,
\psi\left(\frac{\mathbf q-\mathbf q_0}{R}\right),
\label{eq:app_localized_source}
\end{equation}
where $\bar S$ is a characteristic source amplitude and $\psi$
is a dimensionless shape function with compact support.
This is an idealized representation of the effective forcing
defined in Appendix~\ref{app:green}. The source amplitude and
rescaled shape are held fixed when deriving the leading
area-based size estimate; their possible variation in the
full tumor problem is considered below.

The integrated source strength is
\begin{equation}
M
=
\iint_{\Omega_t}
\mathcal{S}_0(\mathbf q)\,\mathrm{d}\mathbf q
=
\bar S R^2 m_0,
\label{eq:app_source_moment}
\end{equation}
where
$m_0=\int_{\mathbb{R}^2}\psi(\boldsymbol{\xi})
\,\mathrm{d}\boldsymbol{\xi}$.
The factor $R^2$ follows from the area element of the
two-dimensional cross-section. Accordingly, $M$ has units
of effective source strength per unit out-of-plane length.
A nonzero $m_0$ permits a leading contribution controlled
by this integrated source moment.

Within the linearized background problem, the surface
perturbation is
\begin{equation}
u_s(y)
=
\iint_{\Omega_t}
G\bigl((0,y),\mathbf q'\bigr)
\mathcal{S}_0(\mathbf q')\,\mathrm{d}\mathbf q'.
\label{eq:app_surface_green_integral}
\end{equation}
If the source is small compared with the spatial scales over
which the kernel varies across the tumor, expansion about
$\mathbf q_0$ gives the leading approximation
\begin{equation}
u_s(y)
\approx
M\,G_s(y-y_0;d),
\label{eq:app_surface_profile_approx}
\end{equation}
where
$G_s(y-y_0;d)=G\bigl((0,y),\mathbf q_0\bigr)$.
This approximation requires the integrated-source contribution
to dominate the terms involving higher spatial moments.

The depth dependence can be made explicit for a laterally
unbounded, homogeneous slab with conductivity $k_{\mathrm{eff}}$,
perfusion coefficient $\beta_{\mathrm{eff}}>0$, and screening
length
$\ell_{\mathrm{eff}}=\sqrt{k_{\mathrm{eff}}/\beta_{\mathrm{eff}}}$,
as in Eq.~\eqref{eq:app_effective_screening_length}.
Impose the homogeneous perturbation conditions
$k_{\mathrm{eff}}\partial_s u=hu$ at $s=0$ and $u=0$ at $s=L$.
Using the lateral Fourier transform
$\widehat G_s(\kappa;d)
=\int_{\mathbb{R}}G_s(\xi;d)e^{-i\kappa\xi}\,\mathrm{d}\xi$,
the transformed Green's equation is
$k_{\mathrm{eff}}(\partial_s^2-\lambda^2)\widehat G
=-\delta(s-d)$, where
$\lambda=\sqrt{\kappa^2+\ell_{\mathrm{eff}}^{-2}}$.
Its surface value is
\begin{equation}
\widehat G_s(\kappa;d)
=
\frac{\sinh[\lambda(L-d)]}
{k_{\mathrm{eff}}\lambda\cosh(\lambda L)
+h\sinh(\lambda L)}.
\label{eq:app_surface_signal_scaling}
\end{equation}
In the half-space limit $L\to\infty$, this becomes
$\widehat G_s(\kappa;d)
=e^{-\lambda d}/(k_{\mathrm{eff}}\lambda+h)$.
Thus, increasing depth attenuates the lateral Fourier
components, with higher spatial wavenumbers undergoing
stronger attenuation.

This depth dependence should be distinguished from the
factors $\exp[-\gamma_n|y-y'|]$ in the modal representation
of Appendix~\ref{app:green}. There, $\gamma_n$ governs
lateral attenuation, while the source-depth dependence is
carried by the eigenfunctions evaluated at $s'=d$.
The corresponding finite-slab depth response is given by
Eq.~\eqref{eq:app_surface_signal_scaling}. For a layered
background or a deformed surface, the appropriate kernel
must incorporate the material interfaces and boundary
geometry.

We now compare two perfusion configurations, $\alpha$ and
$\beta$, with the same tumor geometry and location. Their
surface difference is
\begin{equation}
\Delta u_s^{(\alpha\beta)}(y)
=
u_s^{(\alpha)}(y)-u_s^{(\beta)}(y).
\label{eq:app_surface_difference}
\end{equation}
Let
$\Delta M^{(\alpha\beta)}=M^{(\alpha)}-M^{(\beta)}$.
When this integrated source contrast is nonzero and dominates
the higher-moment contributions, the common-background
approximation gives
$\Delta u_s^{(\alpha\beta)}
\approx\Delta M^{(\alpha\beta)}G_s$.
The profile distance consequently satisfies
\begin{equation}
\begin{aligned}
d_{L^2}^{(\alpha\beta)}
&=
\left[
\frac{1}{L_\Gamma}
\int_{\Gamma_s}
\left|\Delta u_s^{(\alpha\beta)}(y)\right|^2
\,\mathrm{d}y
\right]^{1/2}
\approx
\left|\Delta M^{(\alpha\beta)}\right|
\mathcal K_\Gamma(d),
\end{aligned}
\label{eq:app_dL2_definition_scaling}
\end{equation}
where
\begin{equation}
\mathcal K_\Gamma(d)
=
\left[
\frac{1}{L_\Gamma}
\int_{\Gamma_s}
\left|G_s(y-y_0;d)\right|^2
\,\mathrm{d}y
\right]^{1/2}.    
\end{equation}
Here, $\Gamma_s$ is the lateral comparison interval and
$L_\Gamma=\int_{\Gamma_s}\mathrm{d}y$. The interval and
lateral source position are held fixed when comparing
depths and sizes. The kernel norm $\mathcal K_\Gamma(d)$
retains changes in both the amplitude and spatial extent
of the surface response.

At fixed depth, suppose that the source amplitudes and
rescaled shapes remain independent of $R$, with a nonzero,
dominant integrated contrast. Then
$\Delta M^{(\alpha\beta)}=C_{\alpha\beta}(d)R^2$, yielding
\begin{equation}
d_{L^2}^{(\alpha\beta)}(R)
\propto R^2
\qquad
\text{at fixed depth}.
\label{eq:app_size_scaling}
\end{equation}
Since $D=2R$, the same approximation gives
$d_{L^2}^{(\alpha\beta)}(D)\propto D^2$.
Combining the source and propagation contributions gives
\begin{equation}
d_{L^2}^{(\alpha\beta)}(R,d)
\approx
|C_{\alpha\beta}(d)|R^2\mathcal K_\Gamma(d).
\label{eq:app_joint_scaling}
\end{equation}
The coefficient $C_{\alpha\beta}(d)$ relates the integrated
effective source contrast to the squared tumor radius through
$\Delta M^{(\alpha\beta)}=C_{\alpha\beta}(d)R^2$.
Its possible depth dependence reflects changes in the local
background thermal state.

In the homogeneous half-space approximation, at sufficiently
large $d/\ell_{\mathrm{eff}}$, the surviving low-wavenumber
components give a leading screening factor
$\exp(-d/\ell_{\mathrm{eff}})$.
Accordingly, the kernel norm can be expressed asymptotically as
$\mathcal K_\Gamma(d)\sim
\mathcal P_\Gamma(d)\exp(-d/\ell_{\mathrm{eff}})$,
where $\mathcal P_\Gamma(d)$ retains the algebraic spreading
and observation-window dependence, together with the
dependence on the fixed thermal and boundary parameters.
Equation~\eqref{eq:app_joint_scaling} therefore contains the
familiar $R^2\exp(-d/\ell_{\mathrm{eff}})$ factor only under
these additional assumptions and with its remaining
prefactors retained. The finite posterior boundary and
tissue layering can modify this depth dependence.

Cancellation of the integrated source contrast changes the
leading size estimate. If $\Delta M^{(\alpha\beta)}$ vanishes
or becomes too small to dominate, higher spatial moments
and derivatives of the kernel determine the surface
difference. For a size-independent volumetric source
contrast in two dimensions, dipole and quadrupole moments
scale as $R^3$ and $R^4$, respectively, although symmetry
and the observation geometry determine which contribution
is present. In the perfusion comparisons considered here,
the common conductivity and metabolic-source contrasts
cancel from the leading source difference in
Appendix~\ref{app:green}. The remaining perfusion contrast
is weighted by $T_{\mathrm{bg}}-T_a$, so equality of the
mean perfusion does not generally imply equality of the
effective source moments.

These considerations explain why the finite-range exponents
in Figure~\ref{fig:fig10_new} and
Table~\ref{tab:size_power_law_fit} need not equal two.
The simulated distances depend on source cancellation,
perfusion organization, variation of the temperature field
within the tumor, tissue interfaces, and changes in the
kernel across a finite inclusion. The full numerical
solutions retain these effects beyond the localized,
linearized approximation used here. The fitted exponents
therefore summarize the investigated diameter range.
Within that range, the numerical results show increasing
surface-profile separation with tumor size and decreasing
separation with depth, consistent with the competing roles
of source extent and thermal screening.

\section{Empirical power-law fit statistics}
\label{app:power_law_fit}

The dashed lines in Figure~\ref{fig:fig10_new}(b,d) are empirical
power-law fits used to summarize the tumor-diameter dependence over the
simulated range $D=6$--$24~\mathrm{mm}$. The fits were performed in log--log
space using
\begin{equation}
\log d_{L^2}=\log A+b\log D .
\label{eq:app_power_law_fit}
\end{equation}
Table~\ref{tab:size_power_law_fit} reports the fitted
exponents, 95\% confidence intervals, coefficients of determination, and
log-space residual errors. These values describe the quality of the empirical
regression over the simulated diameter range only. They should
not be interpreted as experimental confidence intervals or as evidence of
universal asymptotic scaling.

\begin{table*}[pos = tbp]
\centering
\caption{Empirical power-law fit statistics for the tumor-diameter
dependence shown in Figure~\ref{fig:fig10_new}, computed using a target
grid spacing of $0.15~\mathrm{mm}$.}
\label{tab:size_power_law_fit}

\begin{threeparttable}
\scriptsize
\setlength{\tabcolsep}{4pt}
\renewcommand{\arraystretch}{1.12}

\begin{tabular*}{\textwidth}{@{\extracolsep{\fill}}llcccc@{}}
\toprule
Parameter set
& Comparison
& Fit range
& $b$ (95\% CI)
& $R^2_{\log}$
& RMSE$_{\log}$ \\
\midrule

Khomsi-based
& Rim--Uniform
& $6$--$24~\mathrm{mm}$
& $1.540$ [$1.410$, $1.670$]
& 0.9894
& 0.0690 \\

Khomsi-based
& Necrotic-core--Uniform
& $6$--$24~\mathrm{mm}$
& $1.779$ [$1.662$, $1.896$]
& 0.9936
& 0.0618 \\

Khomsi-based
& Anisotropic--Uniform
& $6$--$24~\mathrm{mm}$
& $3.341$ [$2.933$, $3.750$]
& 0.9780
& 0.2166 \\

Khomsi-based
& Mean pairwise
& $6$--$24~\mathrm{mm}$
& $2.115$ [$2.011$, $2.219$]
& 0.9964
& 0.0551 \\

Lozano-inspired
& Rim--Uniform
& $6$--$24~\mathrm{mm}$
& $2.965$ [$2.770$, $3.161$]
& 0.9935
& 0.1034 \\

Lozano-inspired
& Necrotic-core--Uniform
& $6$--$24~\mathrm{mm}$
& $2.536$ [$2.519$, $2.552$]
& 0.9999
& 0.0087 \\

Lozano-inspired
& Anisotropic--Uniform
& $6$--$24~\mathrm{mm}$
& $4.023$ [$3.902$, $4.144$]
& 0.9987
& 0.0639 \\

Lozano-inspired
& Mean pairwise
& $6$--$24~\mathrm{mm}$
& $3.410$ [$3.207$, $3.612$]
& 0.9947
& 0.1073 \\
\bottomrule
\end{tabular*}

\begin{tablenotes}[flushleft]
\footnotesize
\item \textit{Note.} The fits are obtained from linear regression in
log--log space using $d_{L^2}(D)=AD^b$. The quantities
$R^2_{\log}$ and RMSE$_{\log}$ are also evaluated in log--log space.
Each regression uses $n=10$ simulated tumor diameters and eight degrees
of freedom. The reported confidence intervals quantify the uncertainty
of the empirical regression over the investigated diameter range; they
should not be interpreted as experimental confidence intervals or as
evidence of universal asymptotic scaling. CI denotes confidence interval, and RMSE denotes root-mean-square error.
\end{tablenotes}

\end{threeparttable}
\end{table*}

\printcredits

\bibliographystyle{model1-num-names}

\bibliography{cas-refs}





\end{document}